\documentclass[final]{iasart}           

\usepackage[pdftex]{graphicx}
\graphicspath{{./Fig/}}
\DeclareGraphicsExtensions{.pdf,.jpg,.png}

\usepackage{amssymb}

\usepackage{color}
\usepackage[utf8]{inputenc}
\usepackage[T1]{fontenc}

\usepackage{algorithm}
\usepackage{algorithmic}

\usepackage[below]{placeins}

\usepackage{latexsym}
\usepackage{lscape}

\usepackage{amsmath}
\usepackage{amsfonts}
\usepackage{amsthm}

\usepackage{url}

\newcommand{\Real}{\mathbb R}

\newcommand{\f}{\mathbf{f}_{\lambda}}
\newcommand{\fr}{\widehat{\mathbf{f}}_{\lambda}}

\newcommand{\ca}{\mathbf{c}^{\mathbf{f}}_{\alpha}}

\newcommand{\ctrain}{\mathbf{c}^{train}_{\be}}
\newcommand{\al}{\alpha}
\newcommand{\be}{\beta}
\newcommand{\la}{\lambda}

\newcommand{\ctab}{\hspace{0.025\columnwidth}}

\DeclareMathOperator*{\argmin}{argmin}

\title{Multivariate Mathematical Morphology for DCE-MRI image analysis
in angiogenesis studies} \shorttitle{Multivariate Mathematical
Morphology for DCE-MRI images} \shortauthors{Noyel G \etal}

\author[1]{Guillaume Noyel}
\author[1]{Jesus Angulo}
\author[1]{Dominique Jeulin}
\author[2]{Daniel Balvay}
\author[2]{Charles Andr\'e Cuenod}

\email{guillaume.noyel@mines-paris.org,
jesus.angulo@mines-paristech.fr,
dominique.jeulin@mines-paristech.fr, daniel.balvay@inserm.fr,
charles-andre.cuenod@egp.aphp.fr}

\affiliation[1]{MINES ParisTech, Centre de Morphologie
Math\'ematique, Math\'ematiques et Syst\`emes - 77305 Fontainebleau,
France}

\affiliation[2]{LRI-PARCC U970 Paris - Descartes University - APHP,
HEGP, Service de Radiologie - Paris, France}

\abstract{We propose a new computer aided detection framework for
tumours acquired on DCE-MRI (Dynamic Contrast Enhanced Magnetic
Resonance Imaging) series on small animals. In this approach we
consider DCE-MRI series as multivariate images. A full multivariate
segmentation method based on dimensionality reduction, noise
filtering, supervised classification and stochastic watershed is
explained and tested on several data sets. The two main key-points
introduced in this paper are noise reduction preserving contours and
spatio temporal segmentation by stochastic watershed. Noise
reduction is performed in a special way that selects factorial axes
of Factor Correspondence Analysis in order to preserves contours.
Then a spatio-temporal approach based on stochastic watershed is
used to segment tumours. The results obtained are in accordance with
the diagnosis of the medical doctors.}


\begin{document}
\begin{paper}

\section{Introduction}
\label{sec:intro}


DCE-MRI (Dynamic Contrast Enhanced MRI) time series is a medical
imaging modality useful to characterise the process of tissue
vascularisation. As tumours correspond to zones of angiogenesis,
where the vascularisation is increased, DCE-MRI series are a
convenient way of identifying or characterising potential tumours.
Hence, DCE-MRI provides an additive and functional information to
more current morphological images. Due to the increasing amounts of
images, the creation of tools to assist medical doctors is of great
interest for the analysis of these images and the detection of
candidate tumour regions. In this paper, our goal is to supply an
automatic tool to help users to localise tumours, based on this
vascular information. An automatic process is required to limit
operator variability during tumour delineation. In current
morphological images such question is very challenging because,
currently, gray level differences between tumour and adjacent
tissues are not sufficient for a confident and stable separation
between tissues. Conversely DCE imaging provides potential richer
information to determine this difference. However this information
is distributed among all the sequence of images, requiring
mathematical development. Here we show the results on DCE-MRI images
of recent developments in mathematical morphology segmentation for
multivariate images. Our developments are particularly useful for a
visual evaluation of the tumour extension, but also as a
pre-processing step before a pharmacokinetic modeling, evaluated
then on images with less noise. The corresponding evaluation of
functional parameters such as blood flow, blood volume,
permeably-surface of capillaries, etc~\citep{Sourbron_2012} should
be improved accordingly.

The considered images are DCE-MRI series of $L= 512$ channels of
size $N\times N$, $N=128$, pixels acquired at a regular step of
$t=1$ second, in time, on mice presenting
tumours~\citep{Balvay_2005}. In DCE-MRI imagery, tumours are regions
corresponding to the accumulation of the contrast product. This
accumulation is characterised by an increasing kinetics of the
temporal signal for each pixel of the tumour. Our aim is to show the
potential of a new method for segmentation purpose in medical
imagery. The tests, made on 25 different series, were used to
develop the new method based on hyperspectral mathematical
morphology. The objective is to achieve computed aided diagnosis of
tumours.

From an image processing point of view, DCE-MRI images are time
series which fulfil a fundamental temporal coherence hypothesis: at
each time step, the MRI image is acquired on the same object of
interest and the images are registered, i.e. any pixel has the same
spatial position for every time step (i.e. for every image channel).
In our case, in each experiment the channels are registered.

Due to these assumptions, the sequences of DCE-MRI images may be
interpreted as multivariate, i.e., hyperspectral images, of the time
evolution of the scene observed. Some images of the sequence,
interpreted as channels of a series, are shown in
figure~\ref{intro:mouse_5_channels}. The tumours (PC3 - human
prostatic) are characterised by an hyper-vascularised ring and an
hypoxic centre or even a necrotic centre which is due to the
distance of the afferent vessels from the centre. This vessels are
repulsed from the centre by tumoural proliferation. More details
about acquisition conditions are given on the appendix section.

\begin{figure}[h]
\centering
\begin{tabular}{@{}c@{}}
\begin{tabular}{ccc}
    \includegraphics[width=0.3\columnwidth]{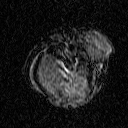}&
    \includegraphics[width=0.3\columnwidth]{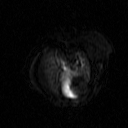}&
    \includegraphics[width=0.3\columnwidth]{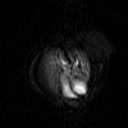}\\
    \footnotesize{$f_{\la_1}$}  &
    \footnotesize{$f_{\la_{12}}$} &
    \footnotesize{$f_{\la_{13}}$}\\
    \footnotesize{0 s} &
    \footnotesize{11 s} &
    \footnotesize{12 s}\\
\end{tabular}\\
\begin{tabular}{cc}
    \includegraphics[width=0.3\columnwidth]{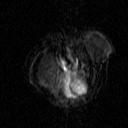}&
    \includegraphics[width=0.3\columnwidth]{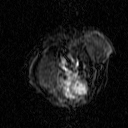}\\
    \footnotesize{$f_{\la_{256}}$} &
    \footnotesize{$f_{\la_{512}}$}\\
    \footnotesize{255 s} &
    \footnotesize{511 s}\\
\end{tabular}
\end{tabular}
  \caption{Five channels of hyperspectral image of a mouse «~serim447~»
  which is a temporal series (128 $\times$ 128 $\times$ 512 pixels)
  with 512 channels acquired every second.}
\label{intro:mouse_5_channels}
\end{figure}

In order to get a better understanding of the images, several
portions are labeled in figure~\ref{intro:mouse_labeled} : i) a
portion of the tumour is in green, ii) a portion of the heart
cavities is in blue, iii) a portion of the background is in red and
iv) a portion of the lungs is in white.

\begin{figure}[!htb]
\centering
    \includegraphics[width=0.5\columnwidth]{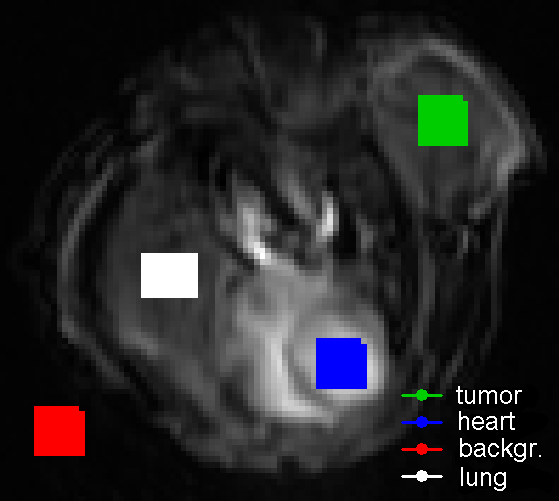}
 \caption{Labeled portions of DCE-MRI series.}
\label{intro:mouse_labeled}
\end{figure}

Hyperspectral images are multivariate discrete functions with
several tens or even hundreds of spectral bands. In a formal way,
for each pixel of a 2D (or a 3D), hyperspectral image is viewed as a
vector, with values associated with a given wavelength, time or any
index $j$. Each wavelength, time or index has a corresponding image,
named channel, in two (or three) dimensions. In the text below, we
use the term of spectrum and spectral channel to describe temporal
phenomena. The segmentation of hyperspectral images by hyperspectral
mathematical morphology has been used for remote sensing
images~\citep{Benediktsson_2005,Noyel_IJRS2011}. However, we show
here the generality of our approach since the methods are also
useful for DCE-MRI series analysis.

\subsection{State-of-the-art on DCE-MRI series analysis}

Dynamic Contrast Enhanced Imaging (DCE-Imaging) is an increasingly
used non-invasive imaging strategy to analyse tissue
micro-vascularisation and perfusion. This technique is based on 1)
an injection of a bolus of contrast agent, 2) an imaging modality
(CT, MRI or Ultrasound imaging) in a sequential mode
\citep{Ivancevic_1996}, and 3) a method to analyse the kinetics of
tissue enhancement over time \citep{Sourbron_2012,Brix_2012}.
DCE-Imaging has been widely demonstrated to be useful in detection
and characterisation of lesions such as ischemia and tumours, in a
variety of tissues such as brain, heart, breast and liver
\citep{van_Dijke_1996}, as well as in prediction and evaluation of
the effects of therapies
\citep{O_Connor_2008,Zahra_2007,Leach_2012}. DCE-Imaging is
especially useful in analysing tumour angiogenesis, i.e. the fast
growth of a new and chaotic capillary network in tumours induced by
growing factors secreted by the fast multiplying tumour cells
\citep{Brasch_2000}. This functional imaging field is therefore
growing rapidly an constitutes a valuable addition to traditional
morphological imaging, signal intensity, shape and size of the
lesions. However, several meta-analyses
\citep{O_Connor_2008,Zahra_2007} have underlined the difficulties in
comparing DCE imaging results between different centres due to
differences in data acquisition and analysis of this technique, as
well as to signal to noise limits. Signal to noise limits are due to
fast dynamic acquisitions and due to motion artefacts induced by
breathing, heart beating, bowel movements as well as involuntary
patient movements. To improve reproducibility, the challenge
consists in minimising the influence of the local experimental
conditions on data.

In \citet{Ding_2009}, a method  based on Karhunen Loeve Transform
has been presented in order to reduce noise on cardiac cine MRI. The
factorial axes are selected according to the autocorrelation
function of each eigenimage. The axes are retained in a contiguous
way by an automatic criterion based on half-maximum height of the
autocorrelation peak. In \citet{Ding_2010}, an extension of this
method for spatially variant noise has been presented. It is based
on the eigenvalue distribution of random matrices.

In \citet{Balvay_2011}, an improvement of Signal to Noise Ratio for
Dynamic Contrast-Enhanced Computed Tomography and Magnetic Resonance
Imaging with PCA is explained. A new criterion, the fraction of
residual information, is proposed to automatically select the factor
axes. It takes into account the temporal order of the images in the
series.

In this paper we present the most interesting results on DCE-MRI
series of a method that reduces the noise by Factor Correspondence
Analysis, followed by a classification and a segmentation stage.
Readers interested in more details and proofs on our method are
invited to consult \citep{Noyel_PhD_2008}. A short overview of some
of the results of this study has been presented in
\citep{Noyel_ISBI_2008}.

\subsection{Paper organisation}

The first part of this paper introduces some pre-requisites on
hyper-spectral image processing. The second part is focused on
filtering and data reduction of hyperspectral images. In the third
part the classification step is explained and in the fourth part the
segmentation by standard and stochastic watershed of DCE-MRI series
is presented. The full image analysis process is illustrated and
validated by an application to the automatic detection of tumours on
animals.



\section{Pre-requisites}
\label{sec:pre}

\subsection{Notations}

In order to analyse DCE-MRI series with methods based on
multivariate image processing, we use a specific notation for
hyperspectral images. By using this generic notation, we consider
the temporal dimension of the image as the ``spectral dimension''.
Let :
\begin{equation}\label{eq_im}
\mathbf{f_{\lambda}}: \left\{
\begin{array}{lll}
 E & \rightarrow & \mathcal{T}^{L}\\
 x & \rightarrow & \mathbf{f}_{\mathbf{\lambda}}(x) = \left( f_{\lambda_{1}}(x), f_{\lambda_{2}}(x), \ldots, f_{\lambda_{L}}(x) \right)\\
\end{array} \right.
\end{equation}
be an hyperspectral image, where:
\begin{description}
  \item[$\bullet$] $E \subset \mathbb{R}^{2}$, $\mathcal{T} \subset
  \mathbb{R}$ and $\mathcal{T}^{L} = \mathcal{T} \times \mathcal{T} \times \ldots \times \mathcal{T}$
  \item[$\bullet$] $x = x_{i} \ \backslash \ i\in\{1,2, \ldots, P \}$ is the spatial coordinates of a vector pixel
$\mathbf{f}_{\lambda}(x_{i})$ ($P$ is the number of pixels in $E$)
  \item[$\bullet$] $f_{\lambda_{j}} \ \backslash \ j \in \{1,2, \ldots, L\}$ is a
  channel ($L$ is the number of channels)
  \item[$\bullet$] $f_{\lambda_{j}}(x_{i})$ is the value of vector pixel
$\mathbf{f}_{\lambda}(x_{i})$ on channel $f_{\lambda_{j}}$.
\end{description}

Several approaches exist to analyse DCE-MRI:
\begin{enumerate}
  \item spatial analysis: channel $f_{\lambda_{j}}$ by channel $f_{\lambda_{j'}}$;
  \item spectral analysis: vector $\mathbf{f}_{\lambda}(x_{i})$ by vector $\mathbf{f}_{\lambda}(x_{i'})$;
  \item spatio-spectral analysis: simultaneous use of both approaches $\mathbf{f_{\lambda}}$.
\end{enumerate}
The methods are illustrated in figure \ref{pre:3analysis}.
Generally, the spatio-spectral approach gives the best results.

\begin{figure}[!htb]
\centering
\begin{tabular}{@{}c@{\ctab}c@{}}
    \includegraphics[width=0.5\columnwidth]{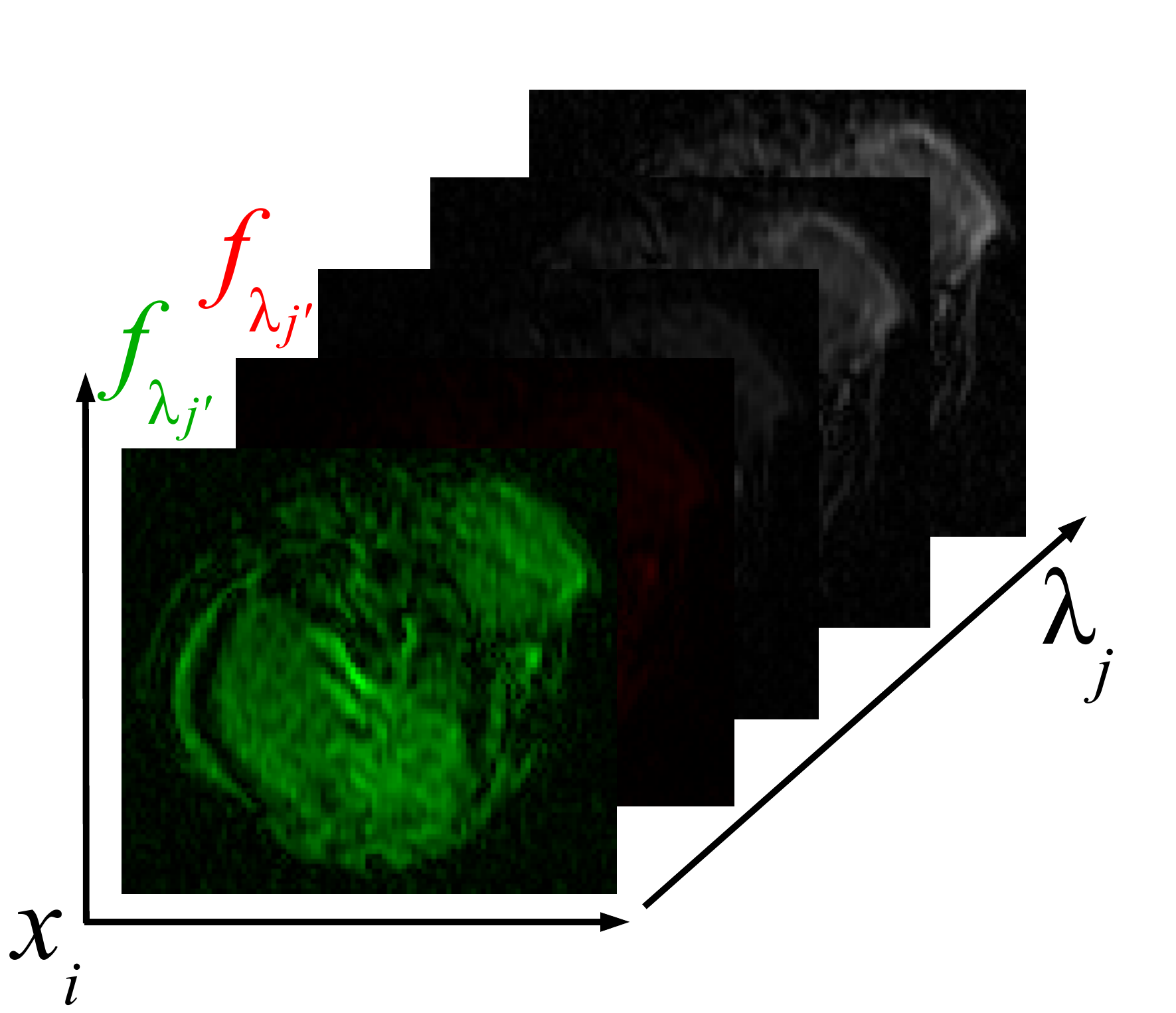}&
    \includegraphics[width=0.5\columnwidth]{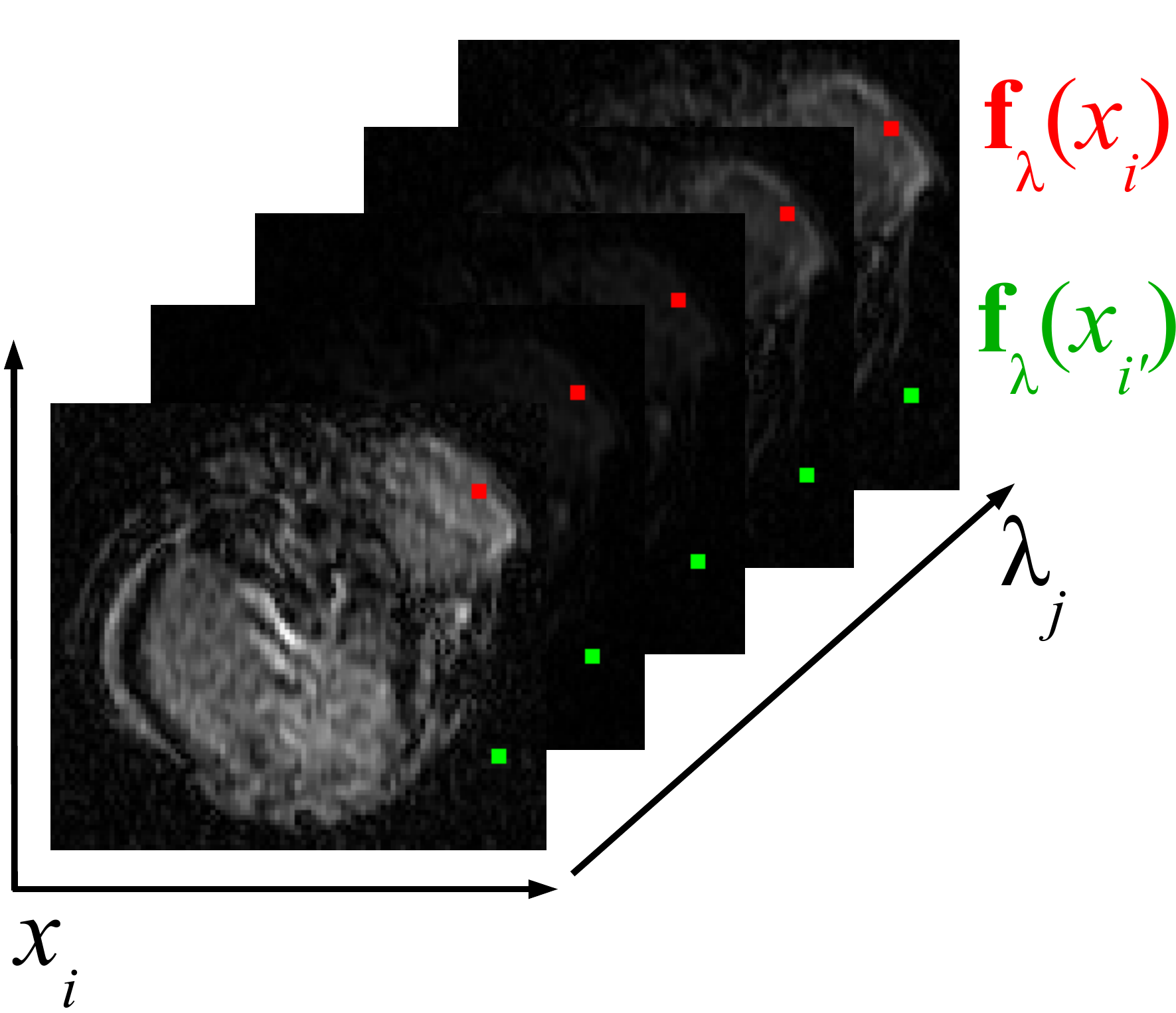}\\
    (a) Spatial  & (b) Spectral\\
    \multicolumn{2}{c}{\includegraphics[width=0.5\columnwidth]{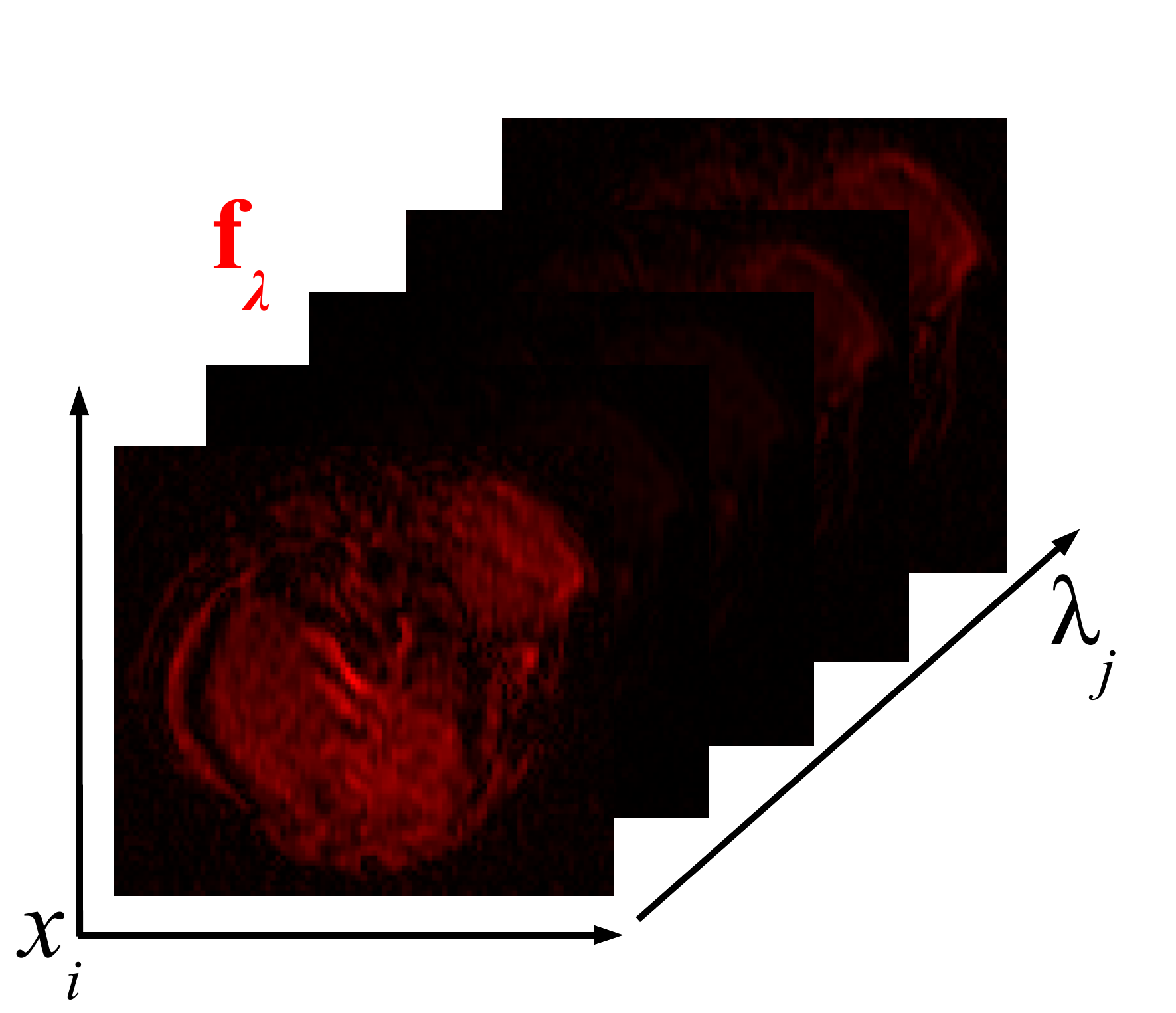}}\\
    \multicolumn{2}{c}{(c) Spatio-spectral}\\
\end{tabular}
  \caption{Three ways to analyse DCE-MRI sequences: (a) spatial, (b) spectral and (c) spatio-spectral approach.}
\label{pre:3analysis}
\end{figure}

\subsection{Necessity of data reduction}

DCE-MRI images are time series composed of several hundreds of time
channels. As channels are not statistically independent, the
reduction of spectral dimension is necessary:

\begin{enumerate}
  \item to reduce ``Hughes phenomenon''
\citep{Hughes_1968} also called the ``curse of dimension'';
  \item to reduce the amount of data and therefore to reduce the
  computational time.
\end{enumerate}

%

Hughes phenomenon has been studied in the case of hyperspectral
images, among others in \citet{Landgrebe_2002,Lennon_2002}. To
tackle this problem several data reduction methods exist, e.g.
Correspondence Analysis, Principal Component Analysis, Independent
Component Analysis, etc. Modeling the spectrum is also useful for
reducing the spectral dimension as the example shown in this paper.

\subsection{Methodology to segment DCE-MRI time series}

The overall proposed methodology to segment DCE-MRI time series is
composed of three steps (fig. \ref{pre:methodology}):
\begin{enumerate}
  \item a filtering of the images and a reduction of their spectral dimension
  \item a classification of the vector pixels to a given number of classes
  \item a segmentation step based on a function to
  flood by a watershed transform. This function combines the spatial and spectral information.
  It consists of a probability density
  function of contours
\end{enumerate}

The first two steps are included  in the pre-processing step, which
in this paper is based exclusively on the spectral information.

\begin{figure}[!htb]
  \centering
    \includegraphics[width=0.7\columnwidth]{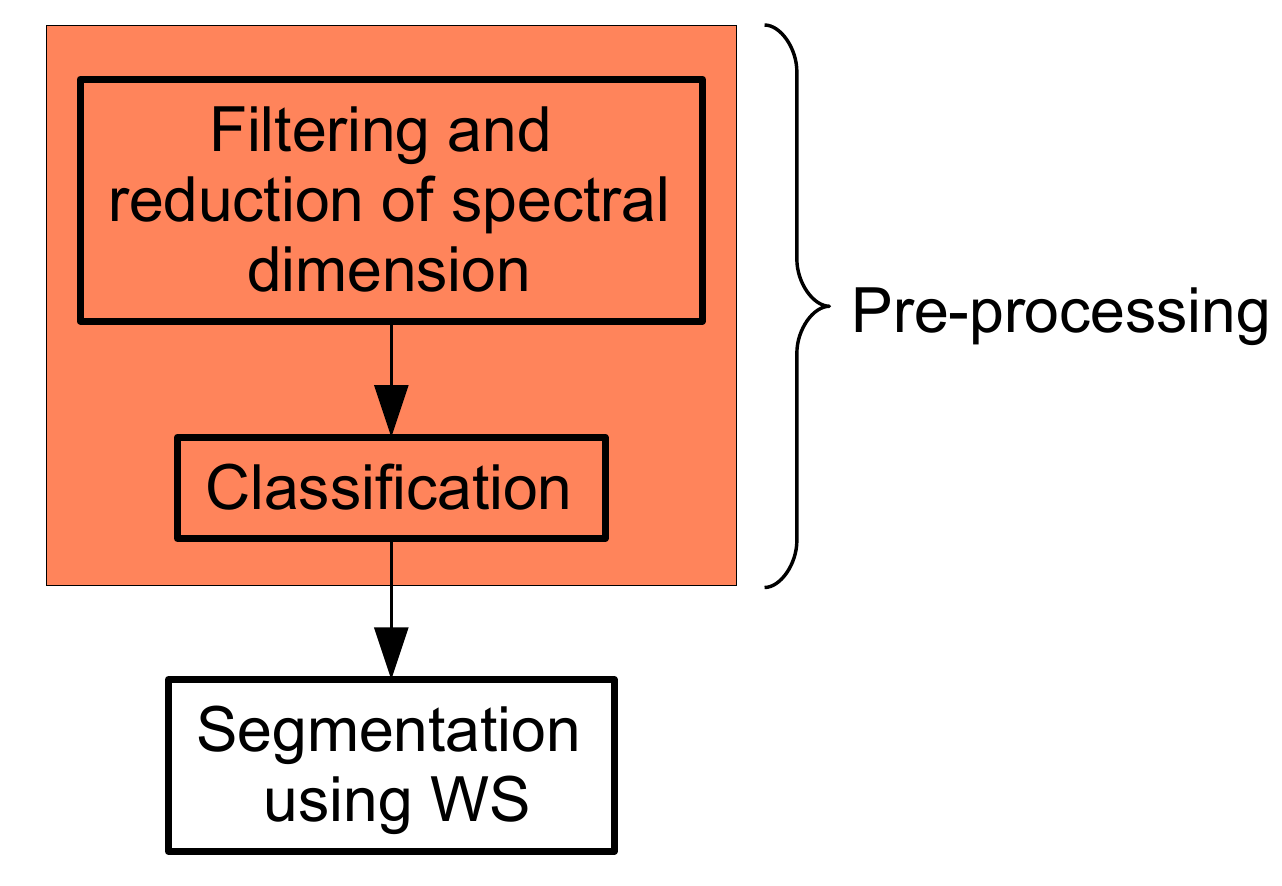}
  \caption{Framework to segment DCE-MRI time series as hyperspectral images.}
  \label{pre:methodology}
\end{figure}

The spectral approach is actually a global analysis on the image,
because we compare all the vector pixels between them. On the other
hand spatial approach is more local, because a given vector pixel is
mainly compared to its nearest neighbours.

\section{Filtering and Data reduction}
\label{sec:filter}



Multivariate image denoising and dimensionality reduction is
addressed in this study with Factor Correspondence Analysis (FCA)
\citep{Benzecri_1973}. A factorial space of reduced dimension is
obtained with respect to the spectral dimension of the original
image space. This reduced space is composed of factor pixels $\ca$
of the hyperspectral image $\f$.

Factor Correspondence Analysis is useful to reduce spectrum
dimension \citep{Noyel_IAS_2007}. Similar results can be obtained
with other methods such as Principal Component Analysis (PCA) or
Independent Component Analysis (ICA), etc. We have given priority to
FCA because it is efficient in segmenting multivariate images with
positive pixels values.

The original image $\f$ can be reconstructed from a limited number
of factors leading to a good approximation $\fr$ of this original
image $\f$. The reconstructed image contains, under certain
conditions, a spectral noise that is smaller than the noise on the
original image. Therefore, a method which reduces the noise on the
factor pixels $\ca$ will be explained in this section.

Additionally, the reduction of the spectral dimension of the image
$\f$ by fitting a spectrum model  on the filtered image $\fr$ will
be presented. This is to take advantage of the prior knowledge of
the spectrum. By modeling the spectra, some maps of the parameters
of the model are obtained. These maps constitute a reduced space
which is useful for further classification and segmentation.

\subsection{Denoising and dimensionality reduction by data analysis}
\label{sec:filter:DA}

\subsubsection{Introduction to FCA}

Data analysis is sa transformation $\zeta$ of the space of the
original image $\f$, with a dimension $L$, into a space of another
hyperspectral image $\ca$, of reduced dimension $K < L$, and a set
of parameters:

\begin{equation}\label{eq:filter:DA:FCA}
    \zeta: \left\{
\begin{array}{lll}
 \mathcal{T}^{L} & \rightarrow & \mathcal{T}^{K} \text{ such that } K < L \\
 \mathbf{f}_{\mathbf{\lambda}}(x) & \rightarrow &
    \left( \begin{array}{l}
        \mathbf{c}^{\mathbf{f}}_{\alpha}(x) =
        \left( c^{\mathbf{f}}_{\alpha_{1}}(x), \ldots, c^{\mathbf{f}}_{\alpha_{K}}(x) \right) \\
        \mathbf{d}^{\mathbf{f}}_{\al \la} =  \left(d^{\mathbf{f}}_{\al_1 \la}, \ldots, d^{\mathbf{f}}_{\al_K \la}\right)\\
        \{\mu_{\al}\}_{\al = 1 \ldots K}\\
        \{\nu_{i.}\}_{i=1 \ldots P}\\
        \{\nu_{.j}\}_{j=1 \ldots L}\\
        f = \sum_i \sum_j f_{ij}
    \end{array}
    \right)
\end{array} \right.
\end{equation}
with:
\begin{itemize}
  \item $\ca$ the factor pixels of the hyperspectral image. It is
  the coordinates of the vector pixels on the factorial axes.
  \item $\mathbf{d}^{\mathbf{f}}_{\al \la}$ the factors of the
channels. They are the coordinates of the channels on the factorial
axes.
  \item $\mu_{\al}$ is the inertia of factorial axis $\alpha$.
  \item $\nu_{i.}$ the marginal frequency of the vector pixel
  $\f(x_i)$:
  $\nu_{i.} = \sum_{j=1}^{L} \frac{f_{\la_j}(x_i)}{\sum_{j=1}^{L}\sum_{i=1}^{P}
  f_{\la_j}(x_i)}$.
  \item $\nu_{.j}$ the marginal frequency of the channel
  $f_{\la_j}$: \\
  $\nu_{.j} = \sum_{i=1}^{P} \frac{f_{\la_j}(x_i)}{\sum_{j=1}^{L}\sum_{i=1}^{P}
  f_{\la_j}(x_i)}$.
  \item $f = \sum_i \sum_j f_{ij} = \sum_i \sum_j f_{\la_j}(x_i)$
  the sum of all the values $f_{\la_j}(x_i)$ of the image $\f$.
\end{itemize}

For data analysis, a limited number $K$ of factors is usually
selected. Therefore, data analysis is a projection of the pixels of
the original image $\f$ into a space of smaller dimension $K < L$
and often $K \ll L$.

The reconstruction $\widehat{\zeta}^{-1}$ of the image $\fr$ is a
pseudo-inverse transform. It is an exact transform if all the axes
are kept ($K = L-1$). It consists in partially reconstructing the
image $\f$ from the pixels factors $\ca$ and some other parameters.
The reconstructed image $\fr$, with a limited number of factors, is
an approximation of the original image:

\begin{equation}\label{eq:filter:DA:FCA_inverse}
    \widehat{\zeta}^{-1}: \left\{
\begin{array}{lll}
 \mathcal{T}^{K} & \rightarrow & \mathcal{T}^{L} \text{ / } K < L \\
 \left(\begin{array}{l}
    \mathbf{c}^{\mathbf{f}}_{\alpha}(x)\\
    \mathbf{d}^{\mathbf{f}}_{\al \la}\\
    \{\mu_{\al}\}_{\al = 1 \ldots K}\\
    \{\nu_{i.}\}_{i=1 \ldots P}\\
    \{\nu_{.j}\}_{j=1 \ldots L}\\
    f = \sum_i \sum_j f_{ij}
 \end{array}
 \right)
 & \rightarrow & \mathbf{\widehat{f}}_{\mathbf{\lambda}}(x)\\
\end{array} \right.
\end{equation}
with $\mathbf{\widehat{f}}_{\mathbf{\lambda}}(x) =
 \left( \widehat{f}_{\lambda_{1}}(x), \ldots, \widehat{f}_{\lambda_{L}}(x)
 \right)$.

\subsubsection{Selection of the factor axes}
\label{sec:filter:DA:selection_axes}

The number of factorial axes to be kept needs to be chosen. It
depends of:
\begin{itemize}
  \item the part of inertia (or variance) that they explained in the data cloud.
   Several tests, based on inertia, exist that allow one to choose the correct number of axes such as
the ``Kaiser criterion'' \citep{Kaiser_1960} or the ``scree test''
\citep{Cattell_1966}.
  \item the amount of information contained in the factor pixels. We
  have introduced a new criterion based on the signal to noise ratio of the
  factor pixels.
\end{itemize}

\begin{figure}[!htb]
  \centering
  \includegraphics[width=0.7\columnwidth]{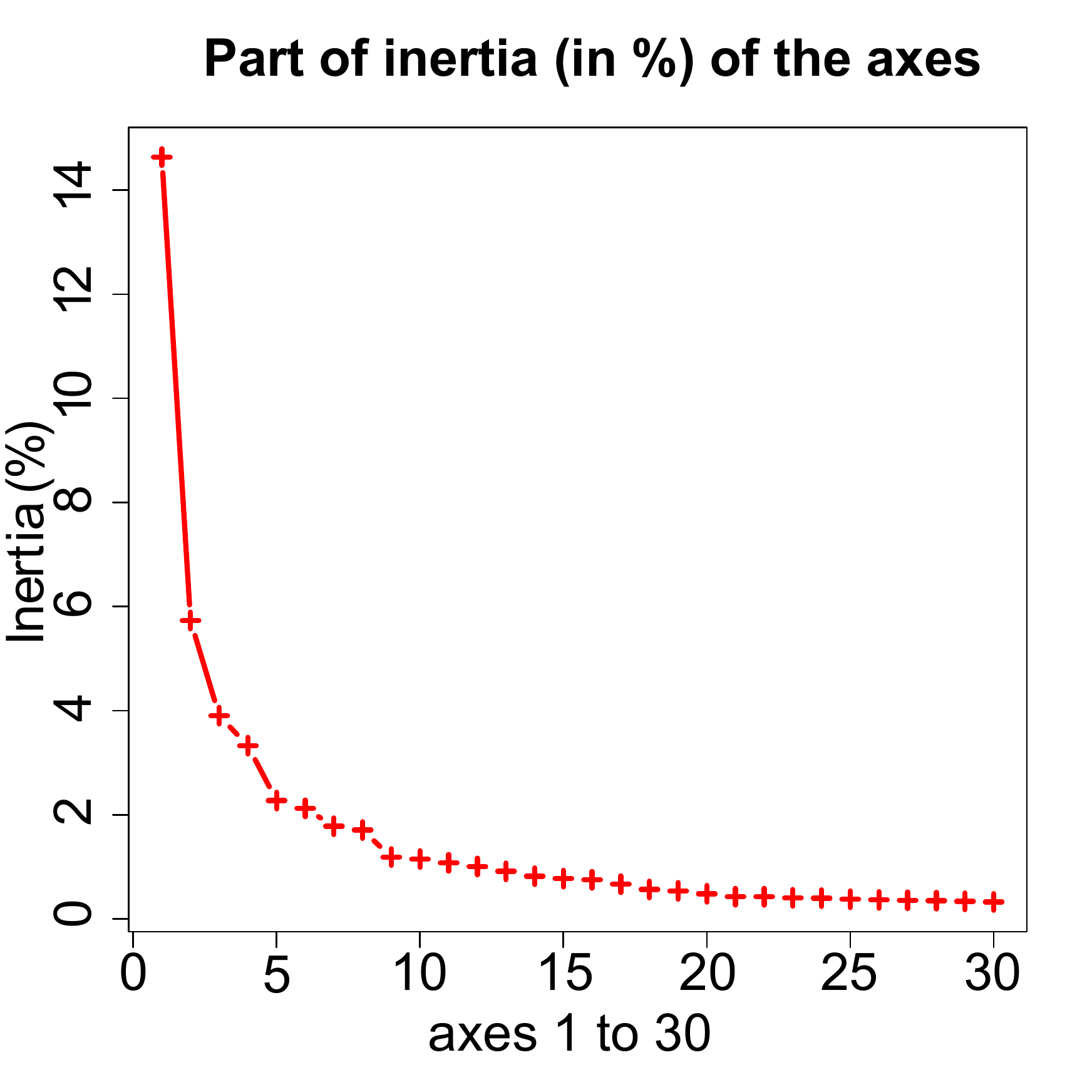}
  \caption{Part of inertia of the thirty first factorial axes.}\label{fig:filter:DA:inertia}
\end{figure}

\begin{figure}[!htb]
\centering
\begin{tabular}{ccc}
    \includegraphics[width=0.3\columnwidth]{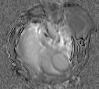}&
    \includegraphics[width=0.3\columnwidth]{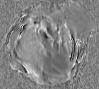}&
    \includegraphics[width=0.3\columnwidth]{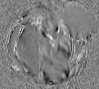}\\
    \footnotesize{$c^{\mathbf{f}}_{\alpha_{1}}$ 14.63\%}&
    \footnotesize{$c^{\mathbf{f}}_{\alpha_{2}}$ 5.73\%}&
    \footnotesize{$c^{\mathbf{f}}_{\alpha_{3}}$ 3.91\%}\\
    \includegraphics[width=0.3\columnwidth]{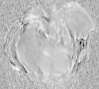}&
    \includegraphics[width=0.3\columnwidth]{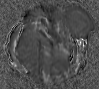}&
    \includegraphics[width=0.3\columnwidth]{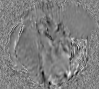}\\
    \footnotesize{$c^{\mathbf{f}}_{\alpha_{4}}$ 3.33\%}&
    \footnotesize{$c^{\mathbf{f}}_{\alpha_{5}}$ 2.27\%}&
    \footnotesize{$c^{\mathbf{f}}_{\alpha_{6}}$ 2.12\%}\\
    \includegraphics[width=0.3\columnwidth]{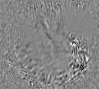}&
    \includegraphics[width=0.3\columnwidth]{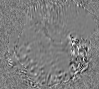}&
    \includegraphics[width=0.3\columnwidth]{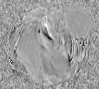}\\
    \footnotesize{\textcolor{red}{$\underline{c^{\mathbf{f}}_{\alpha_{7}} \text{ } 1.78\%}$}}&
    \footnotesize{\textcolor{red}{$\underline{c^{\mathbf{f}}_{\alpha_{8}} \text{ } 1.71\%}$}}&
    \footnotesize{$c^{\mathbf{f}}_{\alpha_{9}}$ 1.19\%}\\
\end{tabular}
  \caption{The factor pixels
on the 9 firsts axes and their inertias. The factors pixels typed in
black are kept; those typed in red are rejected due to a signal to
noise ratio which is below a given threshold.}
   \label{fig:filter:DA:_mouse_factor_pixels}
\end{figure}

In figure \ref{fig:filter:DA:inertia} one can notice that the first
5 axes contains the main part of the total variance or inertia
(about 30\%). However, by observing (fig.
\ref{fig:filter:DA:_mouse_factor_pixels}) the images of the factor
axes $c^{\mathbf{f}}_{\al_k}$ we notice that the factor pixels
$c^{\mathbf{f}}_{\al_7}$ and $c^{\mathbf{f}}_{\al_8}$ contain mainly
noise while the factor pixel $c^{\mathbf{f}}_{\al_9}$ contains
mainly signal. In order to quantify the amount of information
contained in the factor pixels, their signal to noise ratio is
estimated by the method proposed in \citet{Noyel_CGIV_2008}.

The channel $c^{\mathbf{f}}_{\al_k}$ (or $f_{\la_j}$) is considered
as a realisation of a random function. For each factorial axis, the
centred spatial covariance is estimated by assuming that
$c^{\mathbf{f}}_{\al_k}(x)$ is a stationary function:
\begin{equation}\label{eq:filter:DA:_covariance_centree}
    \overline{g}_{\alpha_k}(h) = E[\overline{c}^{\mathbf{f}}_{\alpha_k}(x)
    \overline{c}^{\mathbf{f}}_{\alpha_k}(x+h)]
\end{equation}
with $\overline{c}^{\mathbf{f}}_{\alpha_k}$ the centred channel
$\al_{k}$: $\overline{c}^{\mathbf{f}}_{\alpha_k}(x) =
c^{\mathbf{f}}_{\alpha_k}(x) - E[c^{\mathbf{f}}_{\alpha_k}(x)]$ and
$E[Y]$ the expectation of the random variable $Y$.
$E[c^{\mathbf{f}}_{\alpha_k}(x)]$ corresponds to the mean of the
channel $c^{\mathbf{f}}_{\alpha_k}(x)$.

It is known that the covariance of a noisy random function shows a
discontinuity for $h\rightarrow 0$. This discontinuity, equal to the
variance of the noise, is called the \textquotedblleft
nugget\textquotedblright\ effect
\citep{Matheron_1970,Matheron_1975}. This is illustrated in figure
\ref{fig:filter:DA:Ouverture_autocorr_axes_1_et_100}, where the
covariance is plotted for a channel without much noise and for a
noisy channel. In the latter case, a peak at the origin of the
covariance can be noticed. The variance of the signal can be
estimated by the difference between $\overline{g}_{\alpha_k}(0)$ and
the nugget effect. It can be computed by an automatic extraction of
the peak at the origin of the covariance image after a morphological
opening $\gamma $. The structuring element is chosen as small as
possible. It is a square of size $3 \times 3$ pixels. From this
extraction, a signal to noise ratio is estimated, according to the
following definition:
%
%
%
    \begin{equation}\label{eq:filter:DA:_SNR}
    SNR_{\alpha_k} = \frac{Var(signal)}{Var(noise)} = \frac{ \gamma
    \overline{g}_{\alpha_k}(0) }{ \overline{g}_{\alpha_k}(0) - \gamma
    \overline{g}_{\alpha_k}(0) }
    \end{equation}
    with $\overline{g}$ the centred covariance and
    $\gamma$ the morphological opening.

By observing the signal to noise ratio (SNR) of the different
factors in
figure~\ref{fig:filter:DA:Ouverture_autocorr_axes_1_et_100}, it
appears that the channels $c^{\mathbf{f}}_{\al_7}$ and
$c^{\mathbf{f}}_{\al_8}$ are noisier than others. In what follows,
we propose channels with a SNR greater than 0.3 are retained for
reconstruction.


\begin{figure}[!htb]
\centering
    \begin{tabular}{@{}c@{ }c@{ }c@{}}
    &  \footnotesize{covariance} &  \footnotesize{covariance opening}\\
   \includegraphics[width=0.25\columnwidth]{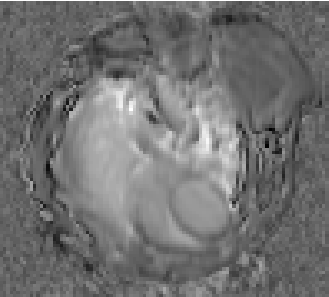}&
   \includegraphics[width=0.35\columnwidth]{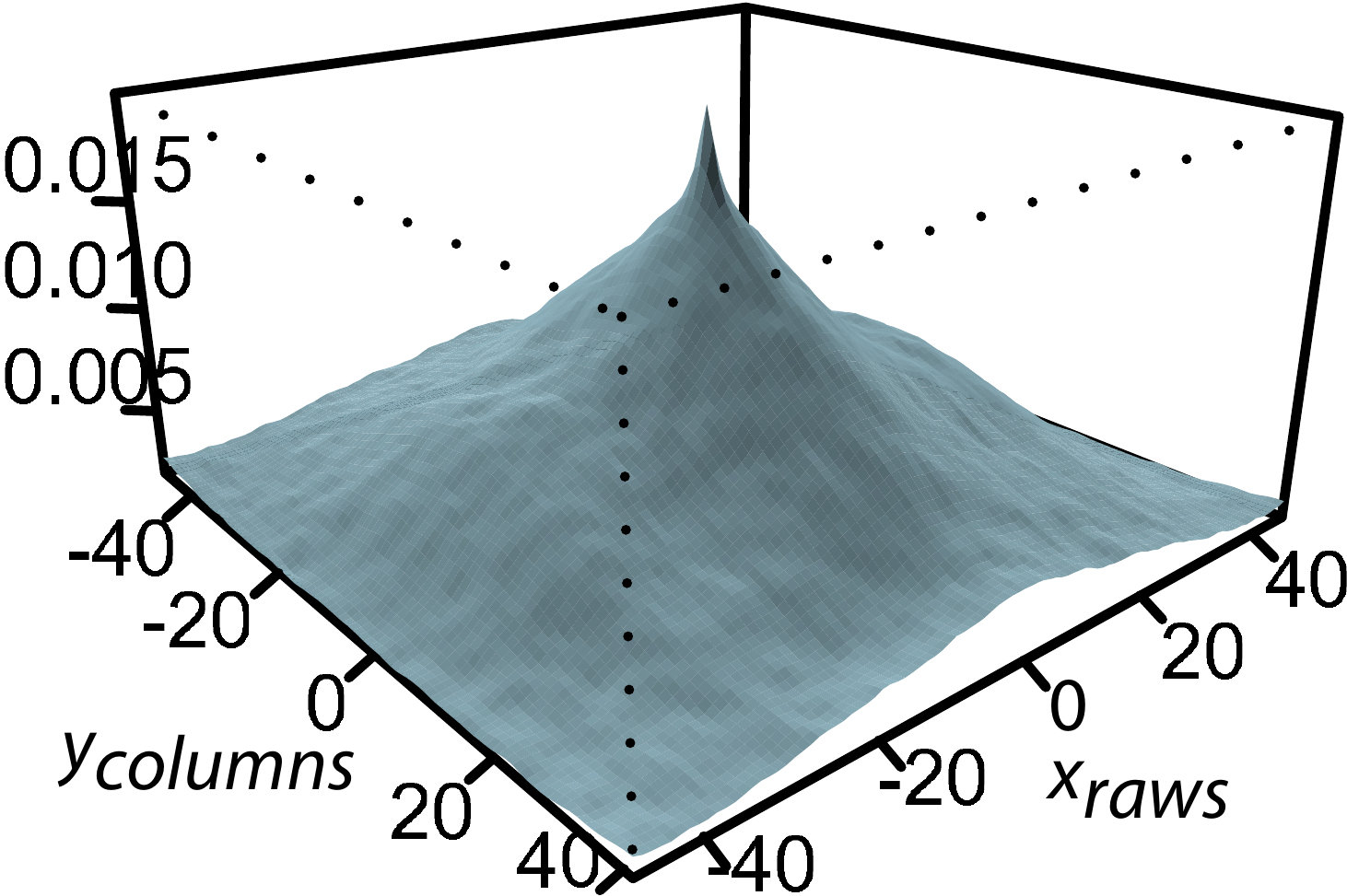} &
   \includegraphics[width=0.35\columnwidth]{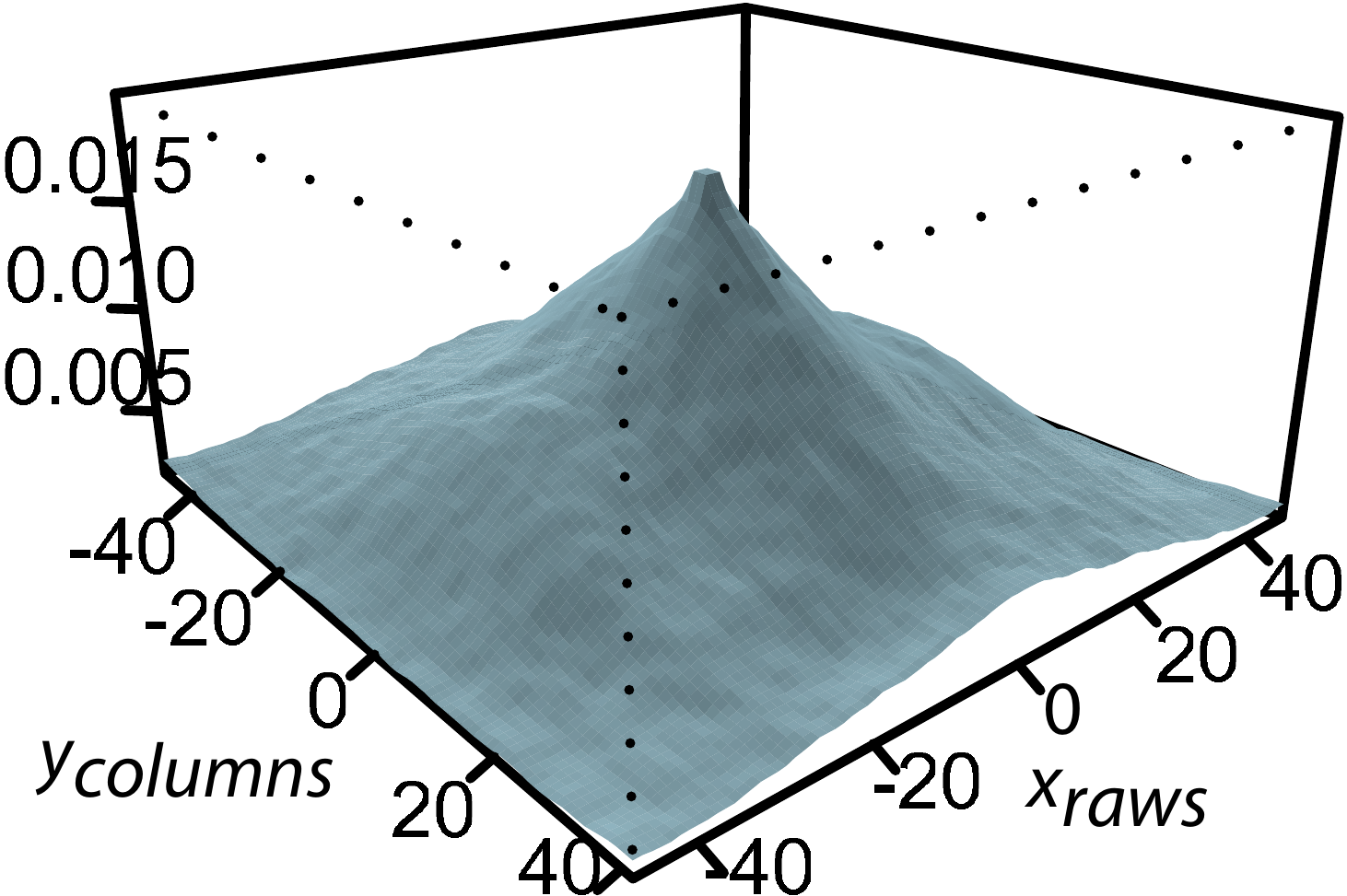} \\
    \footnotesize{$c^{\mathbf{f}}_{\al_1}$}&
    \footnotesize{$\overline{g}_{\al_1}$}&
    \footnotesize{$\gamma \overline{g}_{\al_1}$}\\
   \includegraphics[width=0.25\columnwidth]{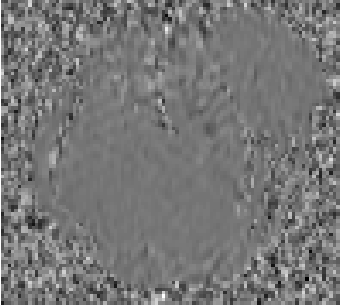}&
   \includegraphics[width=0.35\columnwidth]{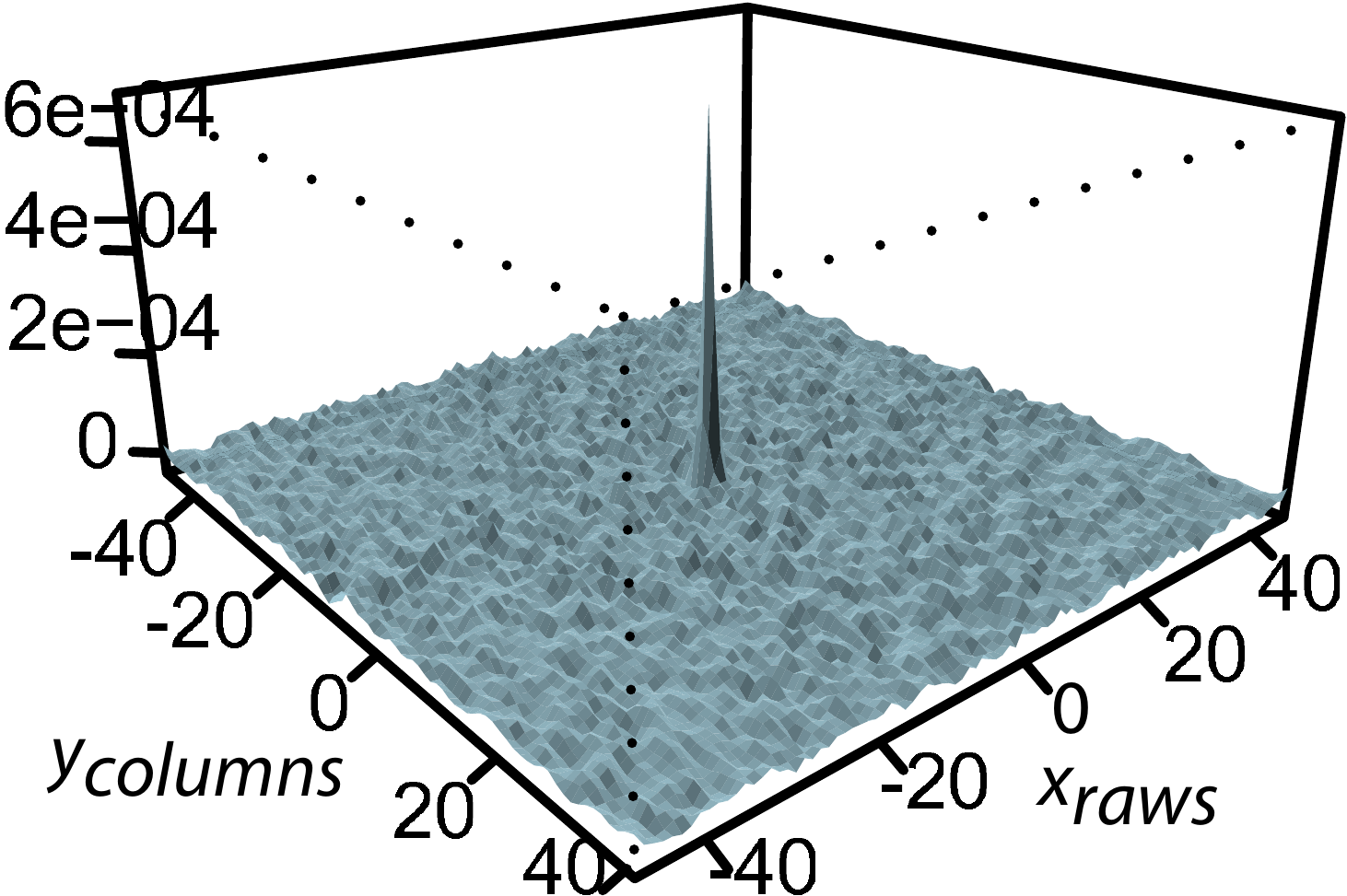} &
   \includegraphics[width=0.35\columnwidth]{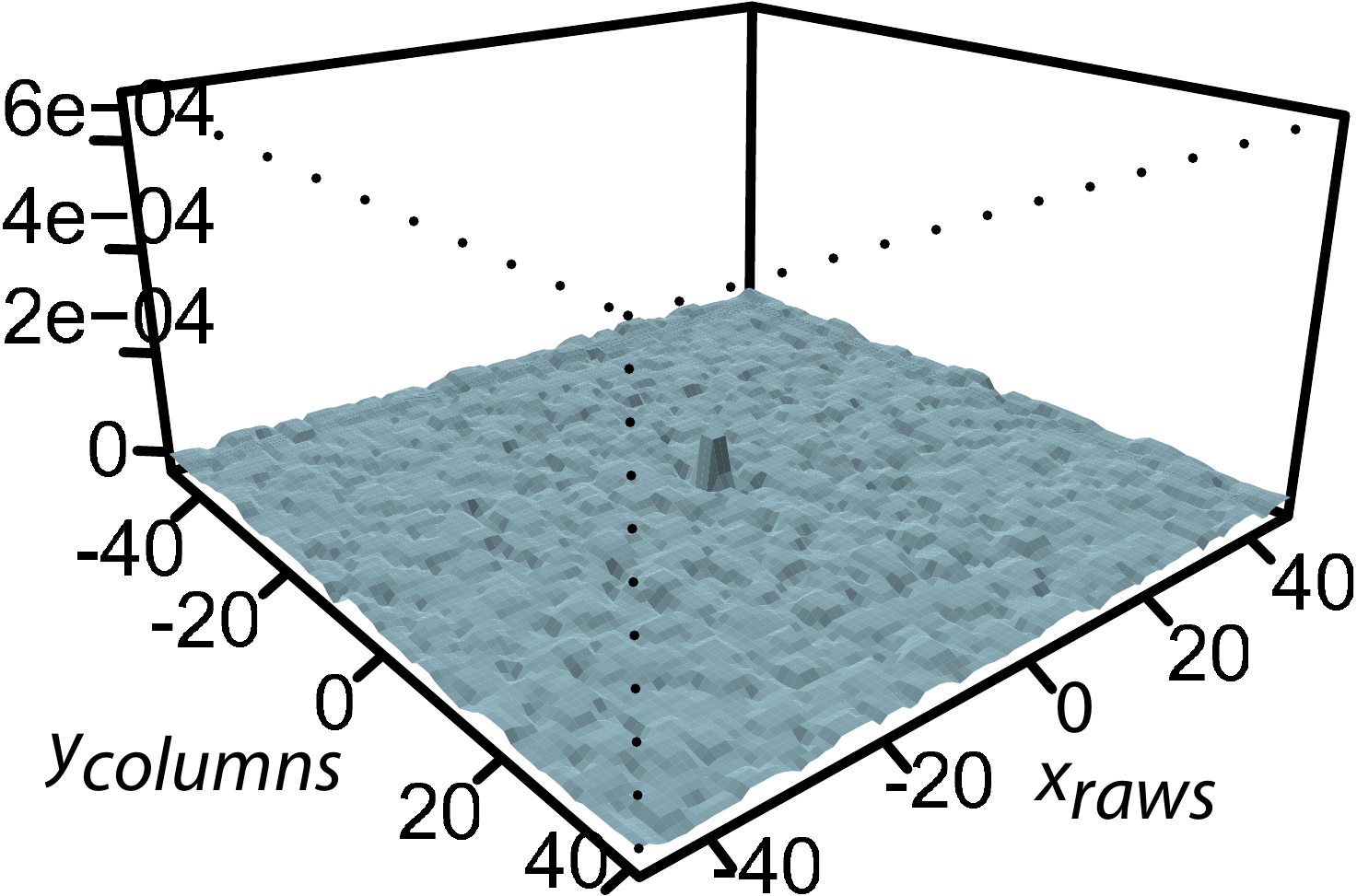} \\
    \footnotesize{$c^{\mathbf{f}}_{\al_{100}}$}&
    \footnotesize{$\overline{g}_{\al_{100}}$}&
    \footnotesize{$\gamma \overline{g}_{\al_{100}}$}\\
\end{tabular}
\caption{Covariance before ($\overline{g}$) and after a
morphological opening
  ($\gamma \overline{g}$) on the factor pixels channels $c^{\mathbf{f}}_{\al_1}$ (without much noise)
   and $c^{\mathbf{f}}_{\al_{100}}$ (noisy channel).}
  \label{fig:filter:DA:Ouverture_autocorr_axes_1_et_100}
\end{figure}

\begin{figure}[!htb]
\centering
    \includegraphics[width=0.7\columnwidth]{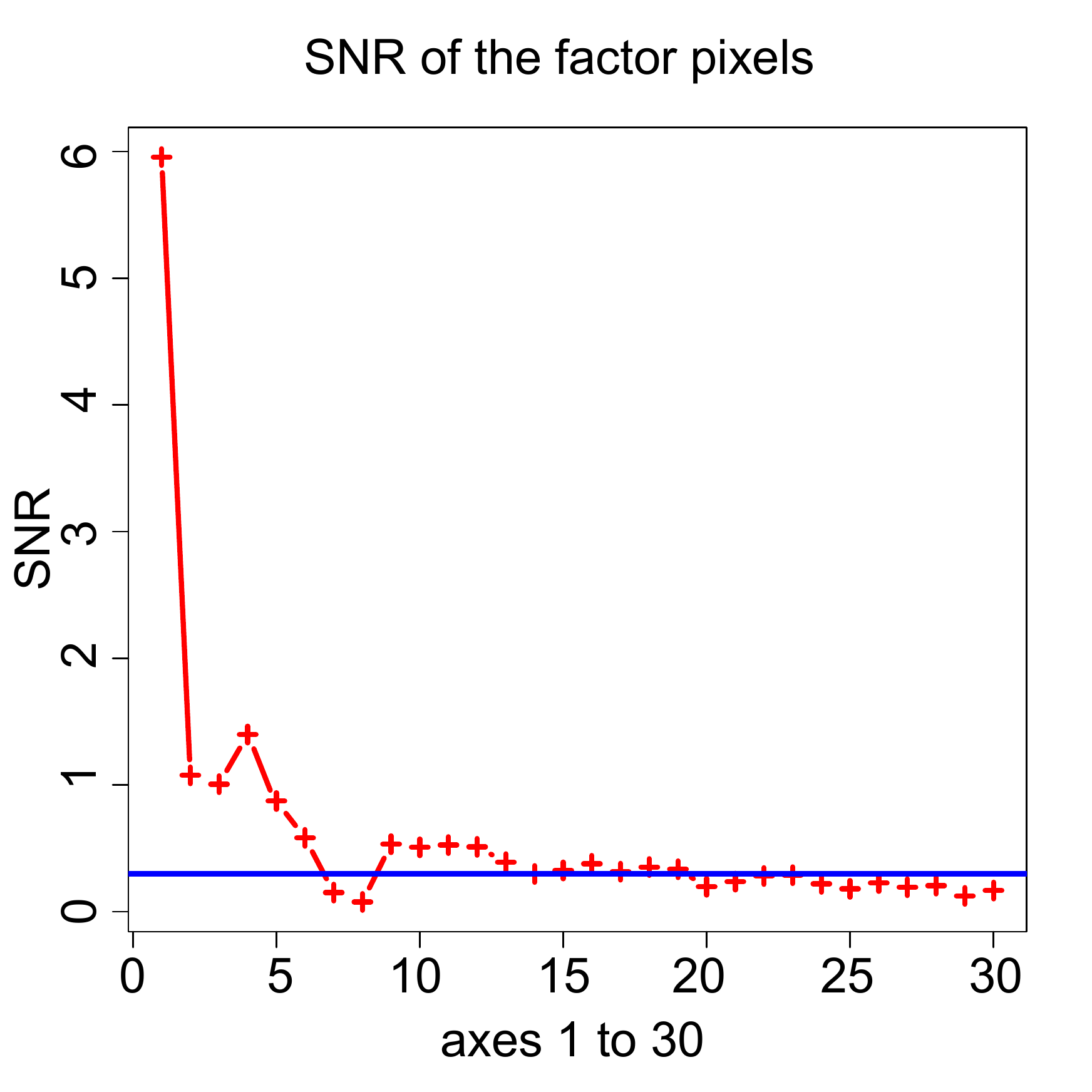}
  \caption{Signal to noise ratio of the factor pixels $\ca$ of the image and a threshold for a SNR of 0.3.}
\label{fig:filter:DA:_RSB_axes}
\end{figure}


The pixels factors in black in table
\ref{tab:filter:DA:_mouse_inertia_SNR_axes} and in figure
\ref{fig:filter:DA:_mouse_factor_pixels} are kept for data analysis,
while those in red are rejected. We notice that the selected
factorial axes, are not contiguous in terms of inertia, as only
components with a SNR $>$ 0.3 are kept.

\begin{table}[!htb]
\begin{tabular}{c|ccccc}  \hline
  Axes & 1 & 2 & 3 & 4 & 5\\
  SNR & 3.91& 1.02 & 0.99 &  1.07 & 0.79\\
  Inertia (\%) & 14.63 & 5.73 & 3.91 & 3.33 & 2.27\\  \hline
\end{tabular}
\begin{tabular}{c|ccccc}  \hline
  Axes & 6 & \color{red}{\underline{7}} & \color{red}{\underline{8}} & 9 & 10 \\
  SNR & 0.58 & \color{red}{\underline{0.15}} & \color{red}{\underline{0.08}} & 0.52 & 0.51\\
  Inertia (\%) &  2.12 & \color{red}{\underline{1.78}} & \color{red}{\underline{1.71}} & 1.19 & 1.15\\  \hline
\end{tabular}
\begin{tabular}{c|ccccc}  \hline
  Axes & 11 & 12 & 13 & \color{red}{\underline{14}} & 15\\
  SNR  & 0.52 & 0.51 & 0.39 & \color{red}{\underline{0.29}} & 0.33\\
  Inertia (\%) & 1.08 & 1.01 & 0.92 & \color{red}{\underline{0.82}} & 0.78\\  \hline
\end{tabular}
\begin{tabular}{c|ccccc}  \hline
  Axes & 16 & 17 & 18 & 19\\
  SNR & 0.38 & 0.32 & 0.35 & 0.33\\
  Inertia (\%) & 0.75 & 0.67 & 0.57 & 0.54\\  \hline
\end{tabular}
  \caption{The SNR of the factor pixels (greater than 0.3 for those kept) and their inertia part.
  The 16 factorial axes which are kept are in black and the 3 rejected are in
  red and underlined.}
  \label{tab:filter:DA:_mouse_inertia_SNR_axes}
\end{table}

\subsubsection{Reconstruction}

In figure \ref{fig:filter:DA:_mouse_5_channels_before_after_1FCA},
some channels of the image $\f$ are displayed before and after
reconstruction. On channel $f_{\la_{12}}$, we notice that one
ventricle of the heart appears dark while the other appears bright.
This opposition is preserved after the reconstruction of the channel
$f_{\la_{12}}$ using the selection of axes by SNR. One can also
notice, that the image $\fr$ is a good reconstruction of the image
$\f$ and that a part of the noise, in the original image $\f$, is
removed by FCA reconstruction. Therefore, the denoising has been
made by a spectral filtering of FCA, which preserves the spatial
structures of the image. This is a crucial improvement before
segmenting the images by mathematical morphology.
%

\begin{figure}[!htb]
\centering
\begin{tabular}{ccc}
    \includegraphics[width=0.3\columnwidth]{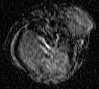}&
    \includegraphics[width=0.3\columnwidth]{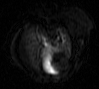}&
    \includegraphics[width=0.3\columnwidth]{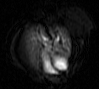}\\
    \footnotesize $f_{\la_1}$  &
    \footnotesize $f_{\la_{12}}$ &
    \footnotesize $f_{\la_{13}}$\\
    \includegraphics[width=0.3\columnwidth]{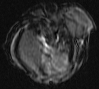}&
    \includegraphics[width=0.3\columnwidth]{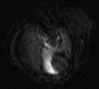}&
    \includegraphics[width=0.3\columnwidth]{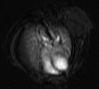}\\
    \textcolor{blue}{\footnotesize $\widehat{f}_{\la_1}$} &
    \textcolor{blue}{\footnotesize $\widehat{f}_{\la_{12}}$} &
    \textcolor{blue}{\footnotesize $\widehat{f}_{\la_{13}}$}\\
    \hline
    \\
    \includegraphics[width=0.3\columnwidth]{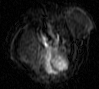}&
    \includegraphics[width=0.3\columnwidth]{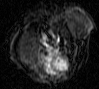}\\
    \footnotesize $f_{\la_{256}}$ &
    \footnotesize $f_{\la_{512}}$\\
    \includegraphics[width=0.3\columnwidth]{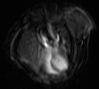}&
    \includegraphics[width=0.3\columnwidth]{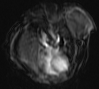}\\
    \textcolor{blue}{\footnotesize $\widehat{f}_{\la_{256}}$} &
    \textcolor{blue}{\footnotesize $\widehat{f}_{\la_{512}}$}\\
\end{tabular}
  \caption{Five channels of the hyperspectral image (i.e. the DCE-MRI series)
   before ($f_{\la_i}$) and after reconstruction (\textcolor{blue}{$\widehat{f}_{\la_{i}}$}) from 16 axes of a FCA.}
  \label{fig:filter:DA:_mouse_5_channels_before_after_1FCA}
\end{figure}

\FloatBarrier
\subsubsection{Noise reduction}
\label{sec:filter:DA:noise}

Like Principal Component Analysis, FCA is useful to perform a
spectral filtering of the data. In
\citet{Benzecri_1964,Benzecri_1973} and in~\citet{Orfeuil_1973}, FCA
is used to filter arrays of \textquotedblleft Euclidean data marred
with errors\textquotedblright . The use of PCA to filter
hyperspectral data was primarily illustrated in~\citet{Berman_1985}
and in~\citet{Green_1988}.

A study about noise reduction by FCA on hyperspectral images is
available in \citet{Noyel_PhD_2008}. In the following text, only the
interesting results obtained in this previous study are used.

In \citet{Noyel_PhD_2008}, we have shown two sequences of FCA and
reconstruction steps reduce the noise more than just a single
sequence.

A sequence of one FCA and a reconstruction $\fr^{(1)}$ is equal to
the reconstructed image. It is written:
\begin{equation}\label{eq:filter:DA:FCA1}
    \fr^{(1)} = \widehat{\zeta}^{-1} \circ
    \zeta (\f)
\end{equation}
Two sequences of FCA-reconstruction are defined as:
\begin{equation}\label{eq:filter:DA:FCA2}
    \fr^{(2)} = t_{\varepsilon} \circ \widehat{\zeta}^{-1} \circ
    \zeta \circ t_{-\varepsilon} \circ \widehat{\zeta}^{-1} \circ
    \zeta (\f)
\end{equation}
with $t_{\varepsilon}$ a spectral translation of the data. Actually
we add the constant $\varepsilon$ to the value of the image $\f$ in
order to recover positive values, as required for FCA:
\begin{equation}\label{eq:filter:DA:translation}
    t_{\varepsilon} : \left\{
    \begin{array}{lll}
    \mathcal{T}^{L} & \rightarrow & \mathcal{T}^{L} \\

    \mathbf{f}_{\mathbf{\lambda}} & \rightarrow &
    t_{\varepsilon}(\mathbf{f}_{\mathbf{\lambda}}) \text{ such as } \\

    &&\forall i=1 \ldots
    P \text{ , } \forall j=1 \ldots L \\

    && t_{\varepsilon}(\mathbf{f}_{\mathbf{\lambda_j}})(x_i)  =
    \mathbf{f}_{\mathbf{\lambda_j}}(x_i) + \varepsilon
    \qquad \varepsilon \in \Real
    \end{array} \right.
\end{equation}

Therefore the second sequence of FCA-reconstruction is not identical
to the first one because the data have been modified by the
translation $t_{\varepsilon}$.

For $\varepsilon$ we propose to use the minimum of all the data in
the first reconstructed image $\fr^{(1)}$:
\begin{equation}\label{eq:filter:DA:valeur_minimum}
    \varepsilon = \min_{i,j} \widehat{f}^{(1)}_{\la_j}(x_i) \qquad
    i = 1 \ldots P \qquad j = 1 \ldots L
\end{equation}

More details about the two sequences of FCA-reconstruction are given
in the appendix section \ref{sec:appendix}.

Notice that the inertia is evaluated on the axes of the original,
i.e. first, FCA.

In order to compare the original DCE-MRI series $\f$ with the series
after 2 sequences of FCA-reconstruction with 16 axes $\fr =
\fr^{(2)}$, the residues are computed channel by channel:

\begin{eqnarray}\label{eq:filter:DA:_residue}
    \nonumber \forall i \in [1 \ldots P] \text{ , } \forall j \in [1 \ldots L]\\
    r_{\lambda_j}(x_i) = |f_{\lambda_j}(x_i) -
    \widehat{f}_{\lambda_j}(x_i)|
\end{eqnarray}

An hyperspectral image of residues $\mathbf{r}_{\lambda}$  may be
defined:
\begin{equation}\label{eq:filter:DA:_residue_2}
    \mathbf{r}_{\lambda} = |\mathbf{f}_{\lambda} -
    \widehat{\mathbf{f}}_{\lambda}|
\end{equation}
of which the channels $r_{\lambda_j}$ are equal to:
\begin{equation}\label{eq:filter:DA:_residue_3}
    \forall j \in [1 \ldots L] \text{\qquad} \\
    r_{\lambda_j} = |f_{\lambda_j} -
    \widehat{f}_{\lambda_j}|.
\end{equation}

The centered spatial covariance $\overline{g}_{r_{\la_j}}$ is also
measured on the residues of the channels:
\begin{equation}\label{eq:filter:DA:cov_residu_canal}
    \forall j \in [1 \ldots L] \text{\qquad}
    \overline{g}_{r_{\lambda_j}} = E[\overline{r}_{\lambda_j} \overline{r}_{\lambda_j}]
\end{equation}
with $\overline{r}_{\lambda_j} = \frac{1}{P}\sum_{i=1}^{P}
r_{\lambda_j}(x_i)$.

In figure \ref{fig:filter:DA:_residues}, some channels after 2 FCA
reconstructions and their residues are presented. Notice that some
noise has been removed. This noise is due to the acquisition process
and is mainly located at the centre of the image. The centered
covariances $\overline{g}_{r_{\lambda_j}}$ contains a peak at their
origin, which means that the residues correspond to noise.

\begin{figure}[!htb]
\centering
\begin{tabular}{@{}ccc@{}}
    \includegraphics[width=0.3\columnwidth]{serim447_reduite001}&
    \includegraphics[width=0.3\columnwidth]{serim447_reduite012}&
    \includegraphics[width=0.3\columnwidth]{serim447_reduite512}\\
    \footnotesize $f_{\la_1}$  &
    \footnotesize $f_{\la_{12}}$ &
    \footnotesize $f_{\la_{512}}$\\
    \includegraphics[width=0.3\columnwidth]{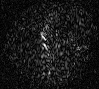}&
    \includegraphics[width=0.3\columnwidth]{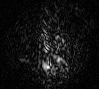}&
    \includegraphics[width=0.3\columnwidth]{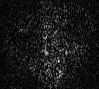}\\
    \footnotesize $r_{\la_{1}}$ &
    \footnotesize $r_{\la_{12}}$ &
    \footnotesize $r_{\la_{512}}$\\
    \includegraphics[width=0.3\columnwidth]{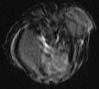}&
    \includegraphics[width=0.3\columnwidth]{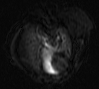}&
    \includegraphics[width=0.3\columnwidth]{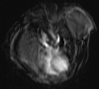}\\
    \footnotesize $\widehat{f}^{(2)}_{\la_1}$  &
    \footnotesize $\widehat{f}^{(2)}_{\la_{12}}$ &
    \footnotesize $\widehat{f}^{(2)}_{\la_{512}}$\\
    \\
    \includegraphics[width=0.3\columnwidth]{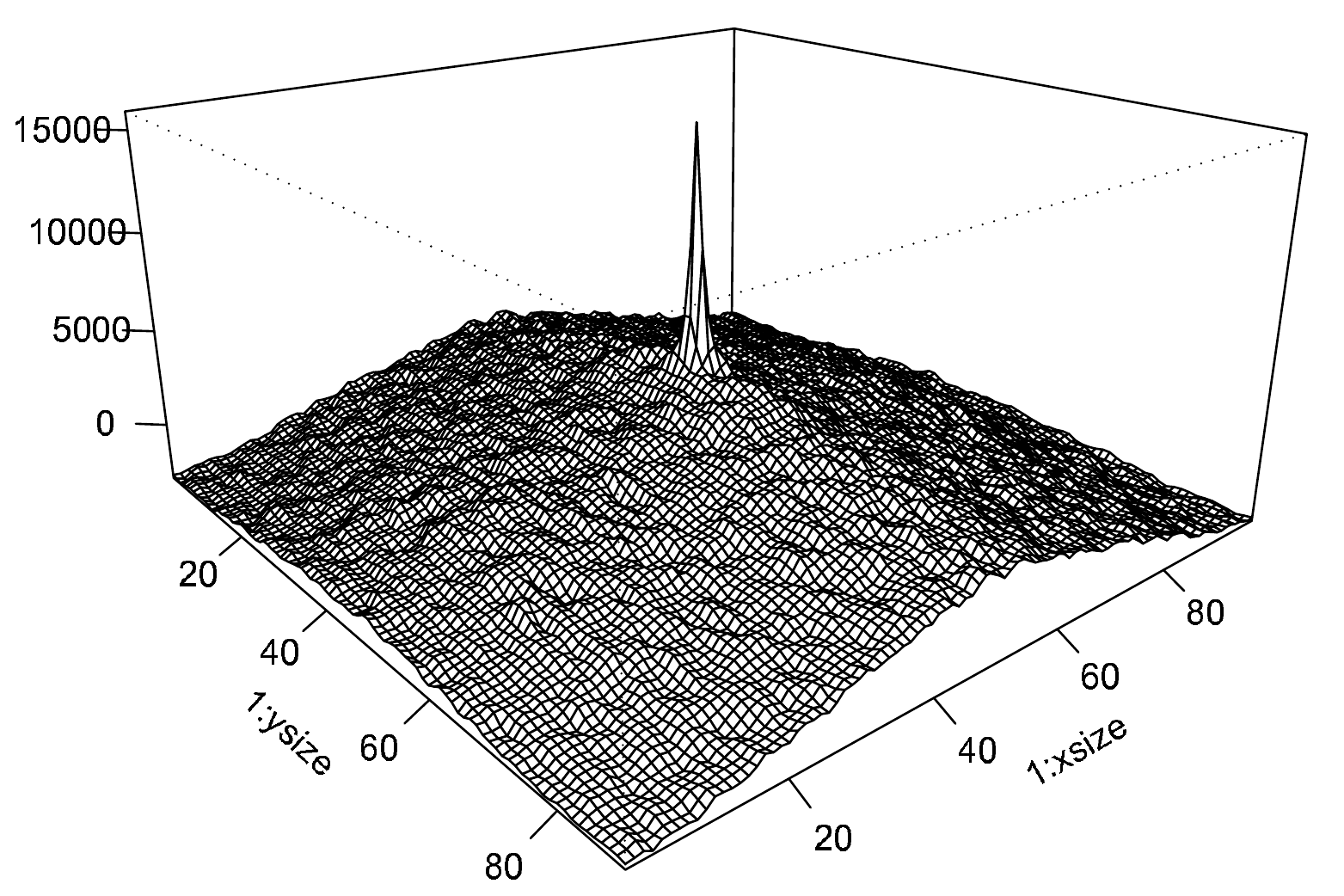}&
    \includegraphics[width=0.3\columnwidth]{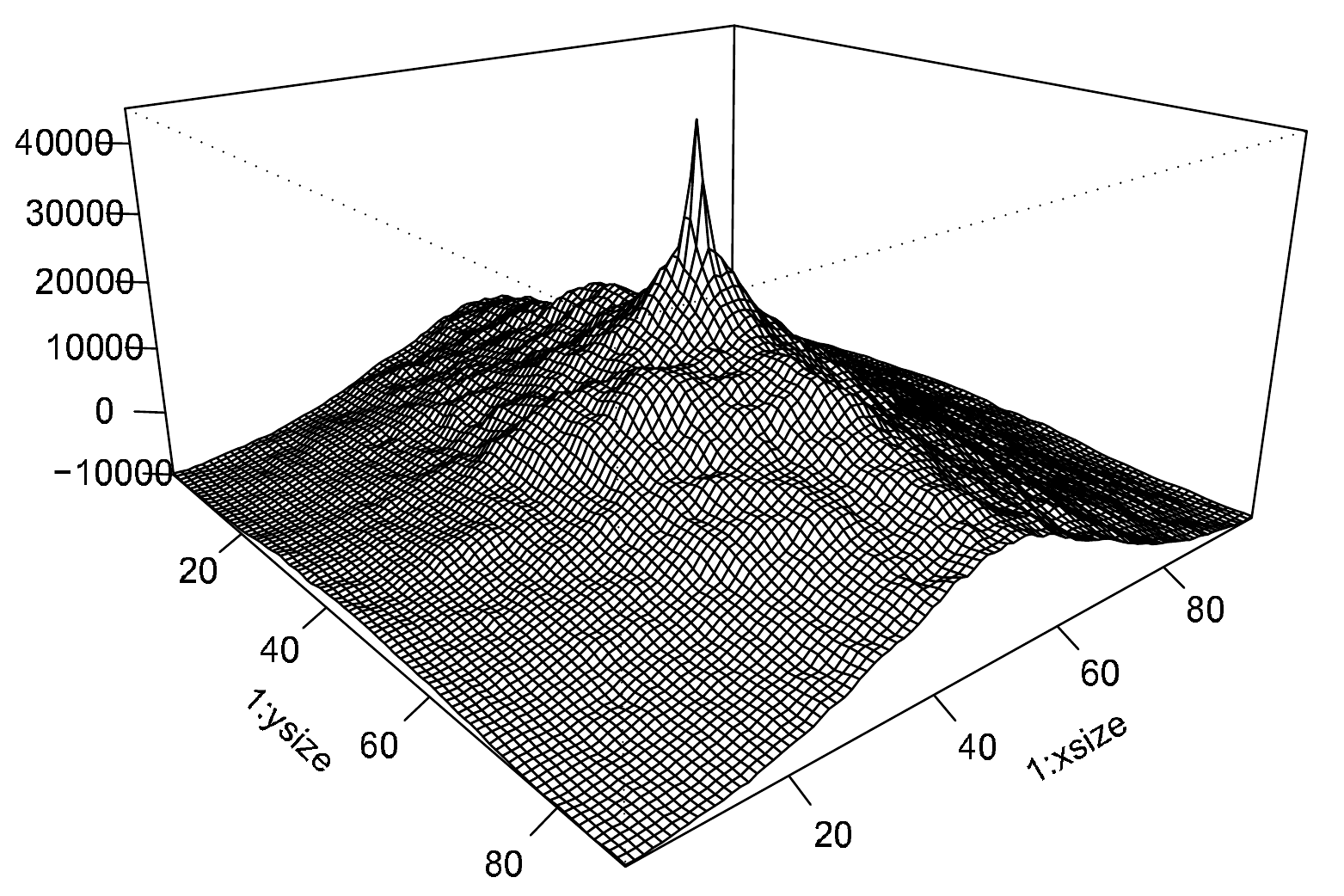}&
    \includegraphics[width=0.3\columnwidth]{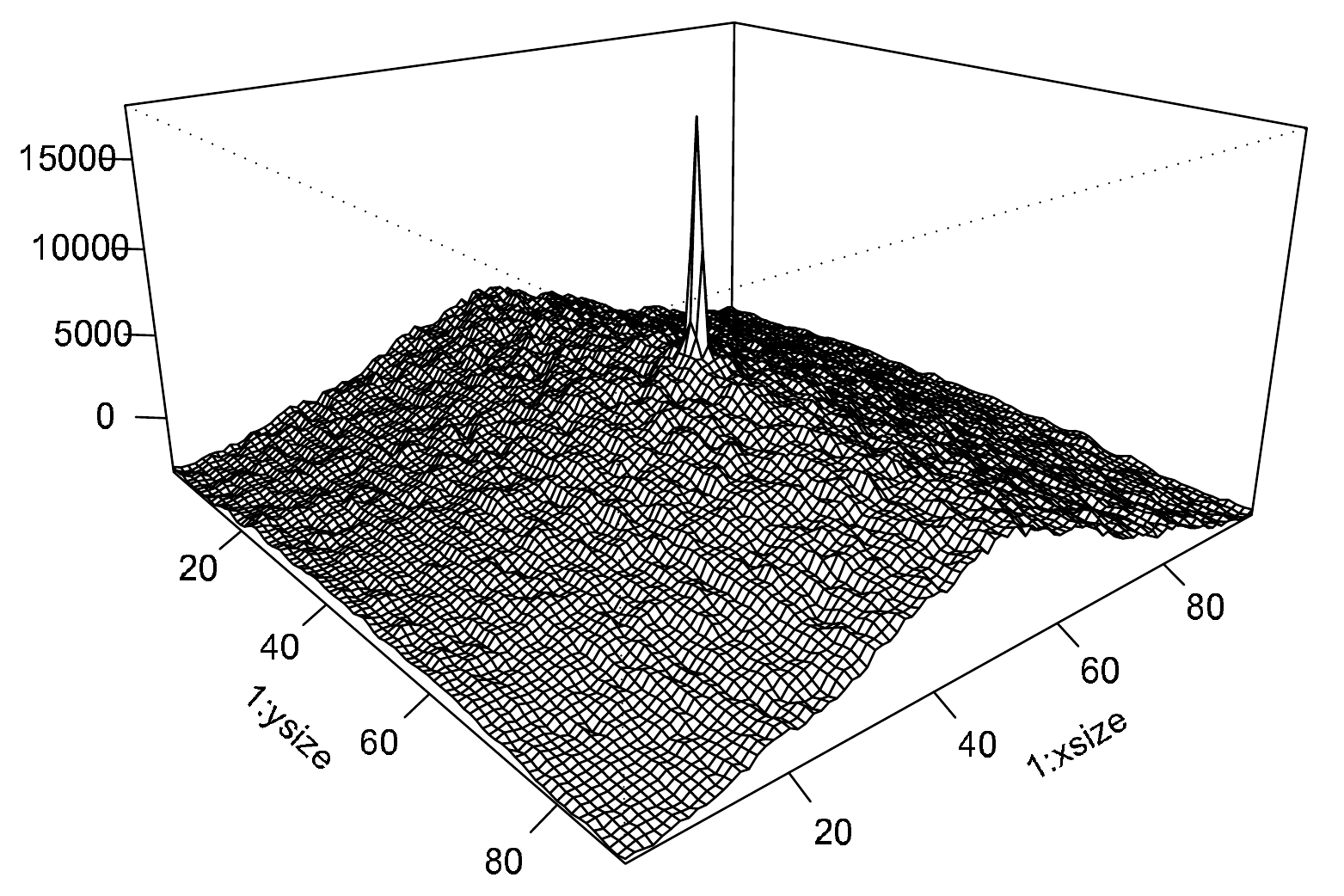}\\
    \footnotesize $\overline{g}_{r_{\lambda_{1}}}$ &
    \footnotesize $\overline{g}_{r_{\lambda_{12}}}$ &
    \footnotesize $\overline{g}_{r_{\lambda_{512}}}$\\
\end{tabular}
  \caption{Three channels of the DCE-MRI series before, $f_{\la_i}$, and after 2
  FCA-reconstructions with 16 axes $\fr^{(2)}$, their residues
  $\mathbf{r}_{\la}$ and the covariances on the residues
  $\overline{g}_{r_{\la}}$. The histogram of the
  residues has been normalised for visualisation.}
  \label{fig:filter:DA:_residues}
\end{figure}

Some spectra of the reconstructed image after 2 FCA are plotted in
figure \ref{fig:filter:DA:_5_spectra}. The filtered spectra by 2
sequences of FCA-reconstruction $\fr(x_{i})$ have a smaller
variability than the original spectra. Moreover after the filtering
stage the general trend of the spectrum is enhanced. For example,
the signal of the spectra corresponding to the tumour signal is
increasing. This corresponds to the accumulation of the product of
contrast inside the tumour.

In conclusion we propose a noise reduction method which consists in
applying two series of FCA-reconstructions. This approach reduces
the temporal (spectral) noise while preserving the contours. This is
an important point for a further segmentation.

\begin{figure}[!htb]
\begin{center}
\begin{tabular}{@{}c|c@{}}
  \includegraphics[width=0.45\columnwidth]{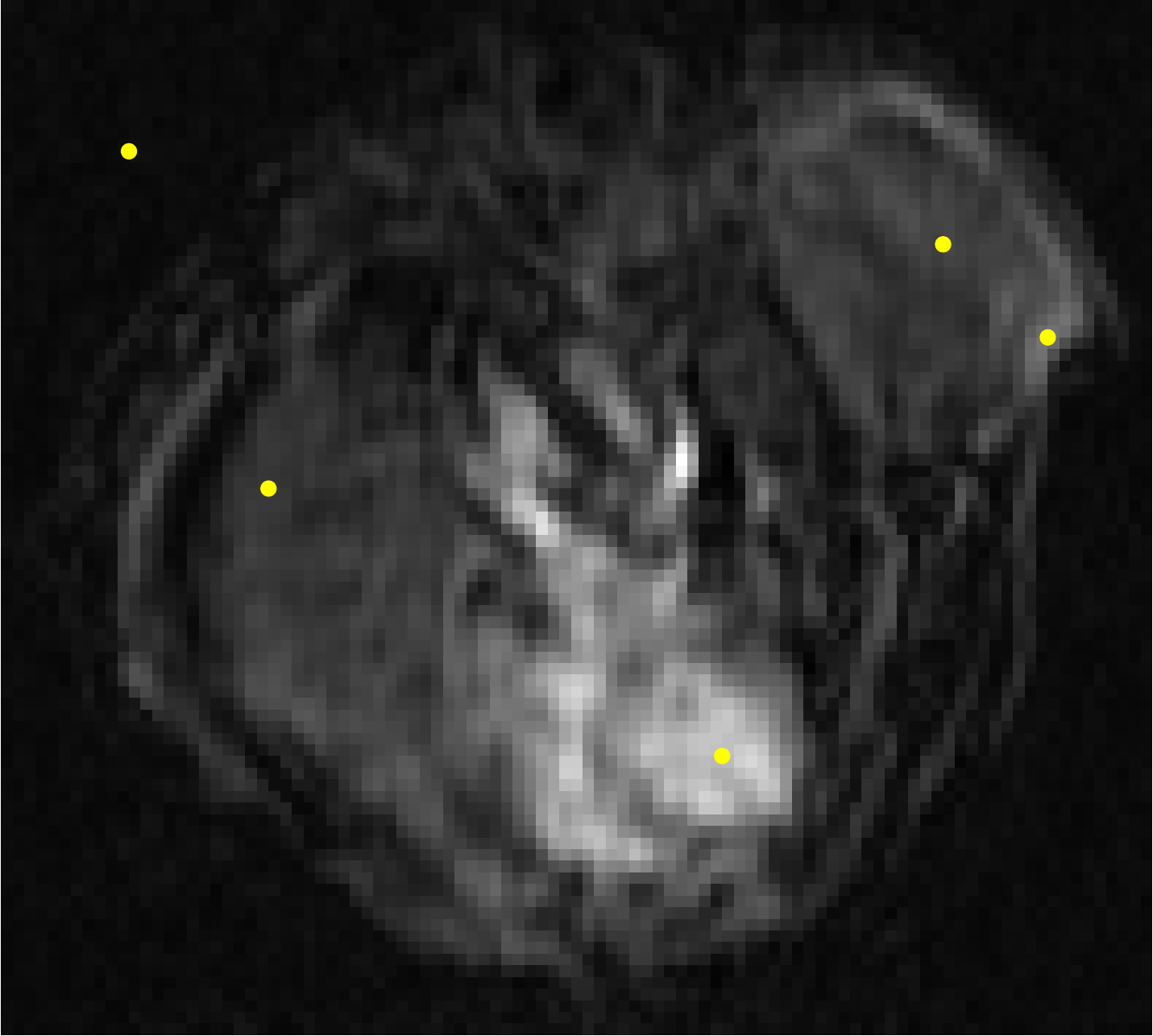}&
  \includegraphics[width=0.45\columnwidth]{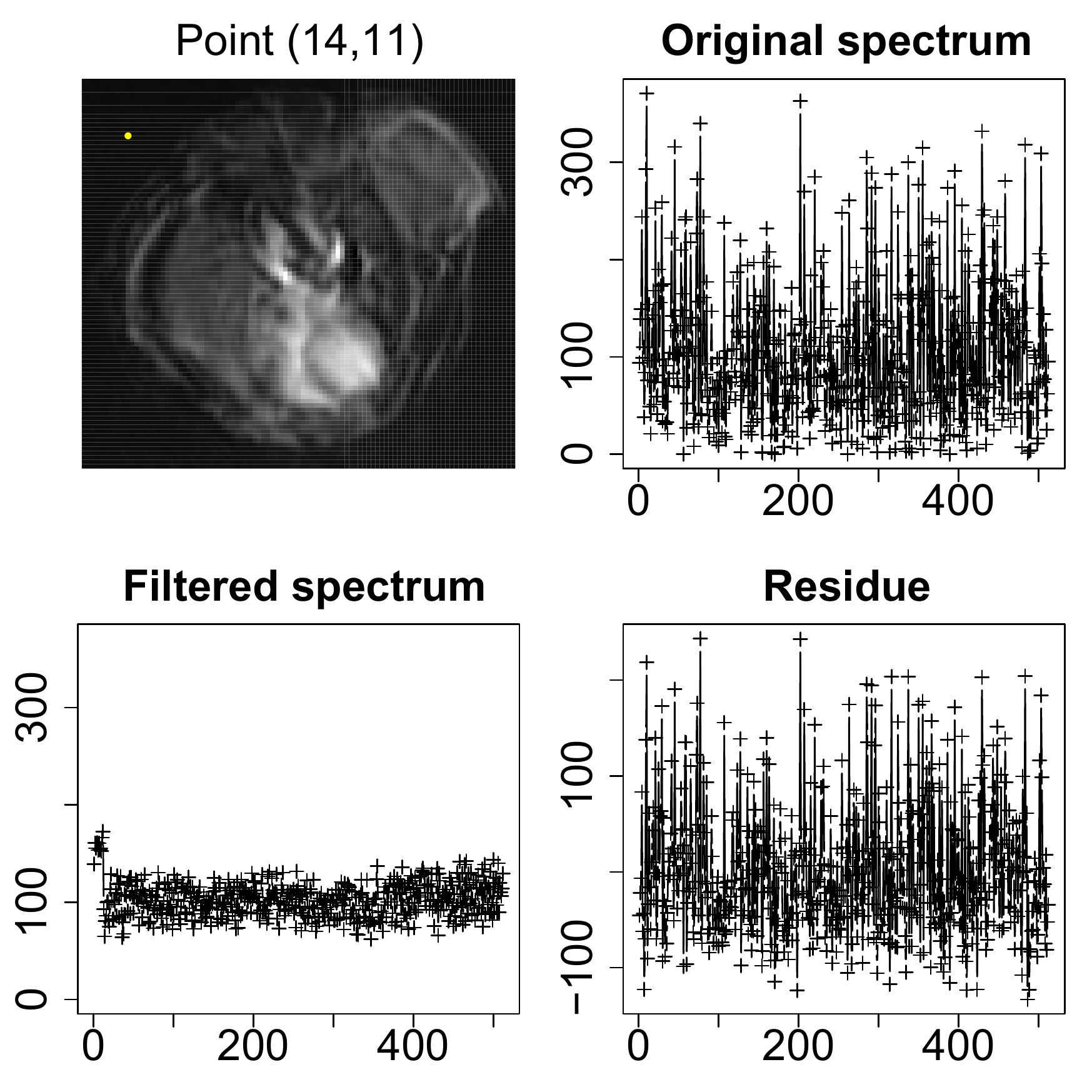}\\
  \hline
  \includegraphics[width=0.45\columnwidth]{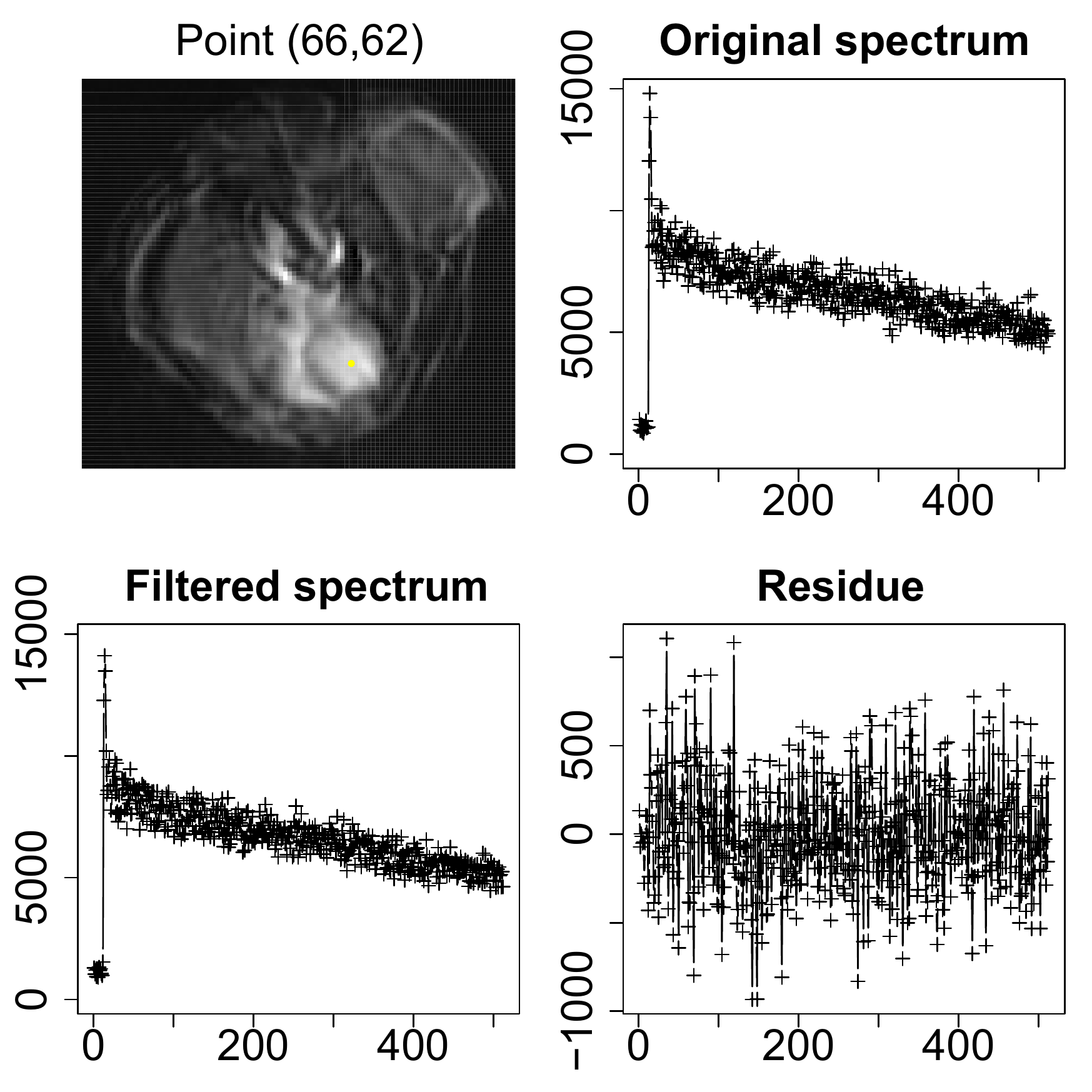} &
  \includegraphics[width=0.45\columnwidth]{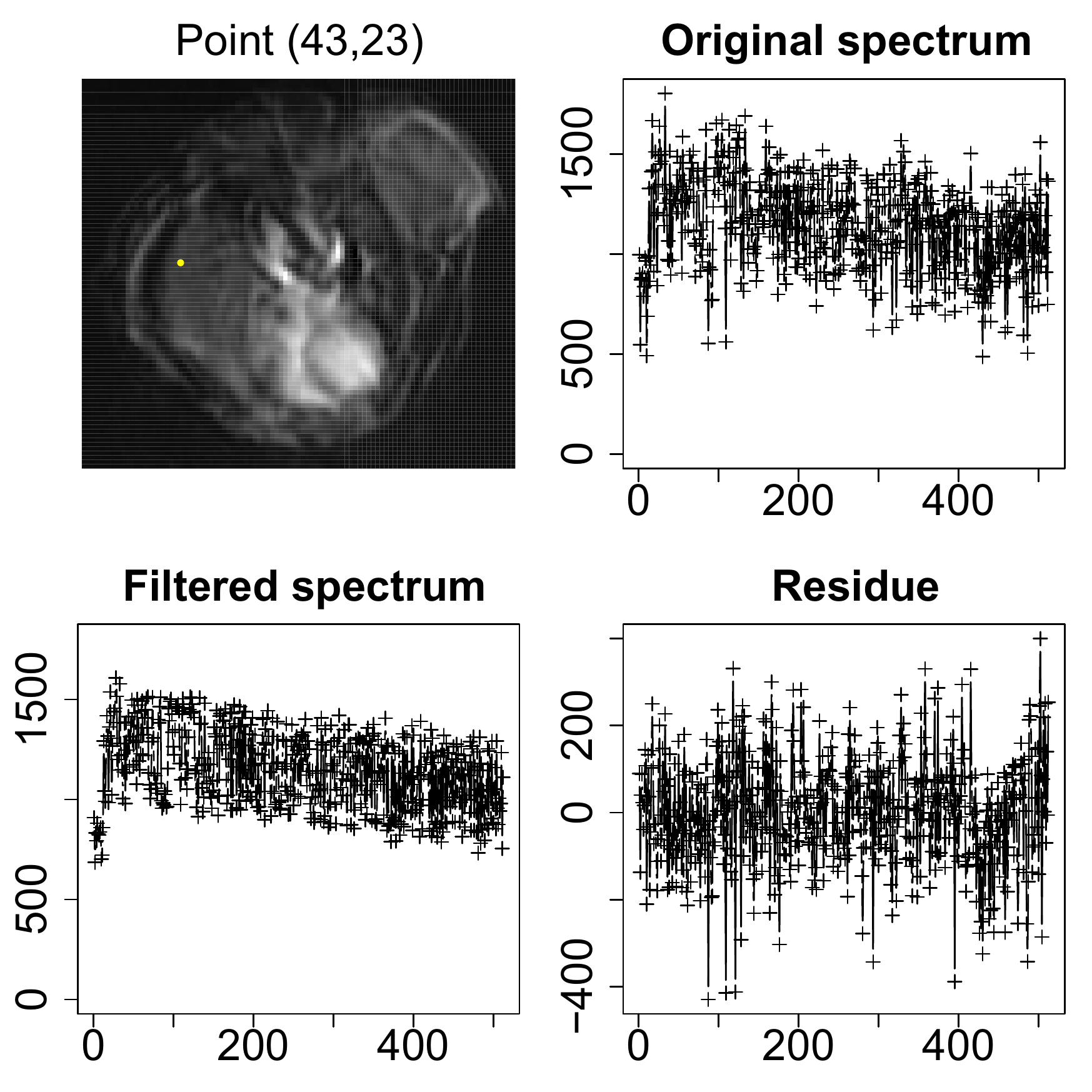} \\
  \hline
  \includegraphics[width=0.45\columnwidth]{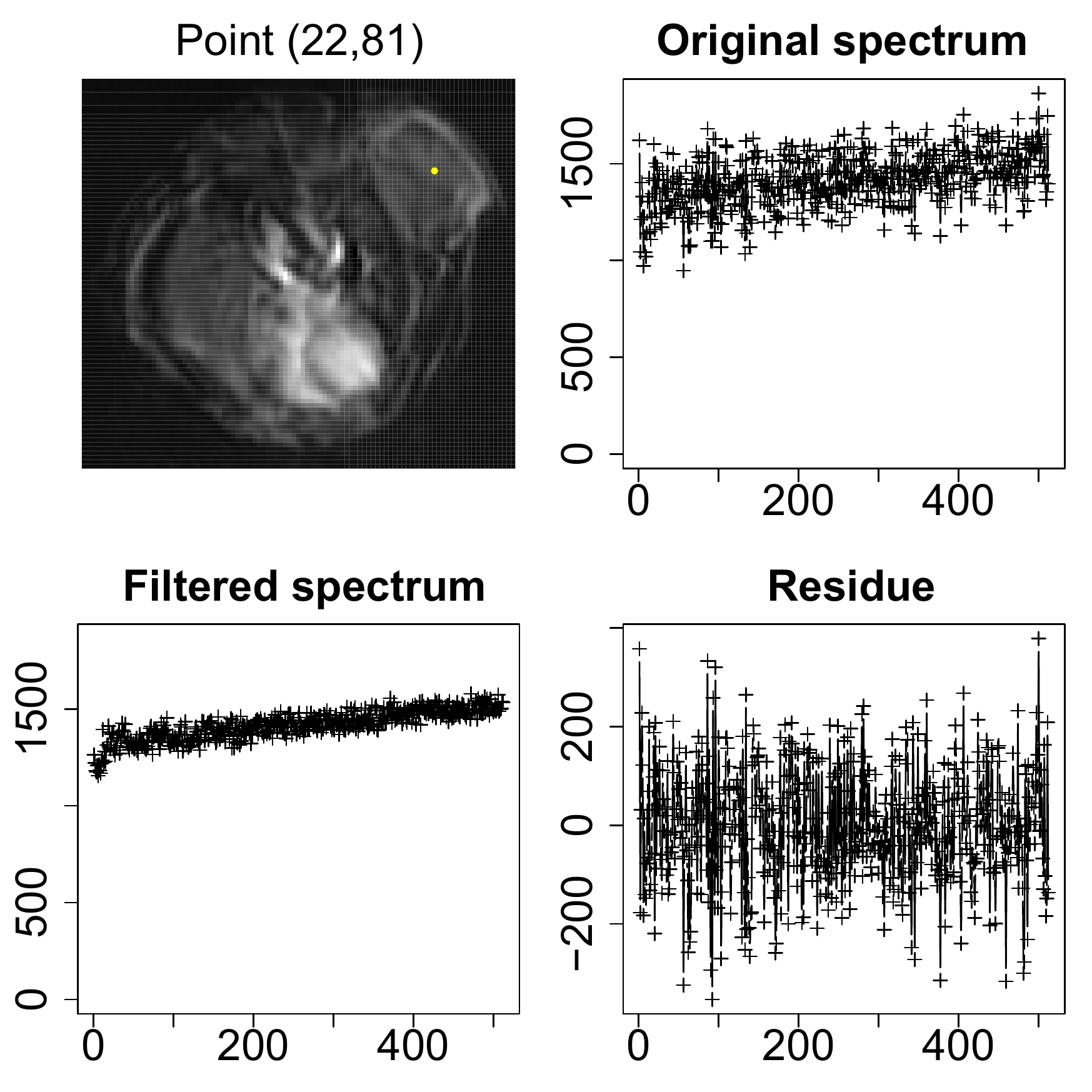}  &
  \includegraphics[width=0.45\columnwidth]{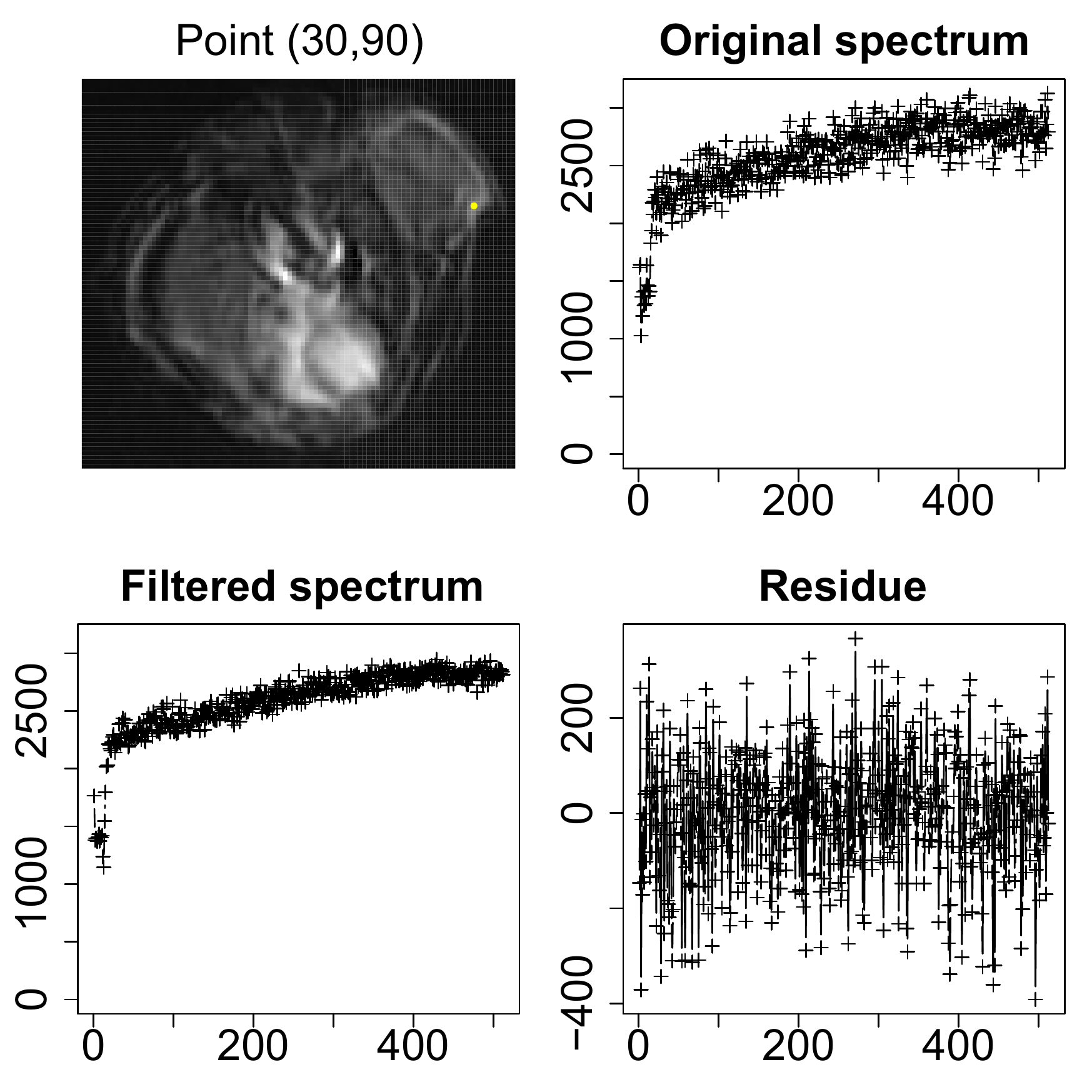} \\
\end{tabular}
  \caption{5 spectra before and after 2 FCA-reconstructions (with 16 axes) and the corresponding residues.}
  \label{fig:filter:DA:_5_spectra}
\end{center}
\end{figure}

\FloatBarrier
\subsection{Strong dimensionality reduction by spectrum modeling}
\label{sec:filter:model}

Following the noise reduction which preserves the spatial
information by data analysis, an additional dimensionality reduction
by spectrum modeling is proposed.

In this approach, a parametric model is fitted to each spectrum and
consequently, the images of parameters can be seen as maps. These
maps are useful for classification and morphological segmentation.
We notice that the set of these maps of parameters
$(p_{1}(x),\ldots,p_{M}(x))$ constitutes a multivariate image with a
reduced dimension:

\begin{equation}\label{eq:filter:model:_im_para}
    \mathbf{p}(x)=(p_{1}(x),\ldots,p_{M}(x))
\end{equation}

The fitted model may take into account the physical phenomena, such
as the physiological pharmacokinetic model used in
\citet{Brochot_2006}. Their model is based on differential equations
on six compartments: arterial and venous plasma, tumour (split into
capillaries and interstitium), and the rest of the body (also split
into capillaries and interstitium). However, in the current study we
adopt a simpler linear model. More precisely, for each spectrum
(time series) of the filtered image $\fr$, a line model is fitted
after removing the first 20 values which correspond to a transitory
phenomenon (fig. \ref{fig:filter:model:_model_fitting})

\setlength{\arraycolsep}{0.0em}
\begin{eqnarray}\label{eq:filter:model:_line_model}
    &&\widehat{\mathbf{f}}_{\la}(x_i) \sim a(x_i) \la + b(x_i) \qquad \forall i =  1 \ldots P\\
    &&\text{with } \la = \la_{j_1} \ldots \la_{512}\nonumber
\end{eqnarray}
\setlength{\arraycolsep}{5pt}


with $j_1$ the first value after the peak ($j_1 = 21$ for our
images).

This model is made with two parameters: the slope $p_1=a$ and the
intercept $p_2 = b$.

In order to take into account some information contained in the
transitory part of the time series, a third parameter is used. It
corresponds to the amplitude of the signal over the twenty first
values of the spectrum. This transitory part is characteristic of
the injection of the contrast agent used in this imaging modality.
This parameter is called the rise $p_3 = m$ and it is defined by:

\begin{eqnarray}
  &&m(x_i) = \max_{\substack{j\in[1;j_1 - 1]}}(\widehat{f}_{\lambda_{j}}(x_i))-
  \min_{\substack{j\in[1;j_1 - 1]}}(\widehat{f}_{\lambda_{j}}(x_i)) \nonumber \\
  &&\qquad \forall i = 1 \ldots P
\end{eqnarray}

\begin{figure}[!htb]
\centering
  \includegraphics[width=0.5\columnwidth]{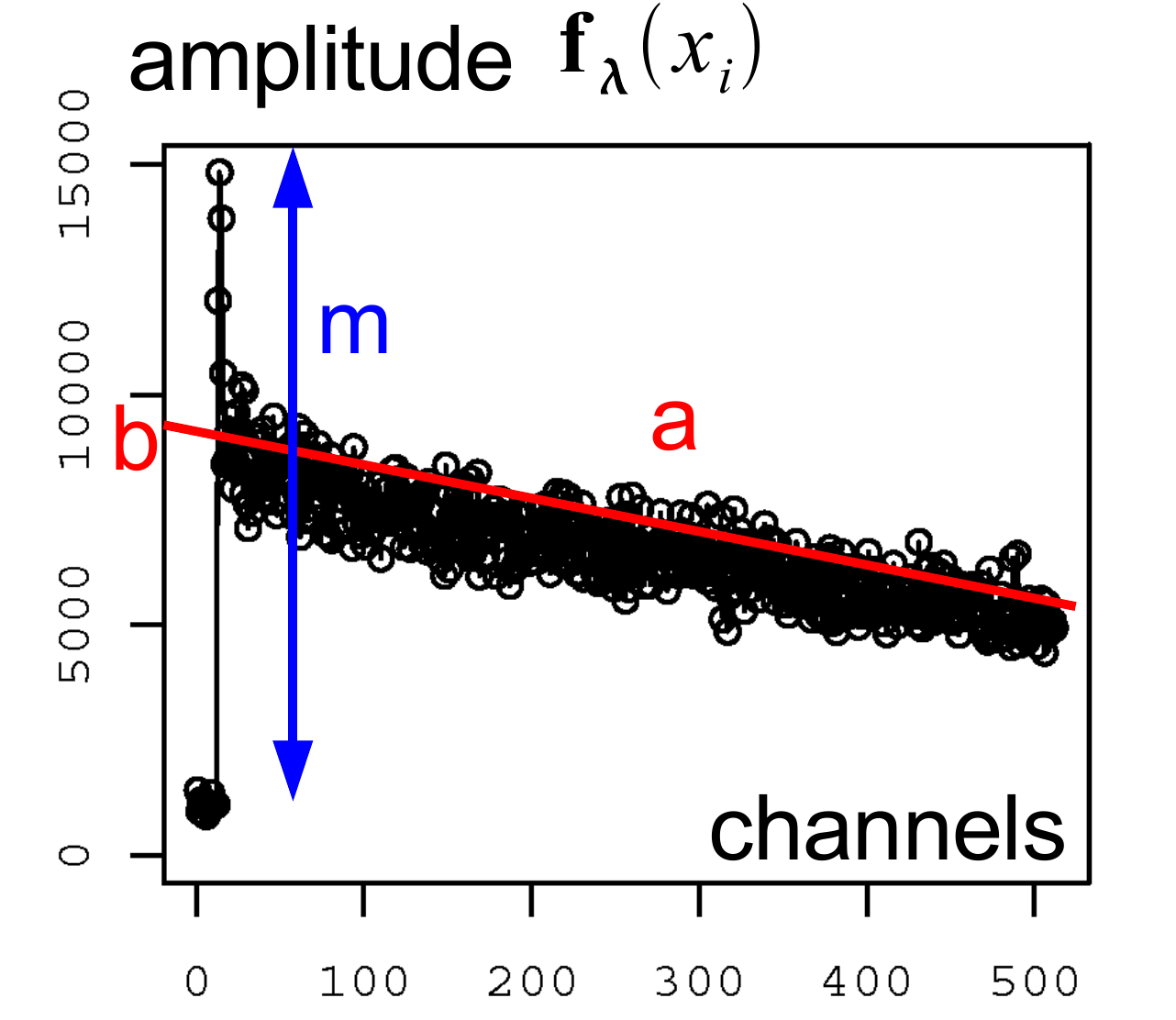}
  \caption{Model fitting on a spectrum $\mathbf{f}_{\la}(x_i)$.}
  \label{fig:filter:model:_model_fitting}
\end{figure}

Hence, by fitting the model for each spectrum $\f(x_i)$, three maps
of parameters are obtained (fig.
\ref{fig:filter:model:_parameters}). We remark that the
dimensionality reduction is very important because the original
multivariate image with 512 channels is transformed into another
multivariate image with only 3 channels. Even with this reduced
amount of information, the main morphological structures of the
mouse (the heart cavities, the tumour and the lungs) are clearly
apparent.

\begin{figure}
\centering
\begin{tabular}{ccc}
  \includegraphics[width=0.3\columnwidth]{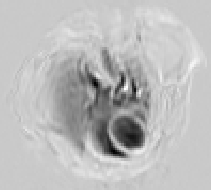}&
  \includegraphics[width=0.3\columnwidth]{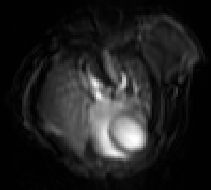}&
  \includegraphics[width=0.3\columnwidth]{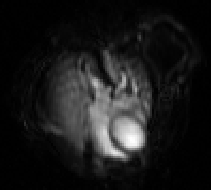}\\
  \footnotesize slope $a$ &
  \footnotesize intercept $b$ &
  \footnotesize rise $m$\\
\end{tabular}
\caption{Parameters maps of the linear model fit for each pixel.}
\label{fig:filter:model:_parameters}
\end{figure}

To conclude, spectrum modeling is very efficient in reducing the
number of channels for tumour segmentation. It could be interesting
in further studies to fit a a model taking into account
physical properties of the injection of the contrast agent. 

%

\section{Temporal classification of DCE-MRI time series}
\label{sec:classif}


After performing a denoising step and a dimensionality reduction,
the classification of pixels is made in the temporal dimension.
Supervised and unsupervised methods are considered.

\subsection{Unsupervised approach: regional improved k-means}

For the unsupervised approach, a k-means classification and a model
classification are compared.

A  k-means classification of pixels is based on the Euclidean metric
\citep{Diday_1971,Hartigan_1979}. This classifier must be used in a
space in which the associated metric is Euclidean. It is the case
for the factor space of FCA or of PCA. However, using k-means in the
image space would overweight channels with large dynamics. So
classifying is done in the factor space instead. For DCE-MRI series,
the k-means classification is performed with 5 classes in the factor
image space $\ca$ of the second sequence of FCA-reconstruction. In
figure \ref{fig:classif:unsupervised_classifications} some
anatomical parts of the mouse have been classified into 5 classes.
The number of classes has been empirically defined. The classes are
as follows: (1) the green class corresponds largely to the tumour -
see top right corner of the image, (2) the red class corresponds
mainly to the background, (3) the blue class corresponds to the
heart cavities, (4) the black class corresponds to the lung and (5)
the cyan class is an intermediate class.

\newpage
\begin{figure}
\centering
\begin{tabular}{@{}c@{ }c@{ }c@{}}
    \includegraphics[width=0.3\columnwidth]{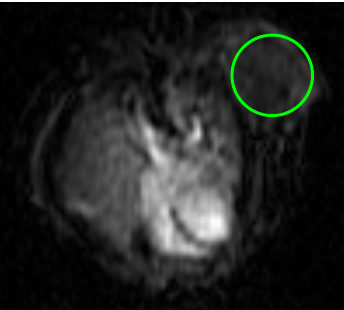}&
    \includegraphics[width=0.3\columnwidth]{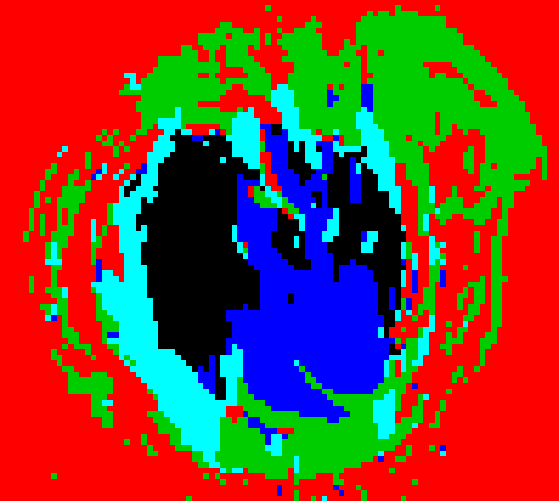}&
    \includegraphics[width=0.3\columnwidth]{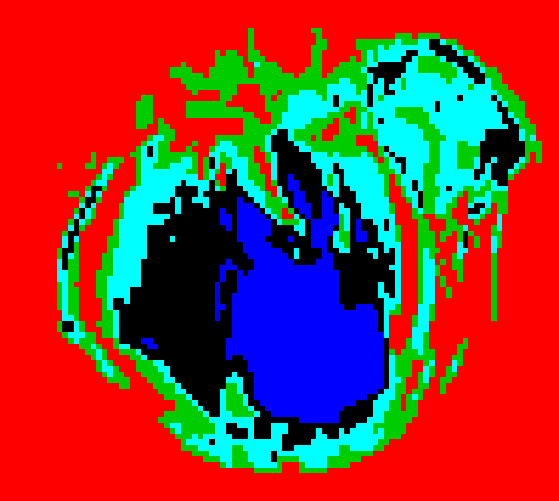}\\
    \footnotesize reference &
    \footnotesize k-means &
    \footnotesize model\\
\end{tabular}
\caption{(a) Reference $ref$,
  (b) k-means classification in 5 classes $\kappa^{kmeans,5}_{\ca}$,
  (c) model classification $\kappa^{mod,5}_{\fr}$ in 5 classes
  on the filtered image $\fr$.}
\label{fig:classif:unsupervised_classifications}
\end{figure}


The results have not been completely satisfactory, so we have
introduced an alternative approach of classification, which clusters
each spectrum in comparison to some reference spectra obtained on
the k-means classification.

\begin{figure}[!htb]
\centering
    \includegraphics[width=0.9\columnwidth]{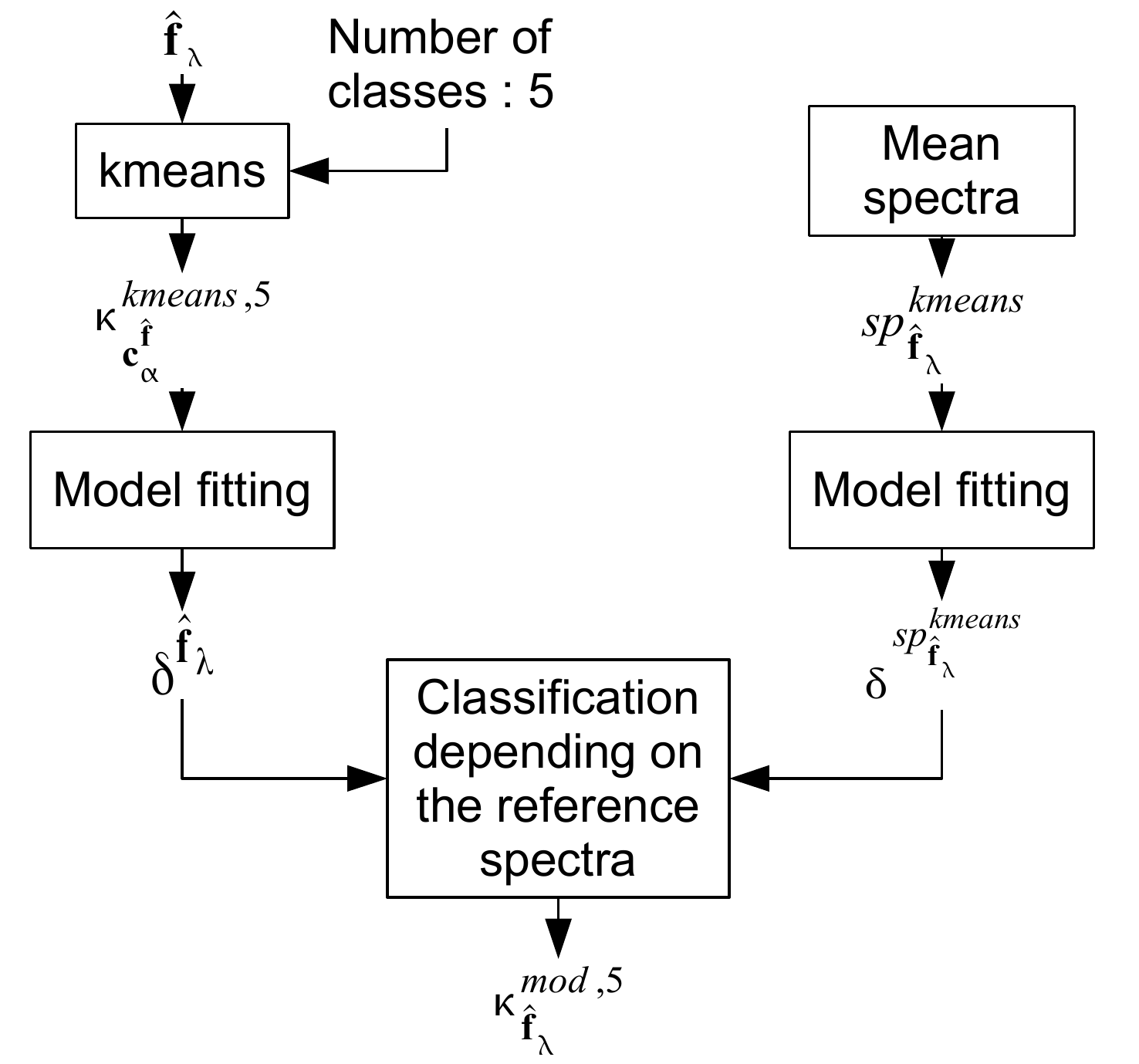}
  \caption{Framework of the classification by model approach.}
  \label{fig:classif:_schema_classif_mod}
\end{figure}

For each class obtained by k-means, a mean filtered spectrum is
computed $sp^{kmeans}_{\fr}$. On these mean spectra, a line model is
fitted $\delta^{sp^{kmeans}_{\fr}} = \{ \delta^{sp^{kmeans}_{\fr}}_k
\}_{k=1}^5$. For each spectrum $\mathbf{\widehat{f}}_{\la}(x_i)$ of
the filtered image by two sequences of FCA-reconstruction the line
model is fitted $\delta^{\fr} = \{ \delta^{\fr}_{x_i} \}_{i=1}^P$.
Then, each point $x_i$ is classified by minimisation of the $L_1$
distance between the model of the mean spectra
$\delta^{sp^{kmeans}_{\fr}}_k$ and the model for each pixel
$\delta^{\fr}_{x_i}$. The class of the point $x_i$ is written
$C(x_i)$ (eq. \ref{eq:classif:_min_L1_distance_spectra} and fig.
\ref{fig:classif:_im_distance_spectra}).

\setlength{\arraycolsep}{0.0em}
\begin{eqnarray}\label{eq:classif:_min_L1_distance_spectra}
C(x_i)&{}={}&\argmin_k d_1 (\delta^{sp^{kmeans}_{\fr}}_k , \delta^{\fr}_{x_i})\nonumber\\
&{}={}&\argmin_k (\sum_{j=1}^L |\delta^{sp^{kmeans}_{\fr}}_k(\la_j) - \delta^{\fr}_{x_i}(\la_j)| )\nonumber\\
\end{eqnarray}
\setlength{\arraycolsep}{5pt}


\begin{figure}[!htb]
\centering
    \includegraphics[width=1\columnwidth]{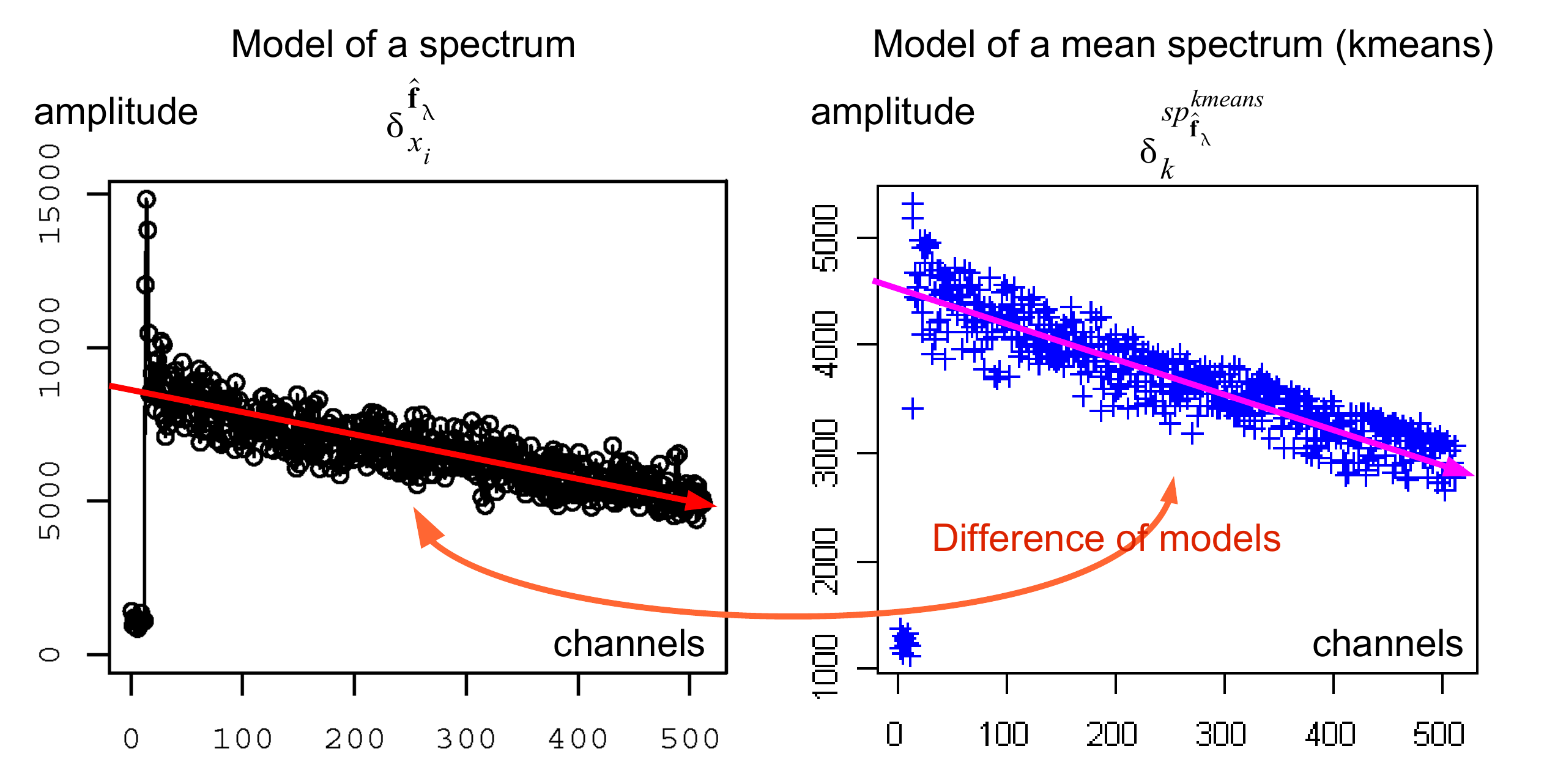}
  \caption{Minimisation of the $L_1$ distance between the model of the mean spectra $\delta^{sp^{kmeans}_{\fr}}_k$ and the model
  of the spectra at each pixel $\delta^{\fr}_{x_i}$.}
  \label{fig:classif:_im_distance_spectra}
\end{figure}

\begin{figure}[!htb]
\centering
\begin{tabular}{@{}c@{ }c@{ }c@{}}
    \includegraphics[width=0.33\columnwidth]{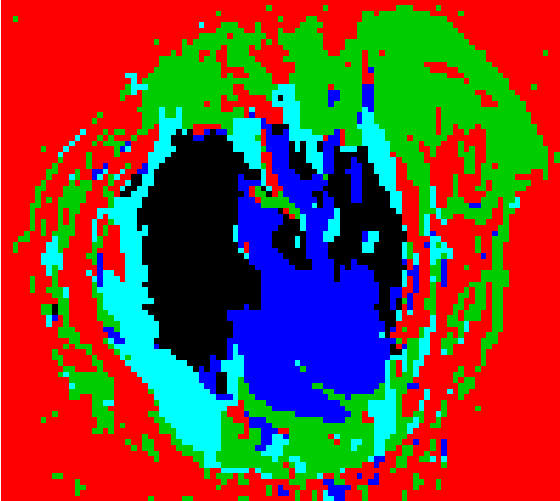}&
    \includegraphics[width=0.33\columnwidth]{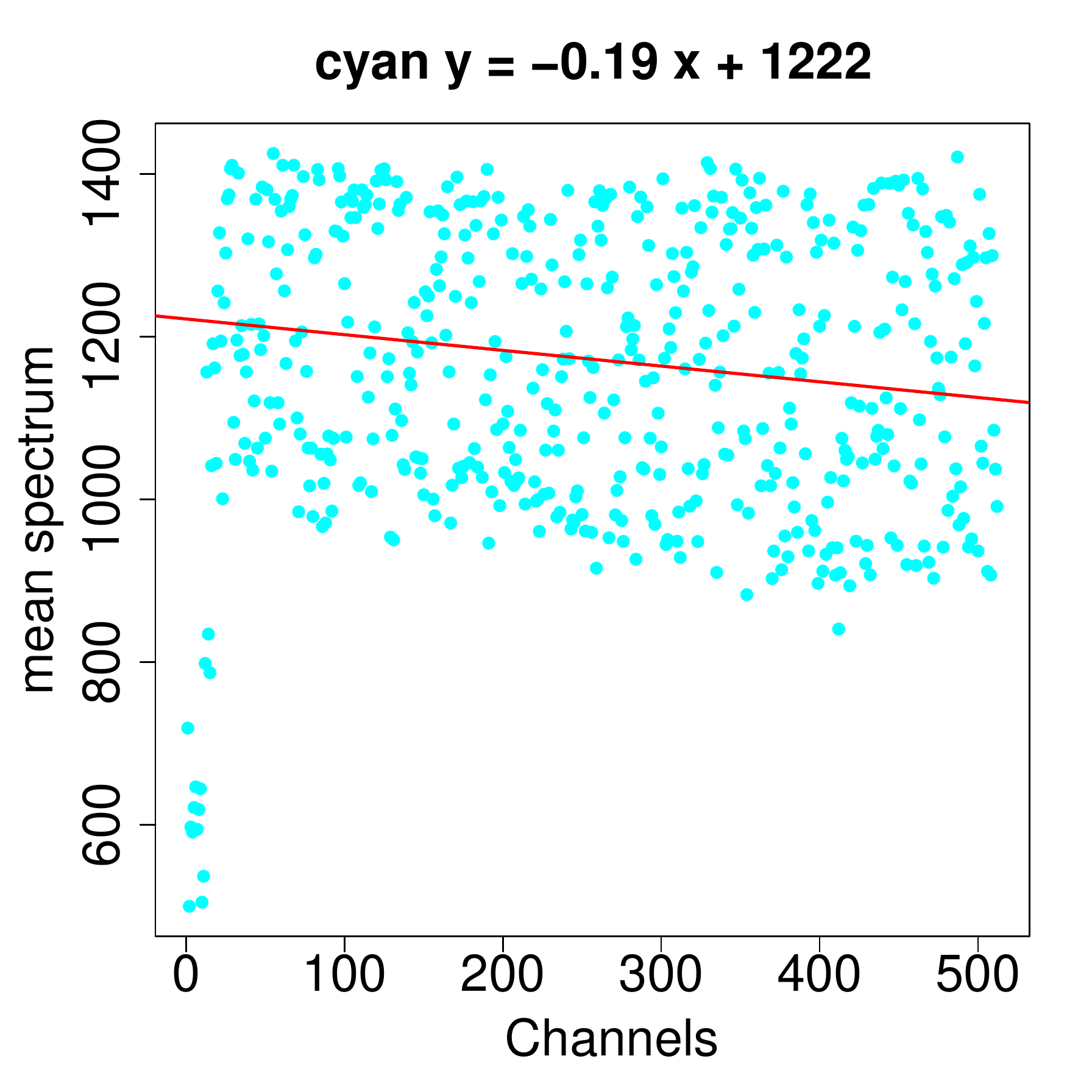}&
    \includegraphics[width=0.33\columnwidth]{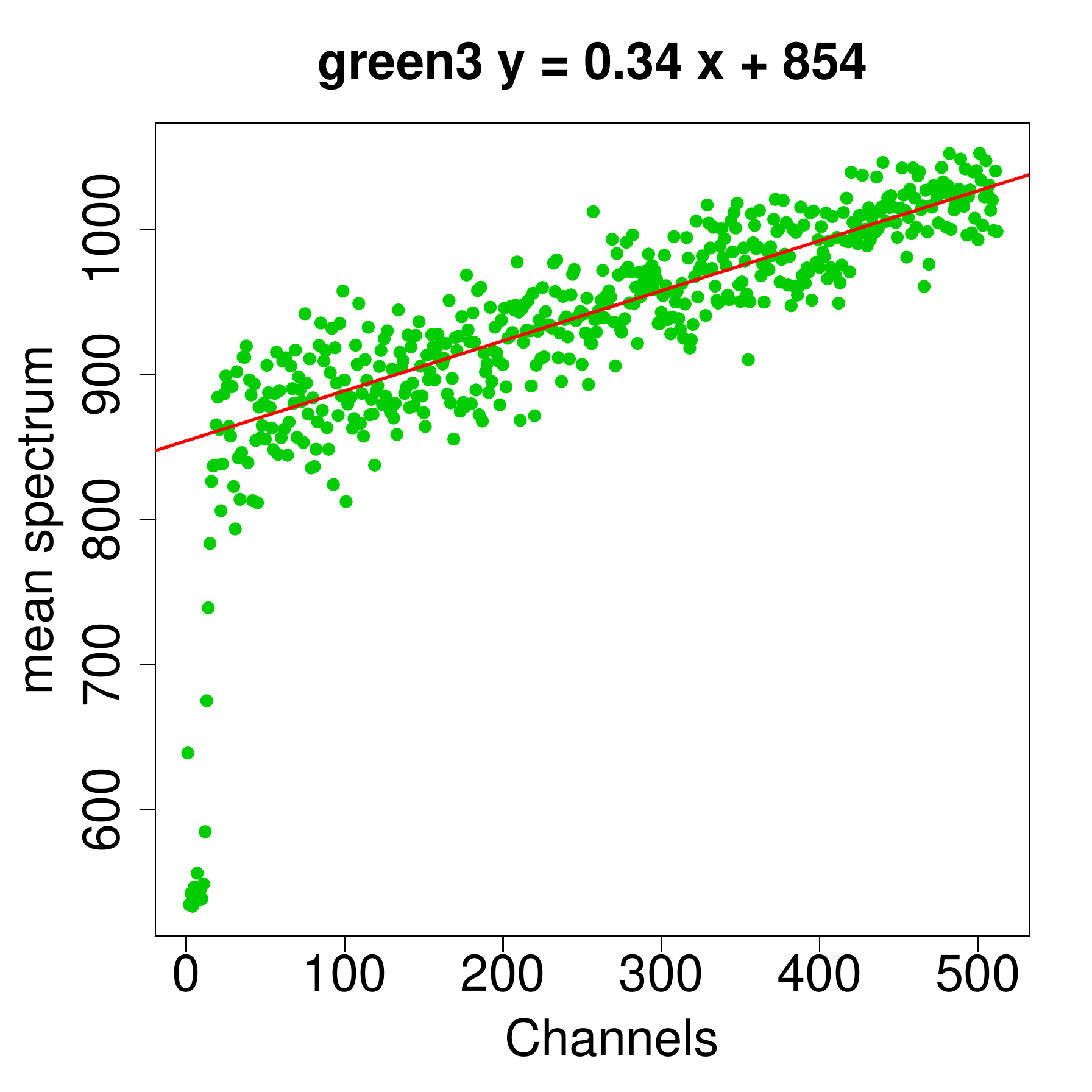}\\
    \includegraphics[width=0.33\columnwidth]{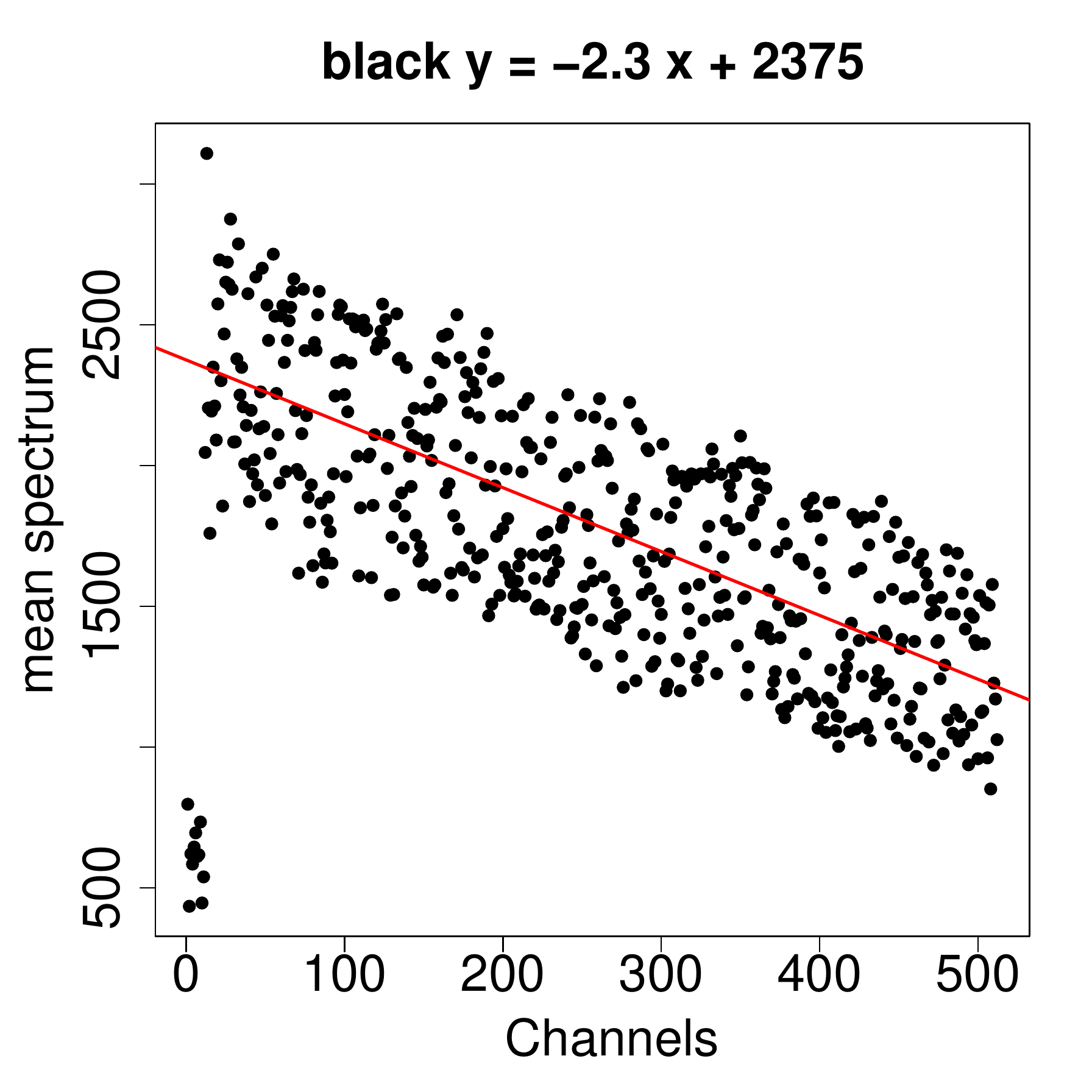}&
    \includegraphics[width=0.33\columnwidth]{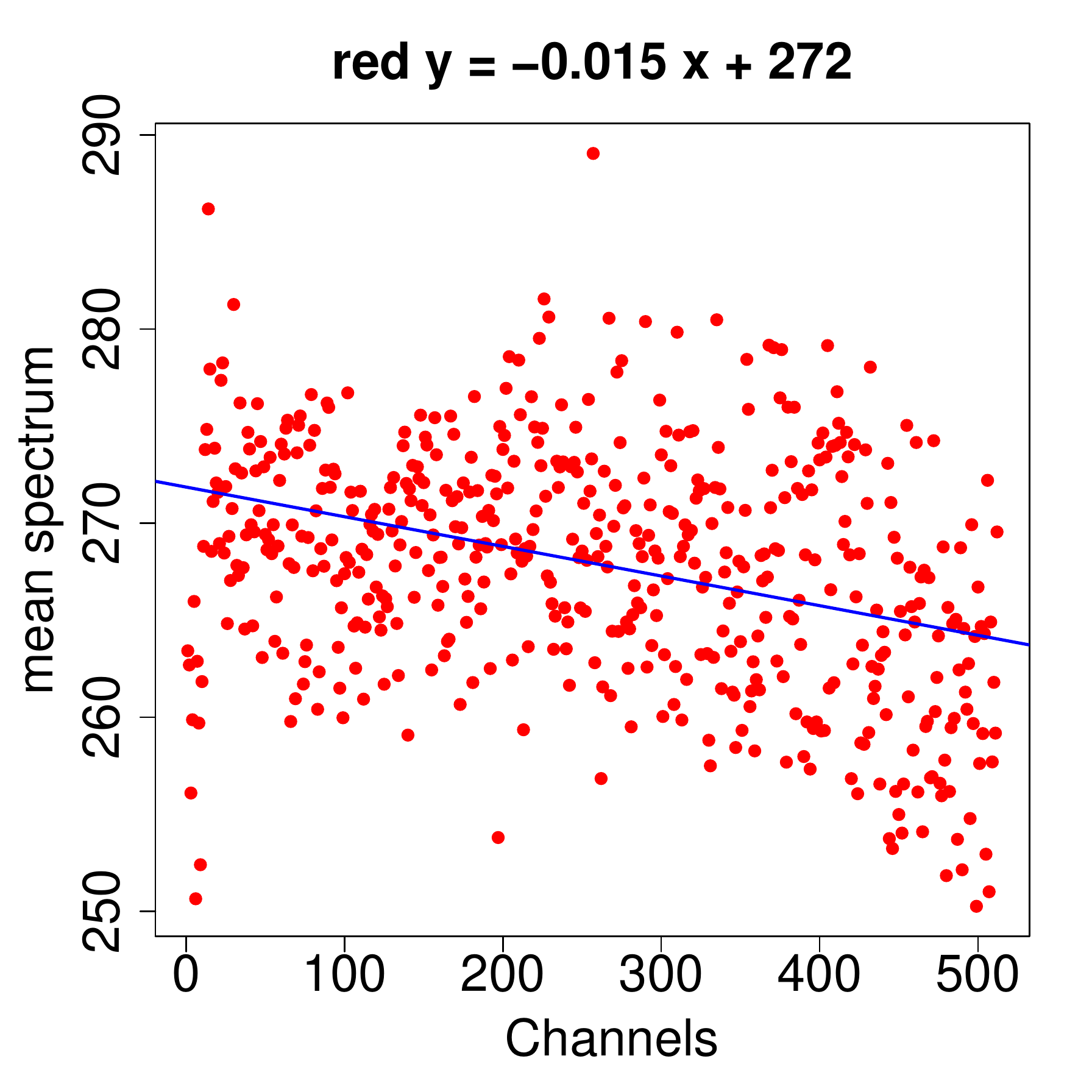}&
    \includegraphics[width=0.33\columnwidth]{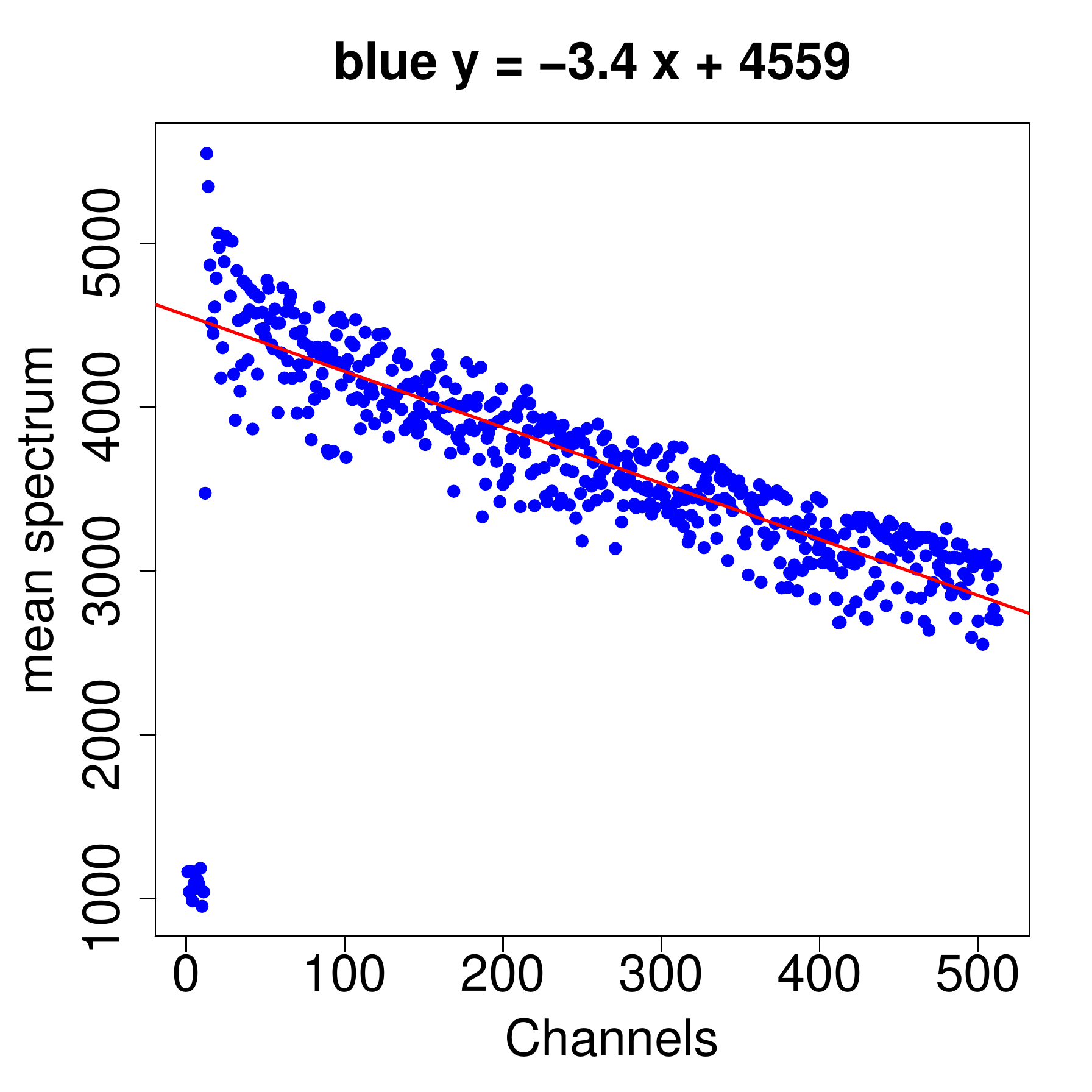}\\
\end{tabular}
  \caption{Mean spectra of the k-means classification $\kappa^{kmeans,5}_{\ca}$.}
  \label{fig:classif:_means_spectra}
\end{figure}

As one can notice in figures
\ref{fig:classif:unsupervised_classifications} and
\ref{fig:classif:classif_LDA_para_series}, the classification
$\kappa^{mod,5}_{\fr}$ by the model approach seems to be more robust
than the k-means classification because a statistical noise
filtering has been made when fitting a model with 3 parameters.
Then, the k-means classification has been computed on 19 channels
while the model classification has been computed on 3 channels with
few noise. Therefore, the model decreases the entropy of the image
by introducing a prior information present in the shape of the
spectra. Nevertheless, the k-means classification is necessary as a
first step, in order to estimate the mean spectra
$\delta^{sp^{kmeans}_{\fr}}_k$.


\FloatBarrier
\subsection{Supervised approach: LDA with histogram
normalisation}

A semi-supervised classification by Linear Discriminant Analysis
(LDA) is also considered for the DCE-MRI series. In particular, LDA
is based on a train set composed of four distinct parts of the
anatomy of the mouse:
\begin{itemize}
  \item the tumour in green $train(green)=t_1$
  \item the heart cavities in blue $train(blue)=t_2$
  \item the background in red $train(red)=t_3$
  \item the lungs in black or white $train(black)=t_4$.
\end{itemize}

Each class of the training set, $T=train=(t_1,t_2,t_3,t_4)$, is made
of 80 vector-pixels , $\mathbf{f}_{\la}(x_i)$ with 512 components,
selected by an operator. By measuring the mean spectra of the
filtered image $\fr$ on these classes $sp^{train}_{\fr}$, we notice
that the kinetics of train classes are different (fig.
\ref{fig:classif:LDA_train}).


\begin{figure}[!htb]
\centering
    \begin{tabular}{c}
    \begin{tabular}{cc}
    \includegraphics[width=0.33\columnwidth]{serim447_selection_train_LDA}&
    \includegraphics[width=0.3\columnwidth]{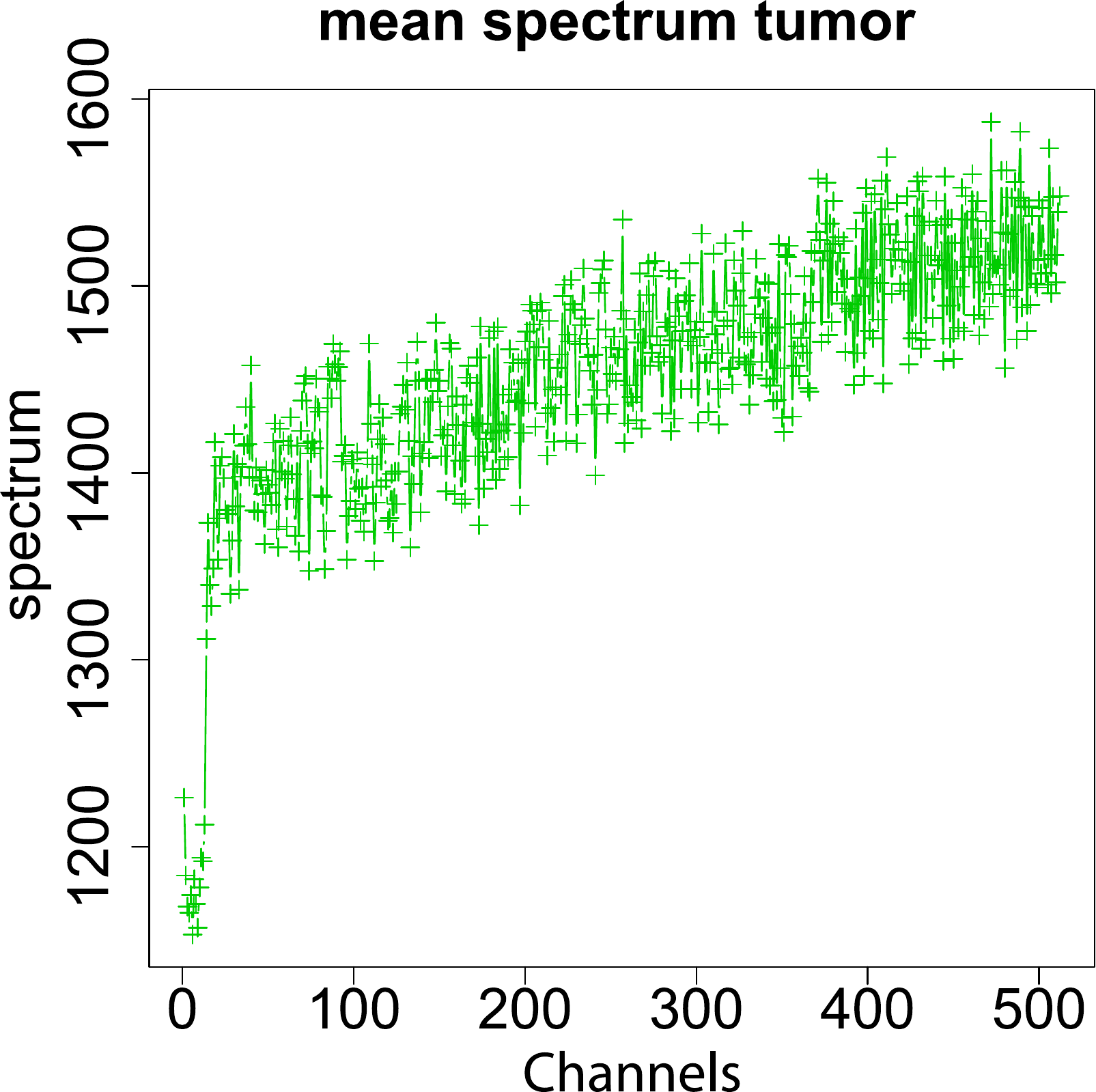}\\
    &\footnotesize  $sp^{train}_{\fr}(green)$\\
    \end{tabular}\\
    \begin{tabular}{@{}ccc@{}}
    \includegraphics[width=0.3\columnwidth]{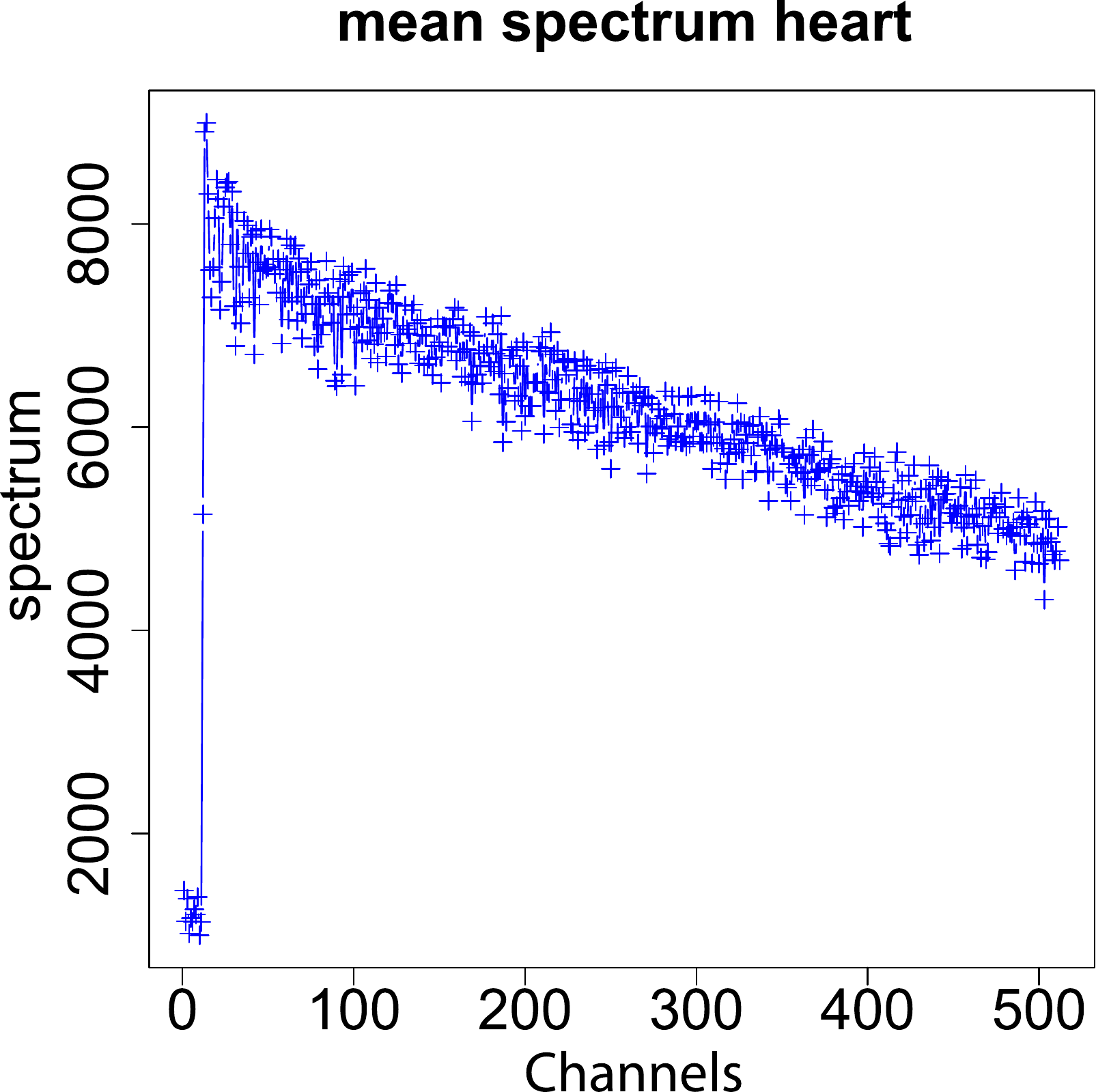}&
    \includegraphics[width=0.3\columnwidth]{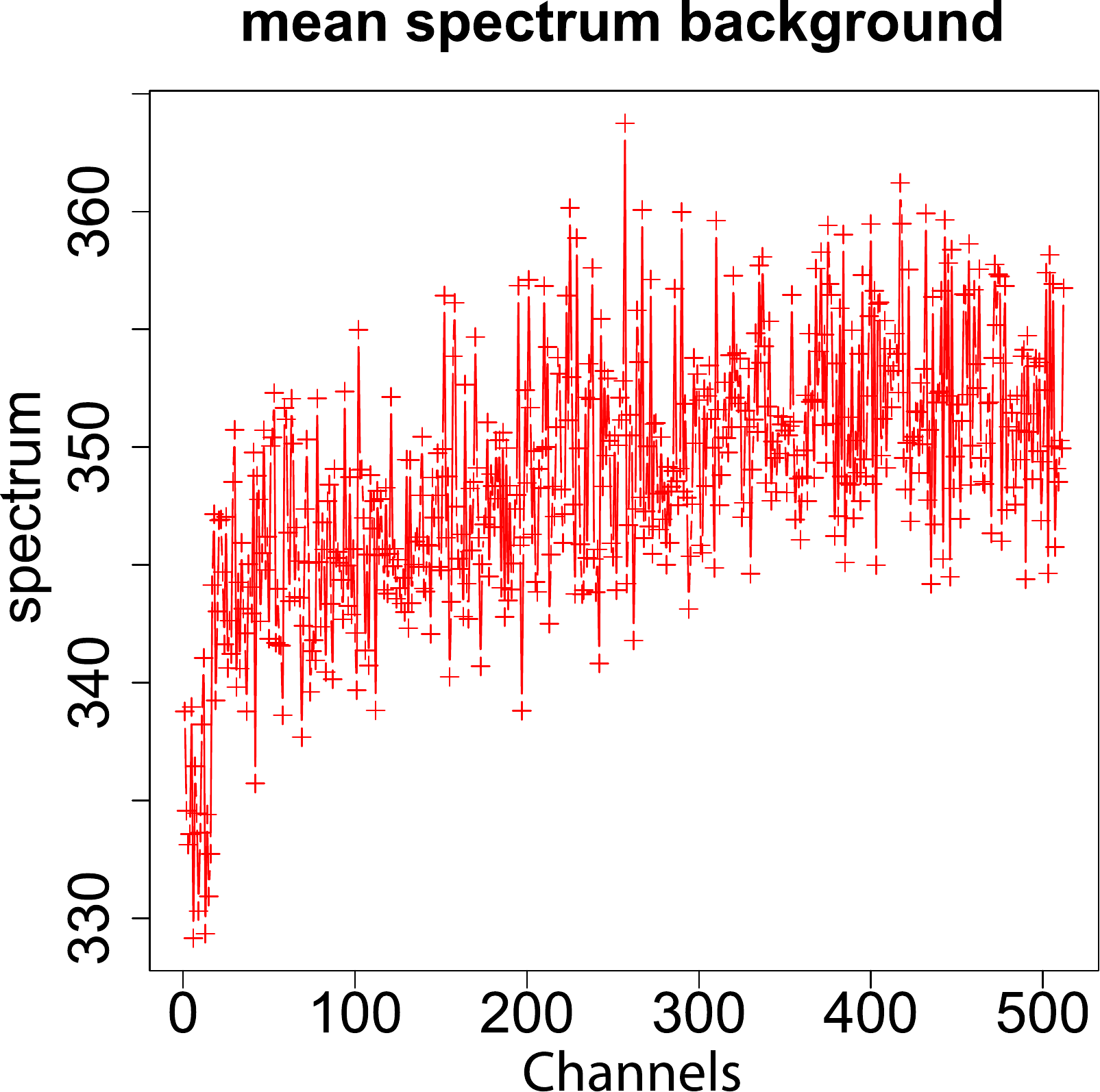}&
    \includegraphics[width=0.3\columnwidth]{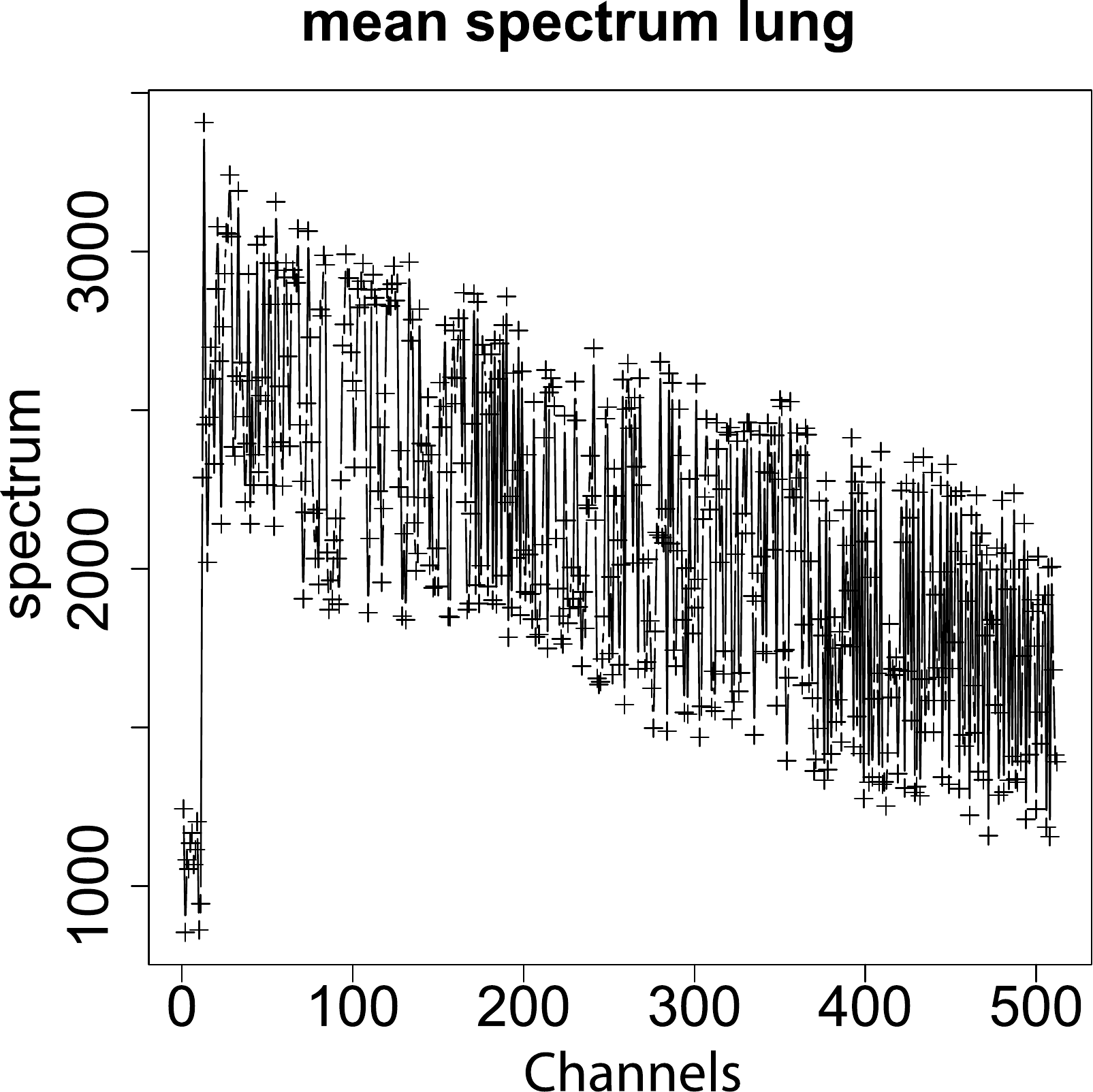}\\
    \footnotesize $sp^{train}_{\fr}(blue)$ &
    \footnotesize $sp^{train}_{\fr}(red)$ &
    \footnotesize $sp^{train}_{\fr}(black)$\\
    \end{tabular}
    \end{tabular}
  \caption{Classes of the training set, $train$, and mean spectra of
  these classes $sp^{train}_{\fr}$ on the filtered image $\fr$.}
  \label{fig:classif:LDA_train}
\end{figure}

The LDA is performed into three different spaces:
\begin{itemize}
  \item the filtered image space: $\{\widehat{\mathbf{f}}_{\la}(x) | x\in T\}$
  \item the PCA factor space of the spectra of the training set: $\zeta(\{\widehat{\mathbf{f}}_{\la}(x) | x\in T\})$
  \item the three parameters space: $\{\mathbf{p}(x) | x\in T\}$
\end{itemize}

In figure \ref{fig:classif:LDA_classifications}, the classifications
to three different spaces are very similar. The LDA in the filtered
image space or the LDA in the PCA factor space of the training set
are a bit better than the LDA in the parameters space. The training
classification error and the test errors are computed on 80 pixels
vector of the training set by a 5-fold cross validation
\citep{Hastie_2003}. Both classification errors are equal to zero.

\begin{figure}[!htb]
\centering
\begin{tabular}{@{}c@{}}
    \includegraphics[width=0.3\columnwidth]{serim447_thread_100_verite_small}\\
    \footnotesize reference\\
    \begin{tabular}{@{}ccc@{}}
        \includegraphics[width=0.3\columnwidth]{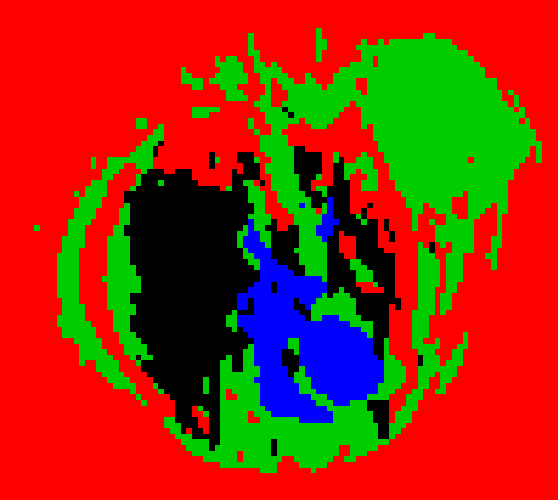}&
        \includegraphics[width=0.3\columnwidth]{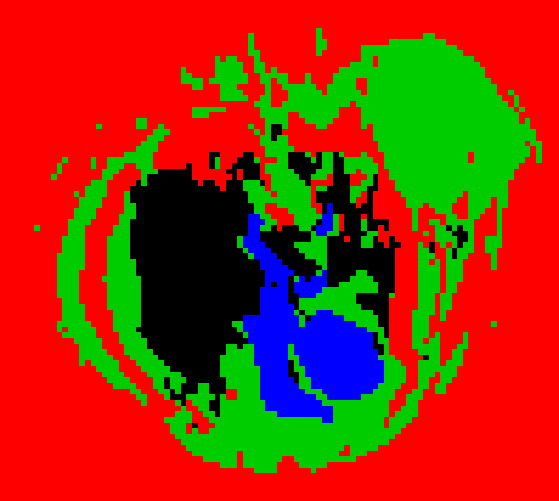}&
        \includegraphics[width=0.3\columnwidth]{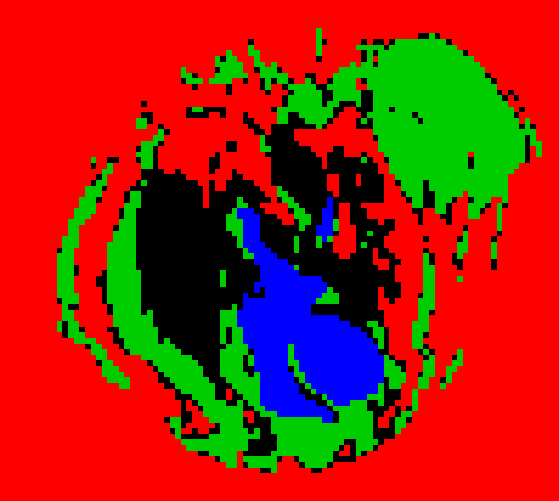}\\
        \footnotesize LDA on $\widehat{\mathbf{f}}_{\lambda}$ &
        \footnotesize LDA on $\zeta_{train}(\widehat{\mathbf{f}}_{\la}(x))$ &
        \footnotesize LDA on $\mathbf{p}$\\
    \end{tabular}
\end{tabular}
\caption{Semi-supervised classification LDA in 4 classes in 3
different spaces: the filtered image space, the PCA factor space of
the spectra of the training set, the parameters space.}
\label{fig:classif:LDA_classifications}
\end{figure}

\FloatBarrier
\subsection{Classification on similar DCE-MRI series}

As we want to classify several DCE-MRI series of large image
databases, it is necessary to develop classification methods which
are robust in all series, without the needs for a training set for
each image series.

By testing our methods on another series called ``serim460'', we can
notice in figure \ref{fig:classif:classif_im_simi_1} that the
unsupervised classifications (k-means and model) are correct
compared to the given reference. However, the LDA classification on
the new series is not correct whatever the image space. In this case
the training set is from another series (``serim447'').

\begin{figure}[!htb]
\centering
\begin{tabular}{ccc}
    \includegraphics[width=0.3\columnwidth]{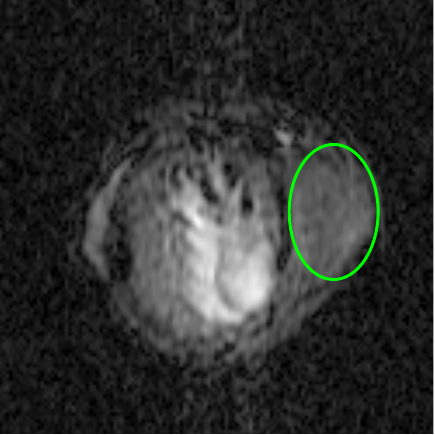}&
    \includegraphics[width=0.3\columnwidth]{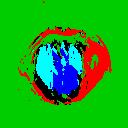}&
    \includegraphics[width=0.3\columnwidth]{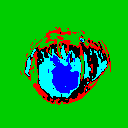}\\
    \footnotesize $ref$&
    \footnotesize $\kappa^{kmeans,5}_{\ca}$&
    \footnotesize $\kappa^{mod,5}_{\fr}$\\
    &
    \includegraphics[width=0.3\columnwidth]{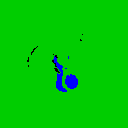}&
    \includegraphics[width=0.3\columnwidth]{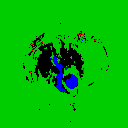}\\
    &
    \footnotesize $\kappa^{LDA,4}_{\zeta (\widehat{\mathbf{f}}_{\la}) | T_{s447}}$&
    \footnotesize $\kappa^{LDA,4}_{\mathbf{p} | T_{s447}}$\\
\end{tabular}
  \caption{Classifications of the series ``serim460'': by k-means $\kappa^{kmeans,5}_{\ca}$, by
  model $\kappa^{mod,5}_{\fr}$, by LDA into the filtered image
  $\kappa^{LDA,4}_{\zeta (\widehat{\mathbf{f}}_{\la}) | T_{s447}}$
  and into the parameter space $\kappa^{LDA,4}_{\mathbf{p} | T_{s447}}$.
  The training set of the LDA is from the series ``serim447''.}
  \label{fig:classif:classif_im_simi_1}
\end{figure}

In the series ``serim447'' and ``serim460'', some mean spectra are
measured into similar zones (tumour, heart cavities, background and
lung). In figure \ref{fig:classif:_mean_spectra_serim447_460} we
notice that the range of the spectra are not the same for both
images. This is the origin of the problem of the classification for
a supervised method such as LDA.

\begin{figure}[!htb]
\begin{center}
\begin{tabular}{@{}c|c@{}}
    serim447 & serim460\\
    \includegraphics[width=0.3\columnwidth]{serim447_selection_train_LDA}&
    \includegraphics[width=0.3\columnwidth]{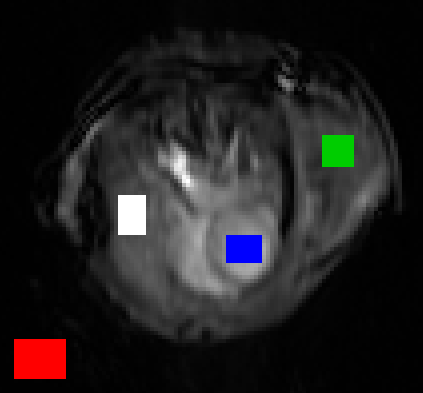}\\
    \includegraphics[width=0.5\columnwidth]{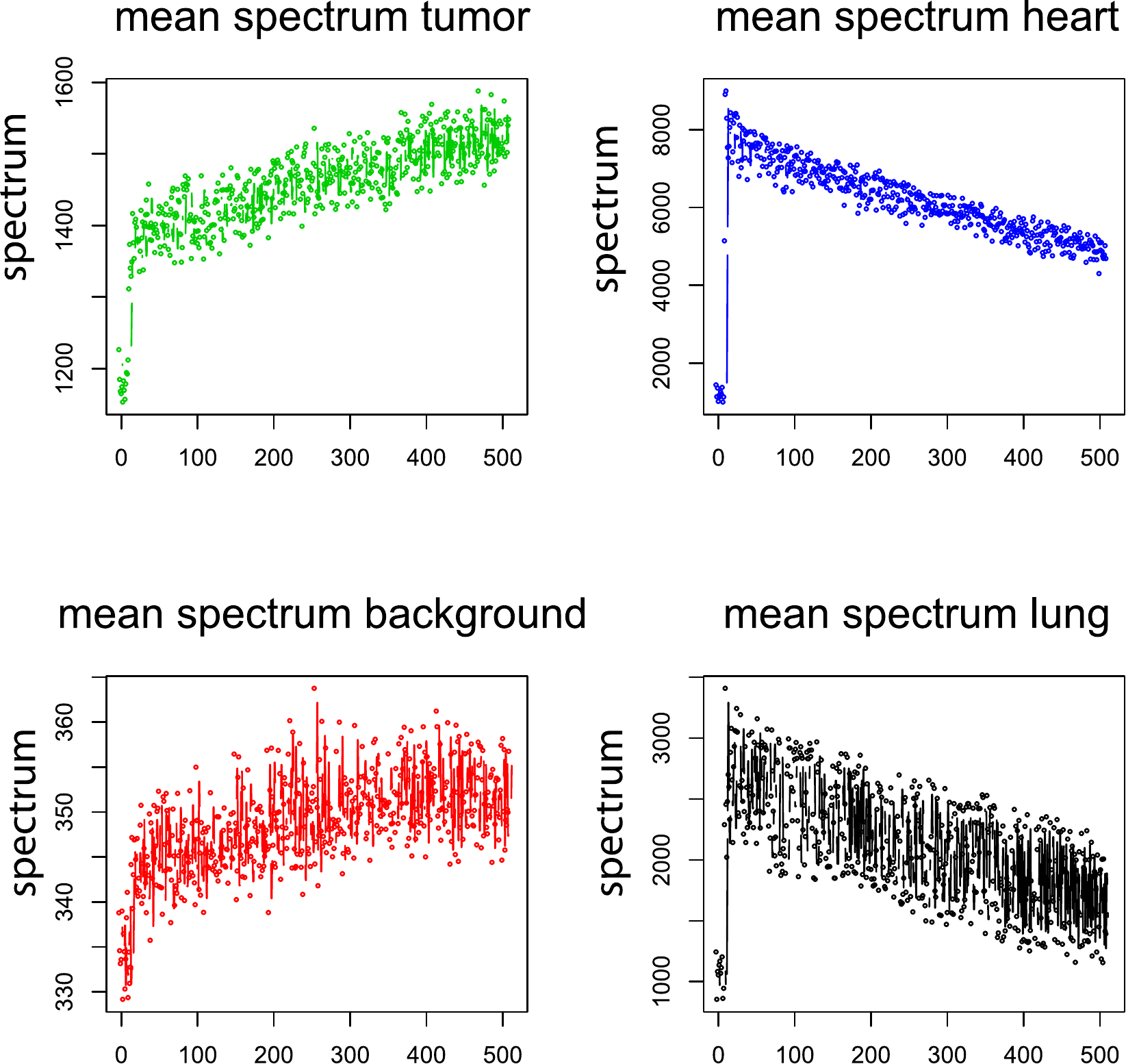}&
    \includegraphics[width=0.5\columnwidth]{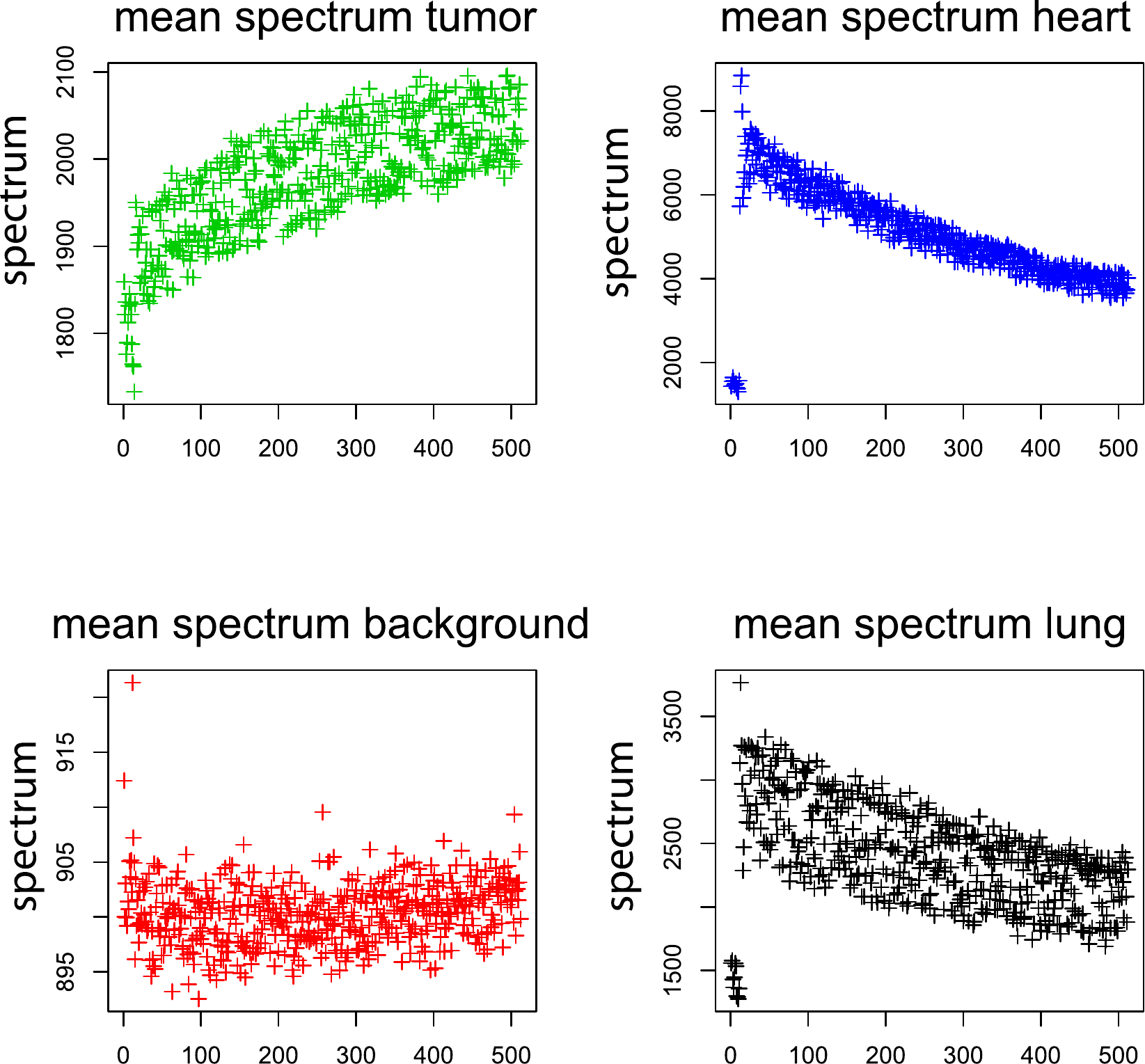}\\
    \footnotesize (a) & \footnotesize (b)\\
\end{tabular}
  \caption{Selected areas and associated mean spectra for the series ``serim447'' and ``serim460''
  after a sequence of 2 FCA-reconstructions (a zoom is made on the series ``serim447''). }
  \label{fig:classif:_mean_spectra_serim447_460}
\end{center}
\end{figure}

In order to use LDA on other image series, using  the initial
training pixels selected on the first image, the range of the grey
levels of the images must be similar. Otherwise the projection of
other series in the classification space of the pixels would be
incoherent. Consequently, we introduced a range normalisation method
based on histogram anamorphosis. To get more robust results, the
multivariate image of parameters $\mathbf{p}$ is used. For each
parameter, the cumulative distribution function (cdf) of the values
is estimated. The cdf is the primitive of the density function
estimated by an histogram. It is composed of 255 classes defined on
each map of parameters of the initial series. The cdf of each image
is transformed by a numerical anamorphosis on the grey-tone values
in order to be similar to the reference cdf of the initial series
``serim447'' (fig. \ref{fig:classif:cdf}). 

\begin{figure}[!htb]
\begin{center}
\begin{tabular}{@{}cc@{}}
    \footnotesize cdf before normalisation &
    \footnotesize cdf after normalisation\\
    \includegraphics[width=0.5\columnwidth]{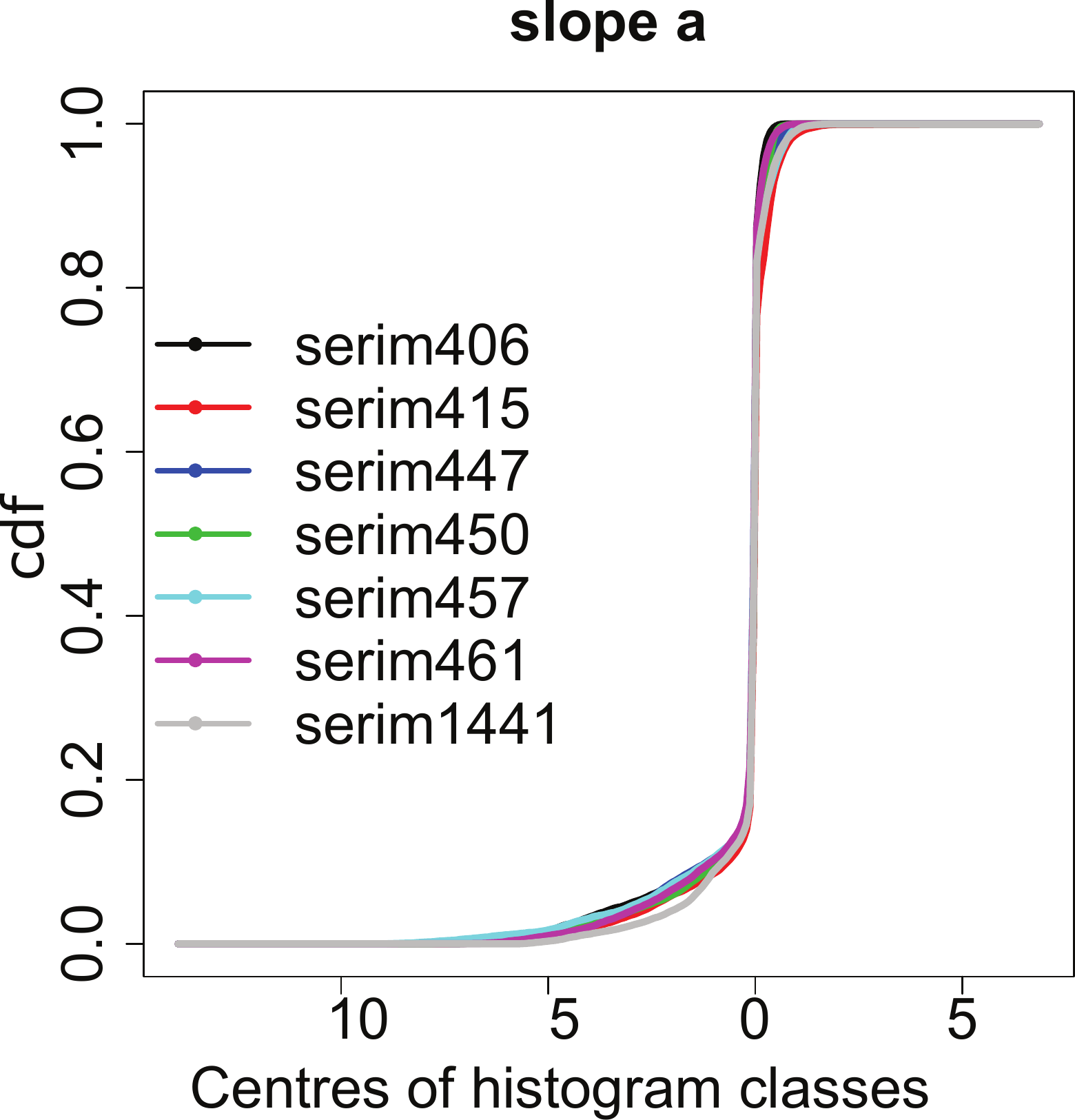}&
    \includegraphics[width=0.5\columnwidth]{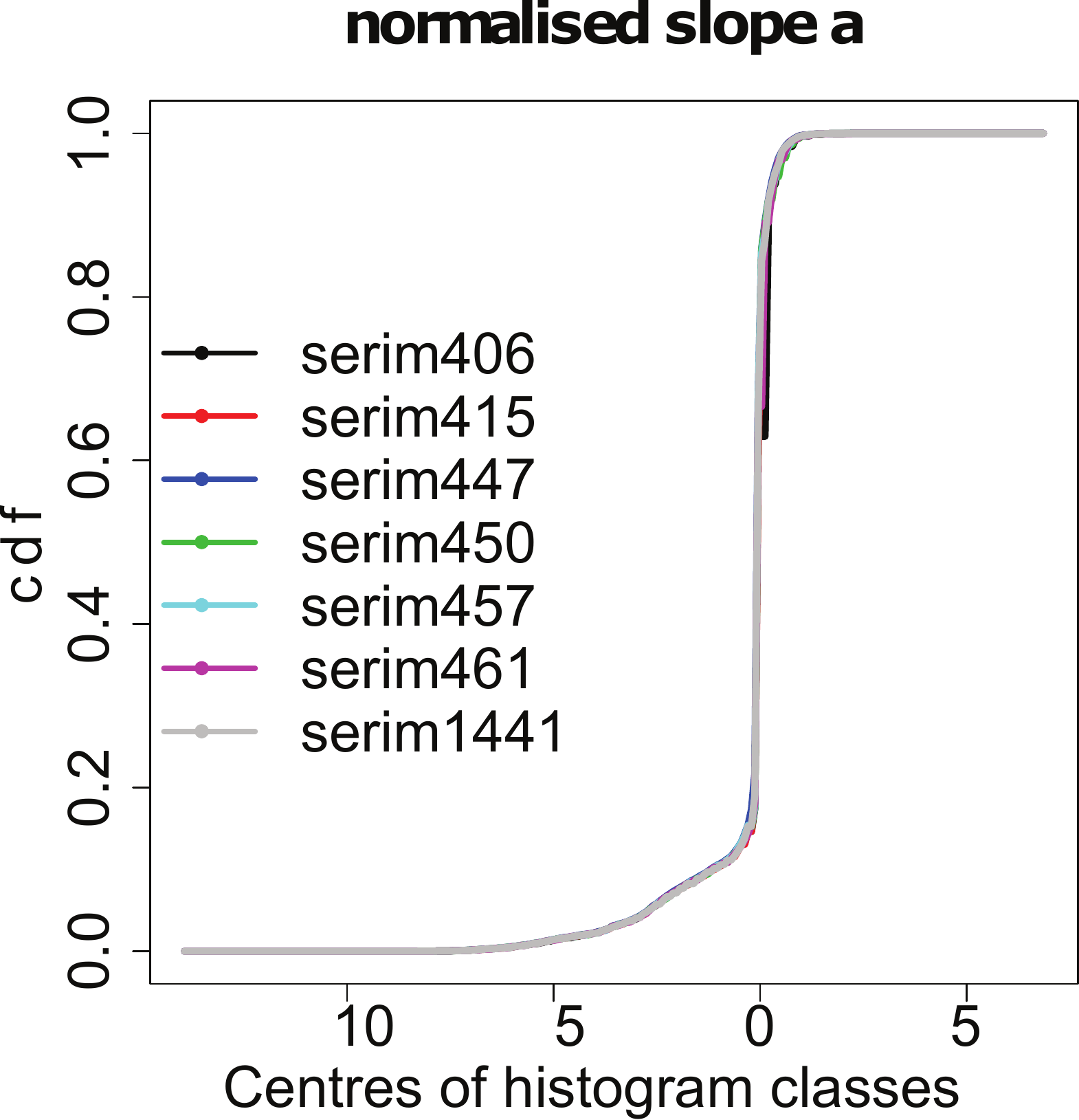}\\
    \includegraphics[width=0.5\columnwidth]{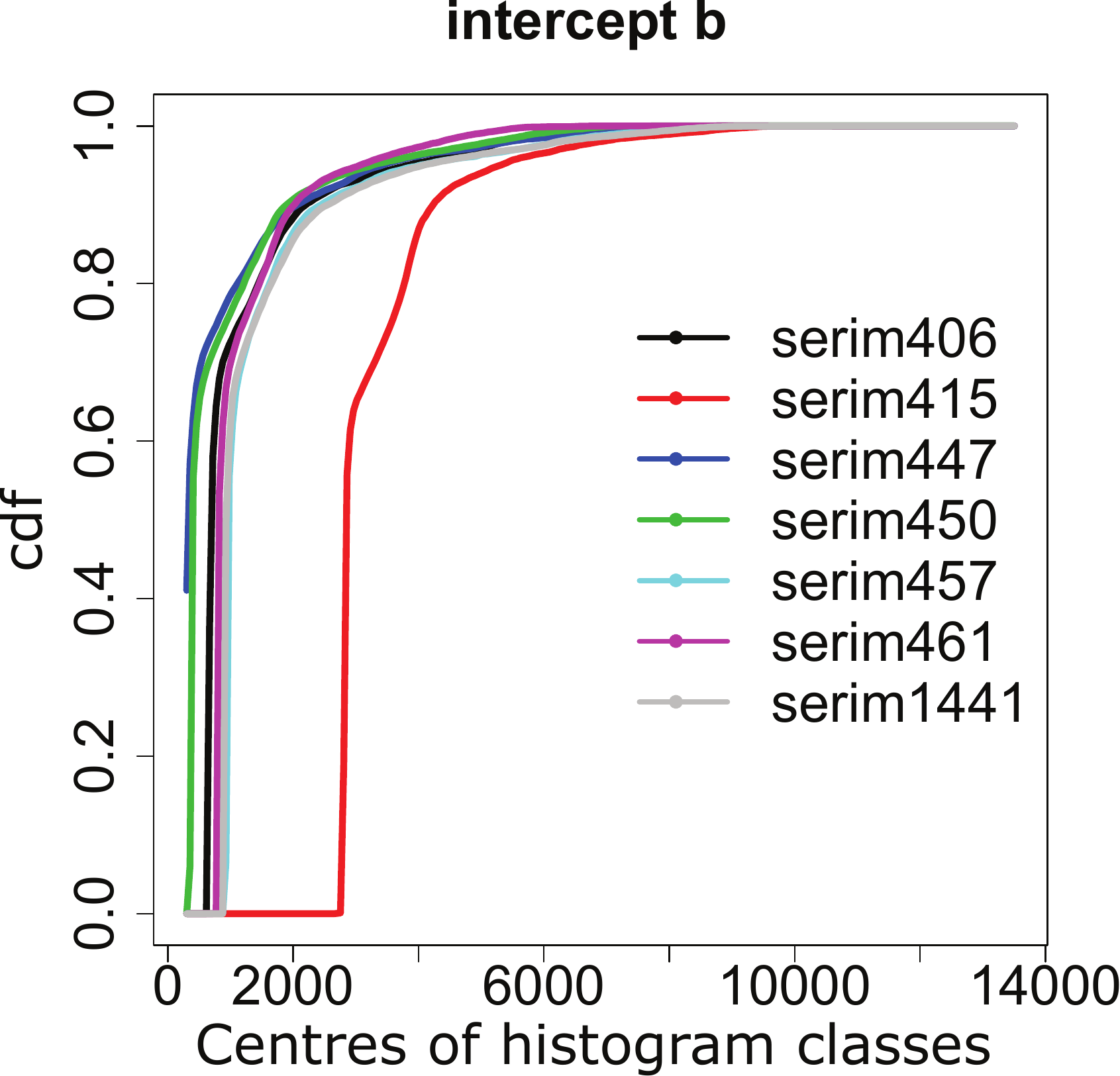}&
    \includegraphics[width=0.5\columnwidth]{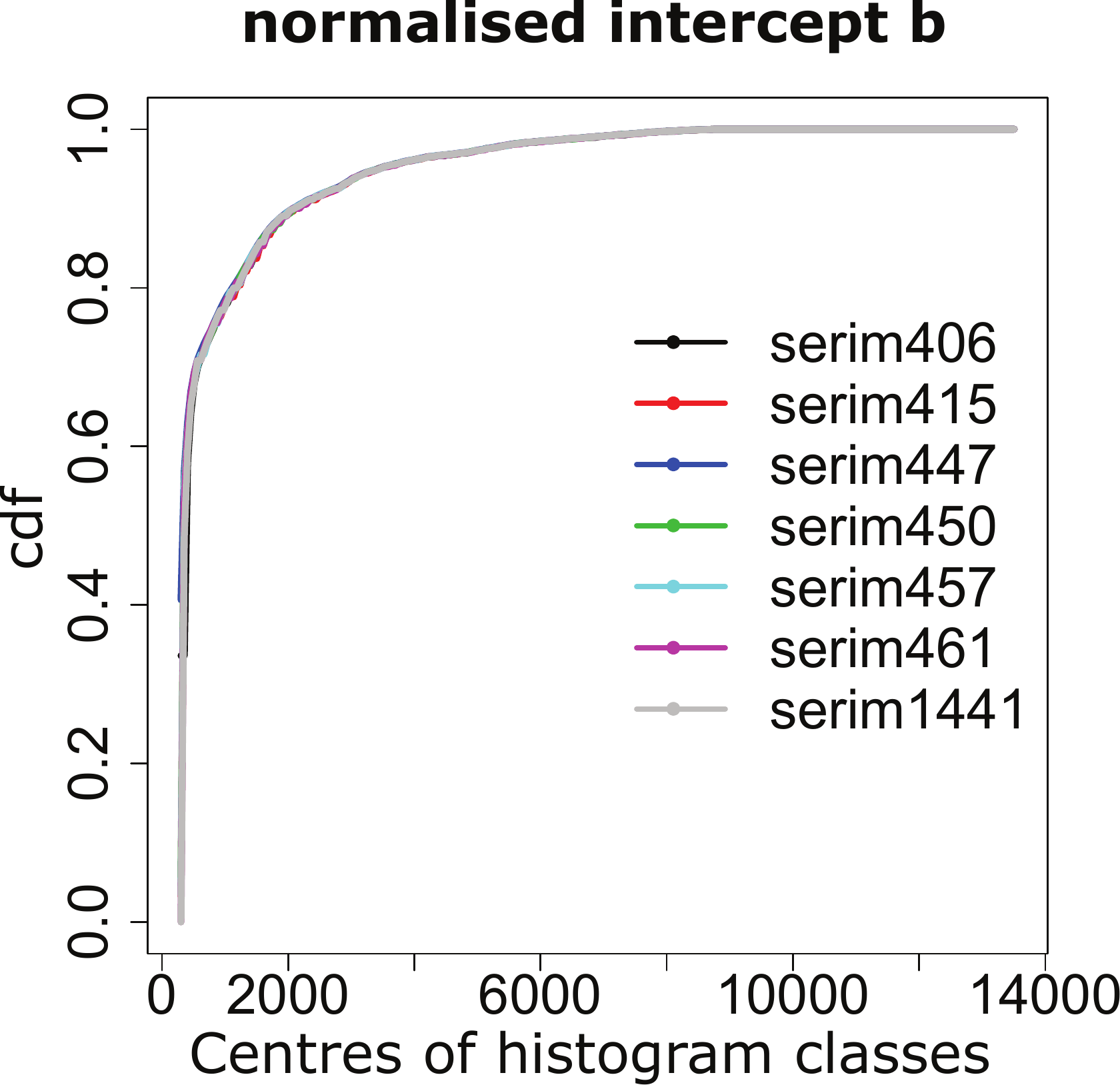}\\
    \includegraphics[width=0.5\columnwidth]{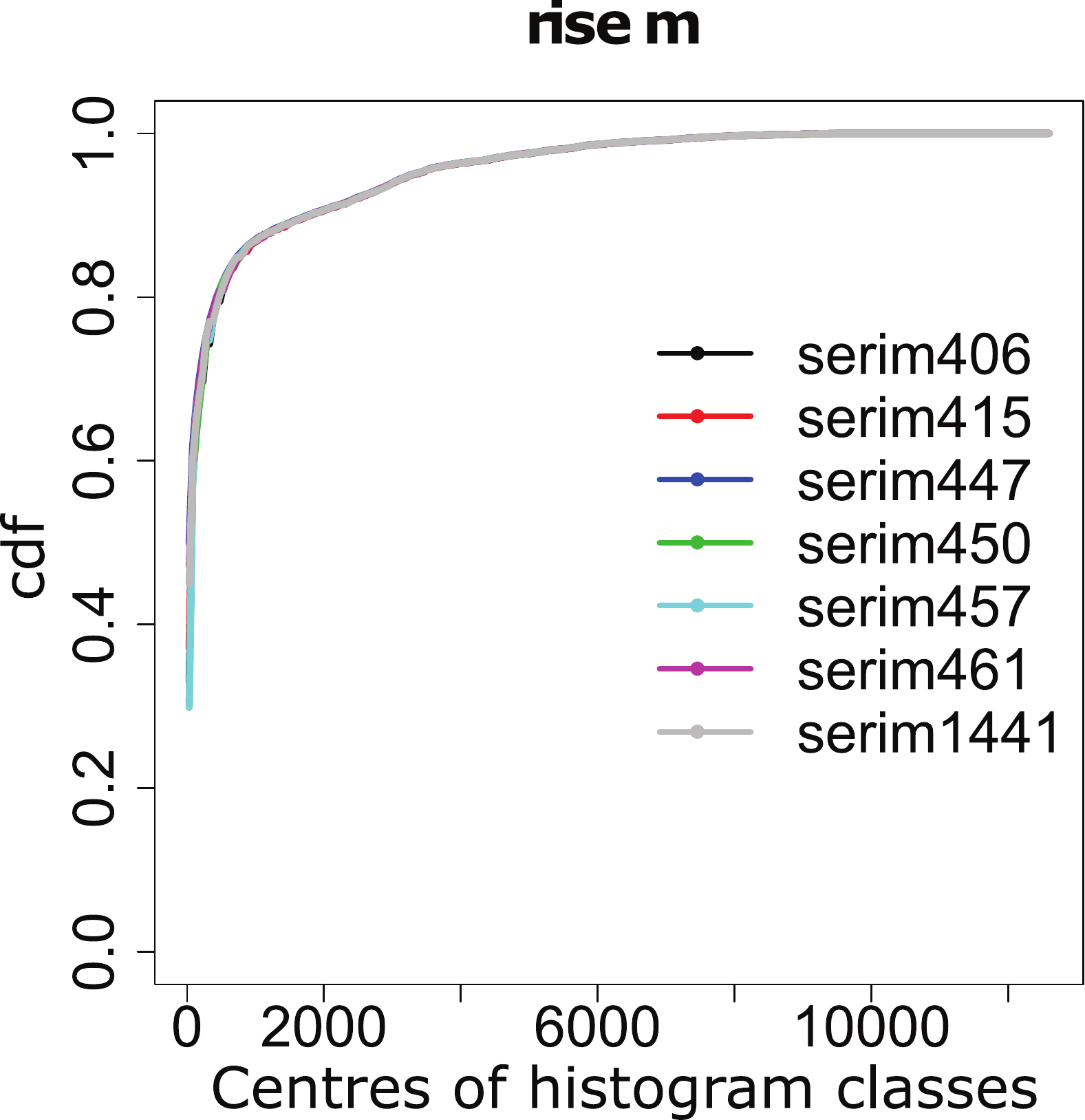}&
    \includegraphics[width=0.5\columnwidth]{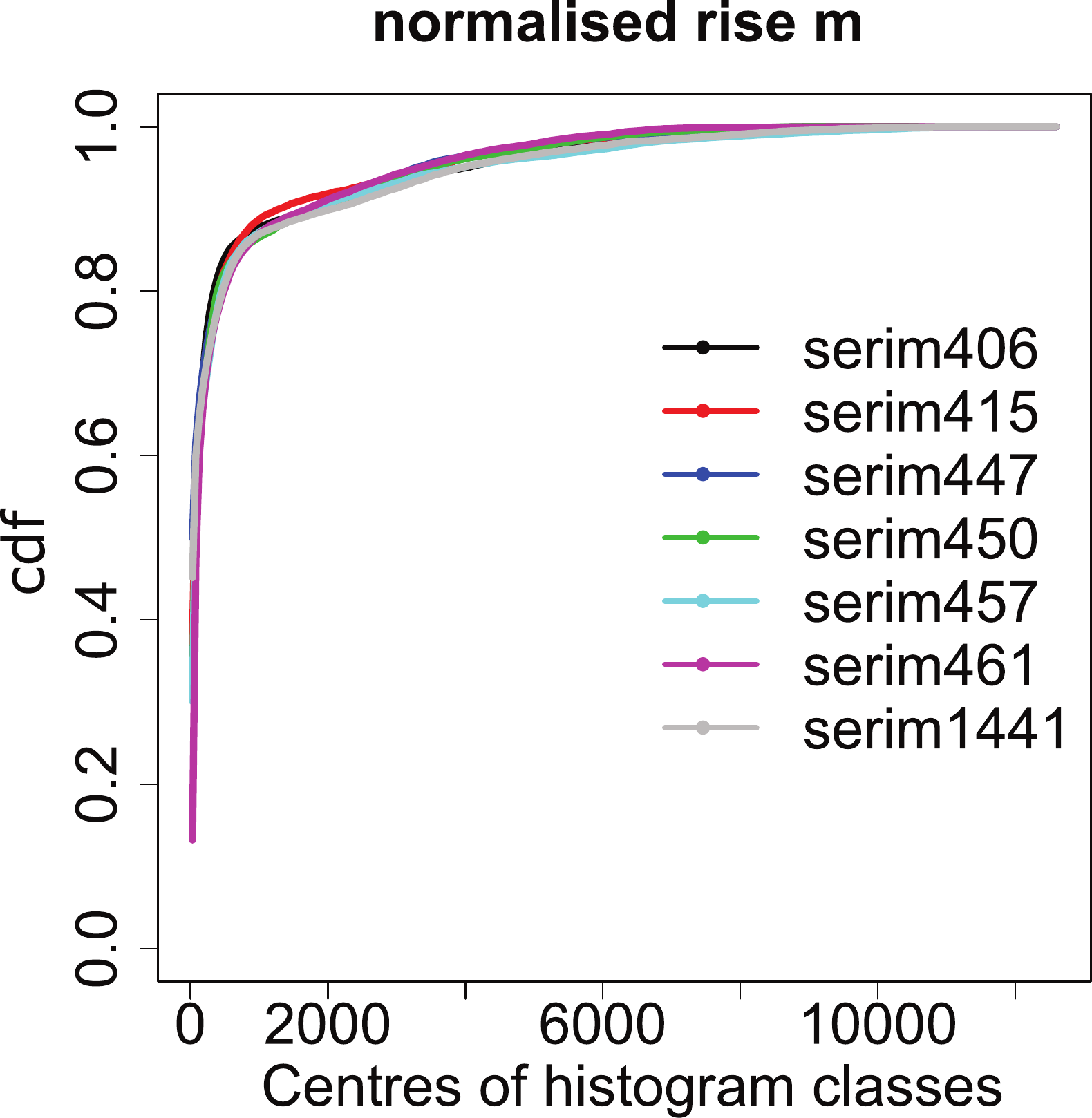}\\
\end{tabular}
  \caption{Cumulative distribution functions computed on the
parameters $a$, $b$ and $m$ of each series before normalisation
(top) and after normalisation (bottom). The references are the cdf
of the parameters of the initial series ``serim447''. To each colour
is associated the cdf of a series} \label{fig:classif:cdf}
\end{center}
\end{figure}

The LDA classification $\kappa^{LDA,4}_{\widetilde{\mathbf{p}}}$ on
series ``serim460'' after cdf normalisation (fig.
\ref{fig:classif:classif_LDA_serim460}) gives much better results
than the same classification without cdf normalisation (fig.
\ref{fig:classif:classif_im_simi_1}).

\begin{figure}[!htb]
\centering
\begin{tabular}{cc}
    \includegraphics[width=0.3\columnwidth]{serim460_thread_100_verite}&
    \includegraphics[width=0.3\columnwidth]{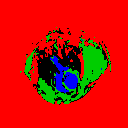}\\
    \footnotesize $ref$ &
    \footnotesize $\kappa^{LDA,4}_{\widetilde{\mathbf{p}} | T_{s447}}$\\
\end{tabular}
  \caption{LDA classification in 4 classes $\kappa^{LDA,4}_{\widetilde{\mathbf{p}}}$ on the maps of parameters
  of the series
  «~serim460~» after cdf normalisation. The training has been made on the series «~serim447~».}
  \label{fig:classif:classif_LDA_serim460}
\end{figure}

\begin{figure*}[!htb]
\centering
\begin{tabular}{@{}c@{\ctab}c@{\ctab}c@{\ctab}c@{}}
    \footnotesize Reference &
    \footnotesize k-means &
    \footnotesize model &
    \footnotesize LDA\\
    \includegraphics[width=0.3\columnwidth]{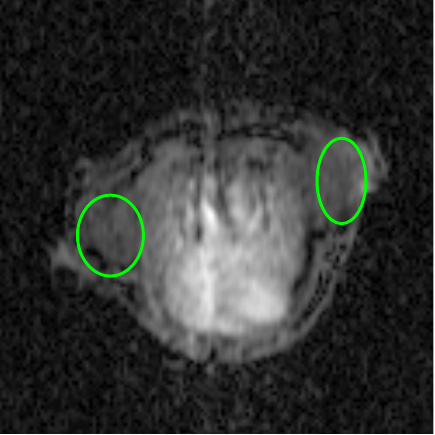}&
    \includegraphics[width=0.3\columnwidth]{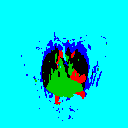}&
    \includegraphics[width=0.3\columnwidth]{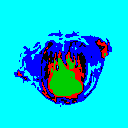}&
    \includegraphics[width=0.3\columnwidth]{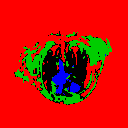}\\
    \multicolumn{4}{c}{\footnotesize \emph{«~serim406~»}}\\
    \includegraphics[width=0.3\columnwidth]{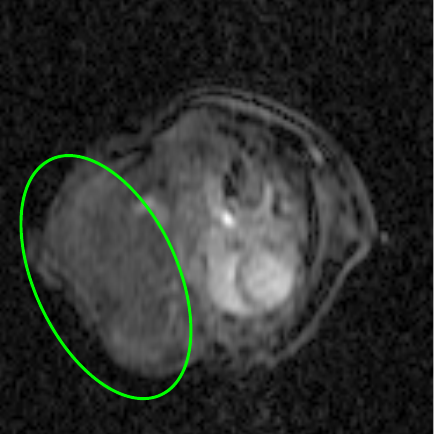}&
    \includegraphics[width=0.3\columnwidth]{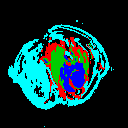}&
    \includegraphics[width=0.3\columnwidth]{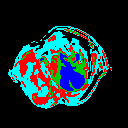}&
    \includegraphics[width=0.3\columnwidth]{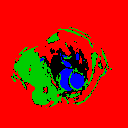}\\
    \multicolumn{4}{c}{\footnotesize \emph{«~serim415~»}}\\
    \includegraphics[width=0.3\columnwidth]{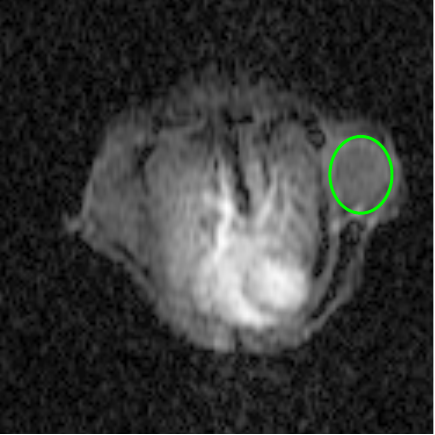}&
    \includegraphics[width=0.3\columnwidth]{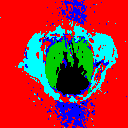}&
    \includegraphics[width=0.3\columnwidth]{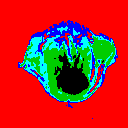}&
    \includegraphics[width=0.3\columnwidth]{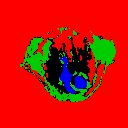}\\
    \multicolumn{4}{c}{\footnotesize \emph{«~serim450~»}} \\
    \includegraphics[width=0.3\columnwidth]{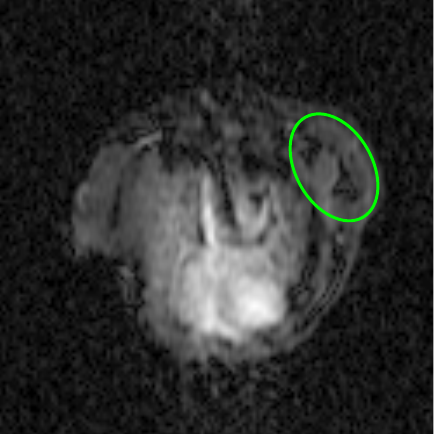}&
    \includegraphics[width=0.3\columnwidth]{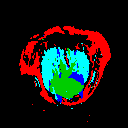}&
    \includegraphics[width=0.3\columnwidth]{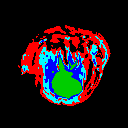}&
    \includegraphics[width=0.3\columnwidth]{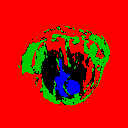}\\
    \multicolumn{4}{c}{\footnotesize \emph{«~serim457~»}}\\
    \includegraphics[width=0.3\columnwidth]{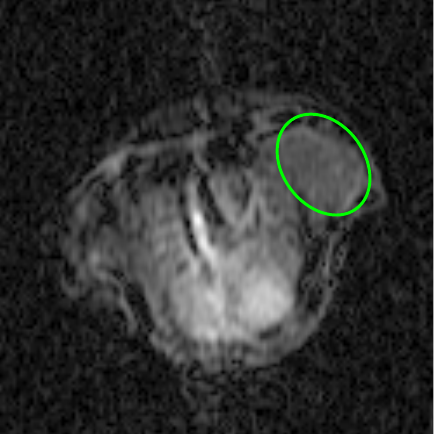}&
    \includegraphics[width=0.3\columnwidth]{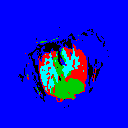}&
    \includegraphics[width=0.3\columnwidth]{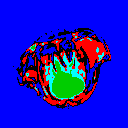}&
    \includegraphics[width=0.3\columnwidth]{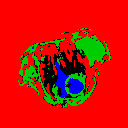}\\
    \multicolumn{4}{c}{\footnotesize \emph{«~serim461~»}}\\
    \includegraphics[width=0.3\columnwidth]{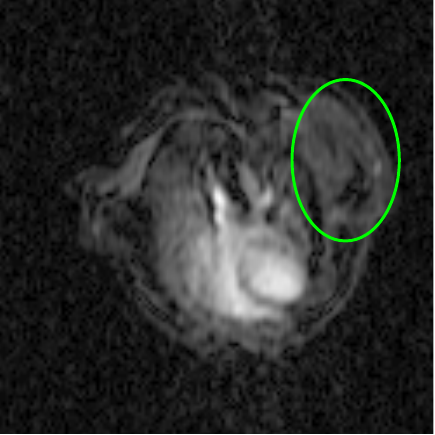}&
    \includegraphics[width=0.3\columnwidth]{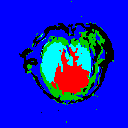}&
    \includegraphics[width=0.3\columnwidth]{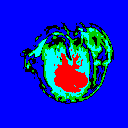}&
    \includegraphics[width=0.3\columnwidth]{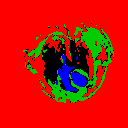}\\
    \multicolumn{4}{c}{\footnotesize \emph{«~serim1441~»}}\\
\end{tabular}
  \caption{Classifications \emph{kmeans} $\kappa^{kmeans,5}_{\ca}$, by model approach
  $\kappa^{mod,5}_{\fr}$ and LDA $\kappa^{LDA,4}_{\widetilde{\mathbf{p}} | T_{s447}}$
  on various series «~serimxxx~».}
  \label{fig:classif:classif_LDA_para_series}
\end{figure*}

For validation purposes, we compare the classifications by k-means,
model and LDA on other series (fig.
\ref{fig:classif:classif_LDA_para_series}). The classifications
based on the model approach are more robust than those obtained with
k-means. LDA classifications also give good results. The heart
cavities are correctly classified and the tumours are characterised
by extended classes in green.

In the case of tumours starting to die in their centre, which would
make them potentially smaller than their real size, our method
identifies viable tissues (i.e. functional tissues). The central
zones which are not included inside the tumours are zones of severe
ischemia or necrosis. Currently, more and more DCE maps are analysed
in the following way: i) identification of the ``necrosis'' ratio
(volume of the necrosis divided by total volume) ; ii)
characterisation of viable tissues of the tumour. Without both
analysis, all the circulatory parameters are underestimated during
the growth of the tumour which leaves in its centre more and more
necrosis (or fibrosis).

\FloatBarrier
\section{Spatio-temporal segmenta-tion by probabilistic\\ watershed}

After having filtered the temporal noise and having reduced the
temporal dimension, the classification produced a partition of the
image into non connected classes only based on temporal information.
Our aim is now to segment the series with smoother contours and
regular classes by combining the spatial and the spectral dimension
in the segmentation process.

In order to segment hyperspectral images, we introduced a general
method based on deterministic watershed (WS)
in~\citep{Noyel_IAS_2007} and another one based on stochastic WS
in~\citep{Noyel_KES_2007,Noyel_IJRS2011}. Based on previous work we
want to present the potential of application of our methods on
DCE-MRI series in order to segment regions to be tumour candidates.

\subsection{Principle of the segmentation of multivariate images by WS}

The watershed transformation (WS) is one of the most powerful tools
for segmenting images and was introduced in~\citet{Beucher_1979}.
According to the flooding paradigm, the watershed lines associate a
catchment basin to each minimum of the landscape to flood (i.e. a
scalar or greyscale image) \citep{Beucher_1992}. Typically, the
landscape to flood is a gradient function which defines the
transitions between the regions. Using the watershed on a scalar
image without any preparation leads to a strong over-segmentation
(due to a large number of minima). There are two alternatives in
order to get rid of the over-segmentation. The first one consists in
first determining markers for each region of interest. Then, using
the homotopy modification, only the local minima of the gradient
function are imposed by the markers of the regions. The extraction
of the markers, especially for generic images, is a difficult task.
The second alternative involves hierarchical approaches either based
on non-parametric merging of catchment basins (waterfall algorithm)
or based on the selection of the most significant minima. These
minima are selected according to different criteria such as
dynamics, area or volume extinction values. Extinction functions
\citep{Meyer_2001} are used to remove the non selected minima.

The general paradigm of WS-based segmentation of multivariate images
(fig. \ref{fig:seg:_WS_framework}) requires two different inputs:
(1) some markers for the regions of interest $mrk$ and (2) a
landscape to flood $g$ which describes the ``likelihood'' of the
frontiers between the regions. The markers can be chosen
interactively by a user, or automatically by means of a
morphological criterion $\xi_{N}$~\citep{Meyer_2001}, or with the
classes of a previous spectral classification. The landscape to
flood is a scalar function (i.e., a greyscale image). For the
deterministic WS, it is usually a gradient (actually its norm), or a
distance function. For the stochastic WS, the function to flood is a
probability density function (pdf) of the contours appearing in the
image. The extracted markers are imposed as sources of the landscape
to flood and the WS is computed. The results denoted $WS(g,mrk)$ or
$WS(g,\xi_{N})$.

\begin{figure}[!htb]
\begin{center}
\includegraphics[width=1\columnwidth]{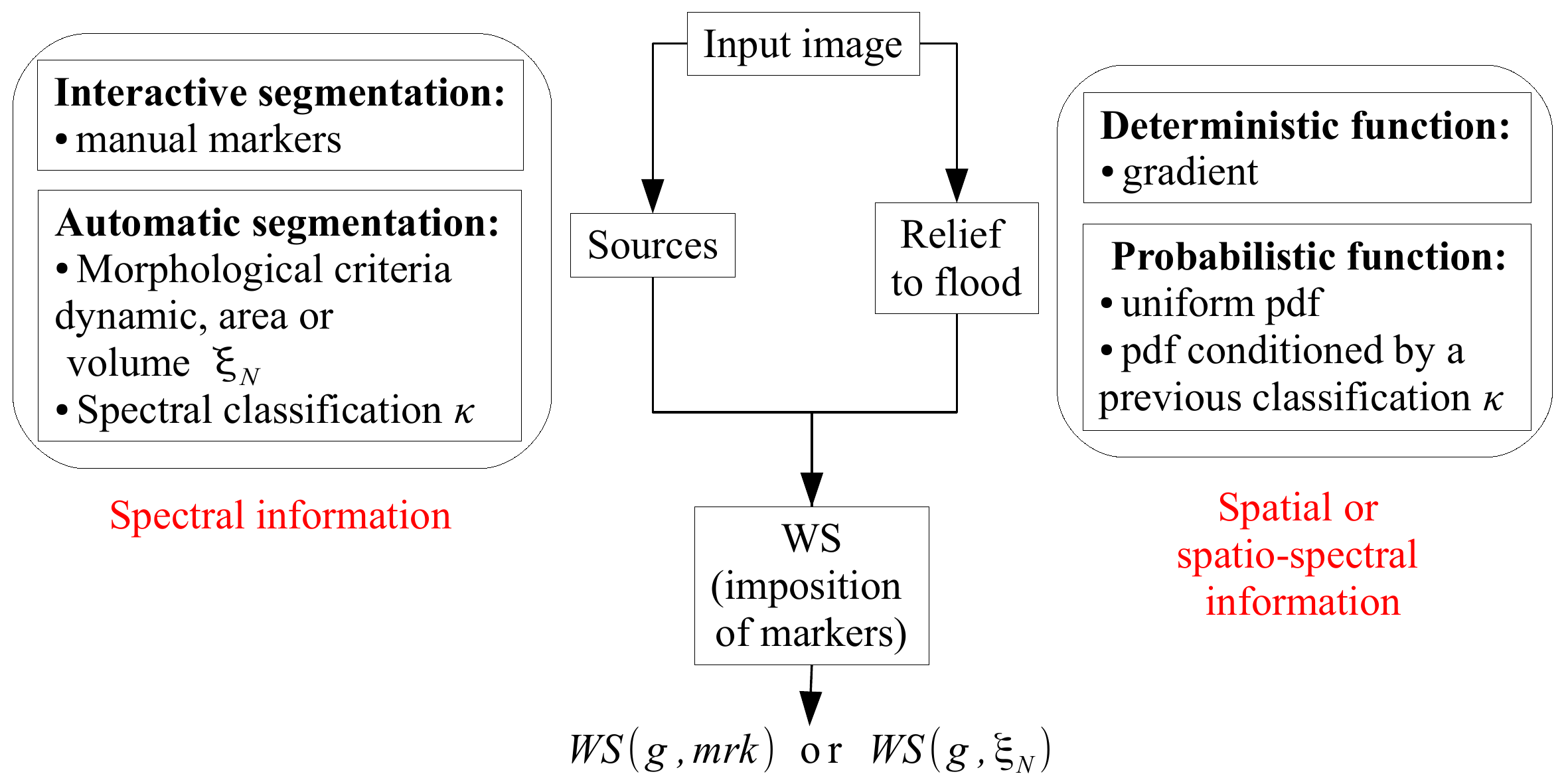}
\end{center}
\caption{General framework of multivariate image segmentation}%
\label{fig:seg:_WS_framework}%
\end{figure}

\subsection{Spectral distances and gradient on multivariate images}

A gradient image, actually its norm, is usually chosen as a function
to flood. After normalisation, the norm of a gradient image is a
scalar function with values in the reduced interval $[0,1]$, i.e.
$\varrho(x):E\rightarrow [0,1]$. In order to define a gradient, two
approaches are considered: the standard symmetric morphological
gradient on each marginal channel and a metric-based vectorial
gradient on all channels \citep{Noyel_IAS_2007}.

The morphological gradient is defined for scalar images $f$ as the
difference between a dilation and an erosion by a unit structuring
element $B$, i.e.,
\begin{eqnarray}\label{eq:seg:morpho_gradient}
  \nonumber \varrho(f_{\lambda_j}(x)) &=& \delta_B(f_{\lambda_j}(x))-\varepsilon_B(f_{\lambda_j}(x)) \\
  \nonumber &=& \vee[ f_{\lambda_j}(y), y \in B(x) ]\\
  && - \wedge[ f_{\lambda_j}(y), y \in B(x) ]\text{ .}
\end{eqnarray}


The morphological gradient can be generalised to multivariate
functions \citep{Hanbury_2001} with the following metric-based
gradient:
\begin{eqnarray}\label{eq:seg:morpho_gradient_multiv}
    \nonumber \varrho^{d}\mathbf{f_{\lambda}}(x) &=& \vee[ d(\mathbf{f_{\lambda}}(x),
    \mathbf{f_{\lambda}}(y)) \text{ } / \text{ } y \in B(x), y \neq x]\\
    \nonumber && - \wedge[ d(\mathbf{f_{\lambda}}(x), \mathbf{f_{\lambda}}(y)) \text{ } /
    \text{ } y \in B(x), y \neq x] \text{ .}\\
\end{eqnarray}

Various metric distances $d(\mathbf{f_{\lambda}}(x),
\mathbf{f_{\lambda}}(y))$ between two vector pixels, useful for
multispectral images, are available for this gradient such as:
\begin{itemize}
 \item the Euclidean distance: $d_{E}(\mathbf{f_{\lambda}}(x), \mathbf{f_{\lambda}}(y)) = \sqrt{
\sum_{j=1}^{L}( f_{\lambda_{j}}(x) - f_{\lambda_{j}}(y) )^2 }$,

\item the Chi-squared distance: $d_{\chi^{2}}( \mathbf{f}_{\lambda}(x_{i}) ,
\mathbf{f}_{\lambda}(x_{i'}) ) = \sqrt{\sum_{j=1}^{L}
\frac{S}{f_{.\lambda_{j}}} \left( \frac{ f_{\lambda_{j}}(x_{i}) }{
f_{x_{i}.} } - \frac{ f_{\lambda_{j}}(x_{i'}) }{ f_{x_{i'}.} }
\right)^{2}}$,

with $f_{.\lambda_{j}} = \sum_{i=1}^{P} f_{\lambda_{j}}(x_{i})$,
$f_{x_{i}.} = \sum_{j=1}^{L} f_{\lambda_{j}}(x_{i})$ and $S =
\sum_{j=1}^{L}\sum_{i=1}^{P}f_{\lambda_{j}}(x_{i})$,

\item the Mahalanobis distance:
$d_{M}(\mathbf{f_{\lambda}}(x), \mathbf{f_{\lambda}}(y)) =
\sqrt{(\mathbf{f_{\lambda}}(x) - \mathbf{f_{\lambda}}(y))^{t}
  \Sigma^{-1}(\mathbf{f_{\lambda}}(x) - \mathbf{f_{\lambda}}(y))}$,
where $\Sigma$ is the covariance matrix between variables (channels)
of $\mathbf{f_{\lambda}}$. If channels are uncorrelated, the
covariance matrix is diagonal. The diagonal values are equal to the
channels variance $\sigma^{2}_{\lambda_{j}} \ \backslash \ j \in
\{1,2, \ldots, L\}$. Therefore, the Mahalanobis distance becomes the
distance inverse of variances:
$d_{1/\sigma^2}(\mathbf{f_{\lambda}}(x), \mathbf{f_{\lambda}}(y)) =
   \sqrt{ \sum_{j=1}^{L} \left( \frac{f_{\lambda_{j}}(x) - f_{\lambda_{j}}(y)}{\sigma_{\lambda_{j}}} \right)^{2}
   }$
\end{itemize}
An important point is to choose an appropriate distance depending on
the space used for image representation: Chi-squared distance
$d_{\chi^{2}}$ and distance of inverse variances $d_{1/\sigma^2}$
are adapted to the image space and Euclidean distance to factorial
space. More details on multivariate gradients are given in
\citep{Noyel_IAS_2007,Noyel_IJRS2011}. Another example of a
multivariate gradient is given in \citet{Scheunders_2002}.

\subsection{Introduction to stochastic WS}

In a classical watershed, small regions strongly depend on the
position of the markers, or on the volume (i.e. the integral of the
grey levels) of the catchment basins, associated with their minima.
In order to improve segmentation results, stochastic watershed aims
at enhancing the contours of significant regions which are
relatively independent of the position of the markers.

Stochastic WS method is described
in~\citep{AnguloJeulin_ISMM_2007,Noyel_KES_2007,Noyel_IJRS2011}.
Starting from a series of $M$ realisations of $N$ uniform or
regionalized random germs (or markers) $\{mrk_{i}(x)\}_{i=1}^{M}$,
series of watershed segmentation $\{sg_{i}^{mrk}(x)\}_{i=1}^{M}$ are
made on a landscape to flood (for example a gradient). With these
$M$ segmentations, the probability density function of contours
$pdf(x)$ is estimated by the Parzen window method \citep{Duda_1973}
with a gaussian kernel (typically with a 3 pixels standard deviation
working on contours of one pixel width). Due to the smoothing effect
of the method, the WS lines with a very low probability, which
correspond to non significant boundaries, are removed.


To obtain closed contours, the pdf image is segmented by a watershed
segmentation into $R$ regions. The stochastic WS needs two
parameters:
\begin{enumerate}

\item $M$ realisations of germs. The method
is almost independent on $M$ if it is large enough (between
20~-~50).
\item $N$ germs (or markers): if $N$ is small, a segmentation in large regions
is privileged; if $N$ is too large, the over-segmentation of
$sg_{i}^{mrk}$ leads to a very smooth $pdf$, which looses its
properties to select the $R$ regions.
\end{enumerate}
As shown in \citet{AnguloJeulin_ISMM_2007}, it is straightforward to
use $N>R$.

The originality of our approach of stochastic WS for DCE-MRI series
is to condition the germs used to build the pdf by a previous
classification.

\FloatBarrier
\subsection{Segmentation by stochastic WS}

\subsubsection{Pre-processing of the temporal classification}

As the classification is based on temporal information we want to
introduce it by conditioning the random markers used to generate the
probability density function of contours.

In order to do this, a pre-processing stage is necessary for two
reasons:
\begin{itemize}
  \item to reduce the ``segmentation noise'' appearing as the smallest connected components of the classification $\kappa$.
  \item to introduce the necessary degrees of freedom to perform
  each WS used to build the pdf.
\end{itemize}
Therefore an anti-extensive transform is performed on each class of
$\kappa$ by a  morphological erosion with a structuring element of
size $3 \times 3$ pixels. An alternative is to make an area opening
\citep{Soille_1999}. Then an extensive transformation such as a
closing by reconstruction \citep{Soille_1999} is performed to fill
the holes inside the largest connected components. As the classes of
the transformed classification are not anymore a partition of the
image, a ''void'' class is introduced. It corresponds to the class
appearing in place of the transformed connected components.

For the DCE-MRI sequences the complete transform $\Upsilon$ is
processed on the LDA classification on the parameters space
$\kappa^{LDA,4}_{\mathbf{p}}$ (fig.
\ref{fig:seg:pre_process_classif}).

\begin{figure}[!htb]
\centering
\begin{tabular}{cc}
  \includegraphics[width=0.3\columnwidth]{souris_classif_LDA_train_complet_image_parametres}&
  \includegraphics[width=0.3\columnwidth]{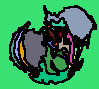}\\
  \footnotesize $\kappa^{LDA,4}_{\mathbf{p}}$ &
  \footnotesize $\Upsilon(\kappa^{LDA,4}_{\mathbf{p}})$ \\
\end{tabular}
  \caption{Classification by LDA on the parameters space $\kappa^{LDA,4}_{\mathbf{p}}$
  and pre-processing transform $\Upsilon$.}
   \label{fig:seg:pre_process_classif}
\end{figure}

\subsubsection{Extension of stochastic WS to multivariate images}

The extension of stochastic WS to multivariate images was introduced
in \citet{Noyel_KES_2007} and detailed in \citet{Noyel_IJRS2011}.

\paragraph{Segmentation of DCE-MRI series by a standard WS on a distance-based gradient}

Before introducing the way to extend stochastic WS to multivariate
images, let us show that the segmentation by a standard WS on a
distance based gradient has limited efficiency.

For the gradient based segmentations, a distance adapted to the
image space is used:
\begin{itemize}
  \item the Euclidean distance for the image of the factor pixels $\ca$
  \item the distance of inverse variances for the image of parameters $\mathbf{p}$
  \item the Euclidean distance for the image of the PCA factor
  space obtained from the spectra of the training set $\mathbf{c}^{train}_{\be}$.
\end{itemize}

\emph{Comments:} The distance-based gradient is computed after a
morphological leveling on each channel of the considered image in
order to get a smoother gradient. A morphological leveling is a
morphological transformation that reduces the positive and negative
``peaks" according to a reference while preserving the transitions
of the objects. See \citep{Meyer_2004} for more details. The
reference for the leveling is obtained by a gaussian filter of size
$11 \times 11$ pixels.

A WS segmentation with a volume criterion is performed in figure
\ref{fig:seg:WS_vol}. A marker-controlled WS is performed in figure
\ref{fig:seg:WS_mark_classif}. The markers are made by a
morphological opening on each connected component of the
classification with an hexagonal structuring element of size 5.

We notice that the segmentation is not perfect, especially for the
tumour. However, the marker based segmentation in the PCA factor
space of the spectra of the training set seems to be the best
segmentation.

\begin{figure}[!htb]
\centering
    \begin{tabular}{ccc}
        \includegraphics[width=0.3\columnwidth]{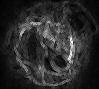}&
        \includegraphics[width=0.3\columnwidth]{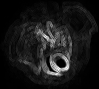}&
        \includegraphics[width=0.3\columnwidth]{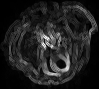}\\
        \footnotesize $\varrho_{E}(\ca)$ &
        \footnotesize $\varrho_{1/\sigma^2}(\mathbf{p})$ &
        \footnotesize $\varrho_{E}(\mathbf{c}^{train}_{\be})$\\
        \includegraphics[width=0.3\columnwidth]{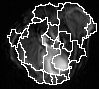}&
        \includegraphics[width=0.3\columnwidth]{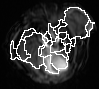}&
        \includegraphics[width=0.3\columnwidth]{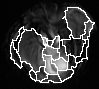}\\
        \footnotesize $sg^{R-vol}(\varrho_{E}(\ca))$ &
        \footnotesize $sg^{R-vol}(\varrho_{1/\sigma^2}(\mathbf{p}))$&
        \footnotesize $sg^{R-vol}(\varrho_{E}(\mathbf{c}^{train}_{\be}))$\\
        \footnotesize (a) &
        \footnotesize (b) &
        \footnotesize (c)\\
    \end{tabular}
    \caption{Gradient-based distances and WS-segmentations with a
    volume criterion in $R = 20$ regions : (a) in the factorial space $\ca$,
    (b) in the parameters space $\mathbf{p}$ and
    (c) in the PCA space of the training set on the parameters $\mathbf{c}^{train}_{\be})$.}
    \label{fig:seg:WS_vol}
\end{figure}

\begin{figure}[!htb]
\centering
    \begin{tabular}{ccc}
        \includegraphics[width=0.3\columnwidth]{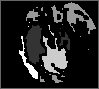}&
        \includegraphics[width=0.3\columnwidth]{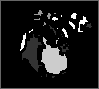}&
        \includegraphics[width=0.3\columnwidth]{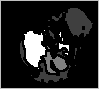}\\
        \footnotesize $mrk_{\kappa}$ &
        \footnotesize $mrk_{\kappa}$ &
        \footnotesize $mrk_{\kappa}$\\
        \includegraphics[width=0.3\columnwidth]{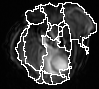}&
        \includegraphics[width=0.3\columnwidth]{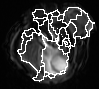}&
        \includegraphics[width=0.3\columnwidth]{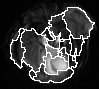}\\
        \footnotesize $sg(\varrho_{E}(\ca),mrk_{\kappa})$ &
        \footnotesize $sg(\varrho_{1/\sigma^2}(\mathbf{p}),mrk_{\kappa})$&
        \footnotesize $sg(\varrho_{E}(\mathbf{c}^{train}_{\be}),mrk_{\kappa})$\\
        \footnotesize (a) &
        \footnotesize (b) &
        \footnotesize (c)\\
    \end{tabular}
    \caption{Segmentations by marker-controlled WS into several spaces:
    (a) in the factorial space $\ca$,
    (b) in the parameters space $\mathbf{p}$ and
    (c) in the PCA space of the training set on the parameters $\mathbf{c}^{train}_{\be}$. The gradient-based
    distances are the same as in figure \ref{fig:seg:WS_vol}. The markers come from the LDA classification for
    the different image spaces.}
    \label{fig:seg:WS_mark_classif}
\end{figure}

\FloatBarrier
\newpage
\paragraph{Probability density function for multivariate images}

In \citet{Noyel_KES_2007}, we studied two ways to extend the
probability density function of contours to multispectral images:

\begin{enumerate}

\item the first one is a marginal approach (i.e. channel by  channel) called
marginal pdf $mpdf$ (alg. in table \ref{alg_mpdf})

\item the second one is a vectorial approach (i.e. vector pixel by vector
pixel) called vectorial pdf $vpdf$ (alg. in table \ref{alg_vpdf}).
\end{enumerate}


\begin{table}[!htb]
\begin{tabular}{@{\vspace{-0.3cm}}p{\columnwidth}}
\hline \vspace{-0.3cm} \caption{Algorithm: $mpdf$}\label{alg_mpdf}\\
\hline
    \begin{algorithmic}[1]
    \STATE For the morphological gradient of each channel
    $\varrho(f_{\lambda_{j}})$, $j \in [1, \ldots, L]$, throw $M$
    realisations of $N$ uniform random germs, i.e. the markers
    $\{mrk^j_i\}_{i=1 \ldots M}^{j=1 \ldots L}$, generating $M \times L$
    realisations. Get the series of segmentations,
    $\{sg^{j}_{i}(x)\}_{i=1 \ldots M}^{j=1 \ldots L}$, by watershed
    associated to morphological gradients of each channel
    $\varrho(f_{\lambda_{j}})$. \STATE  Get the marginal pdfs on each
    channel by Parzen method: $pdf_j(x) =  \frac{1}{M} \sum_{i=1}^M
    sg^j_i(x) \ast G_{\sigma}$. \STATE Obtain the weighted marginal pdf:
    \begin{equation}\label{eq_mpdf}
    mpdf(x) = \sum_{j=1}^L w_j pdf_j (x)
    \end{equation}
    with $w_j=1/L$, $j \in [1, \ldots, L]$ in the image space and $w_j$
    equal to the inertia axes in the factorial space.
    \end{algorithmic}\\
\\ \hline
\end{tabular}
\end{table}


\begin{table}[!htb]
\begin{tabular}{@{\vspace{-0.3cm}}p{\columnwidth}}
\hline \vspace{-0.3cm} \caption{Algorithm: $vpdf$}\label{alg_vpdf}\\
\hline
    \begin{algorithmic}[1]
    \STATE For the vectorial gradient
    $\varrho^{d}(\mathbf{f_{\lambda}})$, throw $M\times L$ realisations
    of $N$ uniform random germs, i.e. the markers $\{mrk_i\}_{i=1 \ldots
    M \times L}$, with $L$ the channels number. Get the segmentation,
    $\{sg_{i}(x)\}_{i=1 \ldots M \times L}$, by watershed associated to
    the vectorial gradient $\varrho^{d}(\mathbf{f_{\lambda}})$, with
    $d=d_{\chi^2}$ in the image space or $d=d_{E}$ in the factorial
    space. \STATE Obtain the probability density function:
    \begin{equation}\label{eq_vpdf}
    vpdf(x) = \frac{1}{M\times L} \sum_{i=1}^{M \times L} sg_i(x) \ast
    G_{\sigma} \text{ .}
    \end{equation}
    \end{algorithmic}\\
\\ \hline
\end{tabular}
\end{table}

A probabilistic gradient was also defined in
\citep{AnguloJeulin_ISMM_2007} to ponder the enhancement of the
largest regions by the introduction of smallest regions. It is
defined as $\varrho_{prob} = mpdf + \varrho^{d}$: after
normalisation in $[0,1]$ of the weighted marginal pdf $mpdf$ and the
metric-based gradient $\varrho^{d}$.

In order to obtain a partition from the $mpdf$, the $vpdf$ or the
gradient $\varrho_{prob}$, these probabilistic functions are
segmented, for instance by a hierarchical WS with a volume
criterion, as studied in \citet{Noyel_KES_2007}. In such a case, the
goal is not to find all the regions. The stochastic WS addresses the
problem of image segmentation in few pertinent regions according to
a combined criterion of contrast and size. In the present study, as
we discuss below, the segmentation of the pdf is obtained from the
WS with a volume criterion.

\subsection{Conditioning of the germs of the pdf by a previous
classification}


The pdf of contours with uniform random germs contains only spatial
information. By conditioning the random germs by the spectral
classification, we introduce a spatio-spectral pdf. These germs are
going to be regionalized  by a pre-segmentation obtained by a
pre-processing of the spectral classification. Several kinds of
germs have been tested:
\begin{enumerate}
  \item uniform random point germs $mrk_i(x)$ ;
  \item random germs regionalized by a pre-segmentation:
  \begin{description}
    \item[a) ] as point-germs $mrk_i^{\kappa-pt}(x)$ ;
    \item[b) ] as ball-germs where :
    \begin{itemize}
      \item each connected class may be hit one time $mrk_i^{\kappa-b}(x)$ ;
      \item each connected class may be hit several times and the union of balls is made in each connected
      class of the pre-segmentation
            $mrk_i^{\kappa-\cup b}(x)$ ;
      \item  each connected class may be hit several times and the union of connected balls is made in each connected
      class of the pre-segmentation
             $mrk_i^{\kappa-\cup b-conn}(x)$.
    \end{itemize}
  \end{description}
\end{enumerate}

The pdf of contours with uniform random germs is constructed without
any prior information about the spatial/spectral distribution of the
image. Spectral information is introduced in the pdf by conditioning
the germs by the previous transformed classification
$\widehat{\kappa}$. To do this, it is possible to use point germs or
random ball germs whose location is conditioned by the
classification. An exhaustive study of the germs is presented in
\citep{Noyel_PhD_2008,Noyel_ECMI_2008}. Below, we present random
ball germs regionalized by a classification where each connected
class may be hit one time, $mrk_{i}^{\kappa-b}(x)$. For the
detection of tumours in DCE-MRI series we prefer to use the last
kind of regionalized random balls germs $mrk_i^{\kappa-\cup
b-conn}(x)$.

The procedure is as follows: the transformed classification
$\widehat{\kappa}$ is composed of connected classes,
$\widehat{\kappa}=\cup_{k}C_{k}$ with $C_k \cap C_{k'} = \emptyset$,
for $k \ne k'$. The new void class, which appears after the
transformation of $\kappa$, is written $C_{0}$. Then random germs
are drawn conditionally to the connected components $C_{k}$ of the
filtered classification $\widehat{\kappa}$. To do this, the
following rejection method is used: random point germs are uniformly
distributed. If a point germ $m$ falls inside a connected component
$C_{k}$ of minimal area $S$ and not yet marked, then it is kept,
otherwise it is rejected. Therefore not all the germs are kept.
These point germs are called random point germs regionalized by the
classification $\kappa$. However, these regionalized point germs are
sampling all the classes, independently of their prior estimate of
class size/shape. In order to address this limitation, we propose to
use random balls as germs.

The centres of the balls are the random point germs and the radii
$r$ are uniformly distributed between $0$ and a maximum radius
$Rmax$: $\mathcal{U}[1,Rmax]$. At each step, only the intersection,
$\mathrm{B}(m,r) \cap C_k$, between the ball $B(m,r)$ and the
connected component $C_{k}$ is kept as a germ. Then the union is
made with the previous germs. At the end of the ``fall'' of the
random germs, the connected classes are considered as markers for
the watershed used to build the pdf of contours.

These balls are called random balls germs regionalized by the
classification $\kappa$ and noted $mrk_i^{\kappa-\cup b-conn}(x)$.

The algorithm in table \ref{alg_balls_germs_region_U_balls_c}
sketches the process. Note that if $N$ is the number of random germs
to be generated, the effective number of implanted germs is lower
than $N$.

\begin{figure}
\centering
    \begin{tabular}{ccc}
        \includegraphics[width=0.3\columnwidth]{souris_classif_LDA_train_complet_image_parametres}&
        \includegraphics[width=0.3\columnwidth]{serim447_marqueurs_classif_LDA_filtres_color_fond_noir}&
        \includegraphics[width=0.3\columnwidth]{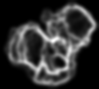}\\
        \footnotesize $\kappa^{LDA,4}_{\ctrain}$ &
        \footnotesize $\widehat{\kappa}^{LDA,4}_{\ctrain}$ &
        \footnotesize $mpdf(\mathbf{p} , mrk^{\kappa})$ \\
        \includegraphics[width=0.3\columnwidth]{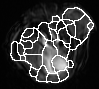}&
        \includegraphics[width=0.3\columnwidth]{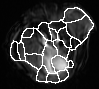}&\\
        \footnotesize $sg^{R-vol}(mpdf)$&
        \footnotesize $sg^{R-vol}(mpdf)$\\
        \footnotesize with $R=30$&
        \footnotesize with $R=20$&
        \\
    \end{tabular}
    \caption{Segmentations by stochastic pdf with a volume criterion in $R$ = 30 (or 20) regions.
    The pdf $mpdf$ is conditioned
    by the transformed classification  $\widehat{\kappa}^{LDA,4}_{\ctrain}$
    on the parameters space $\mathbf{p}$.
    The parameters used to build the $mpdf$ with regionalized
random balls-germs $mrk_i^{\kappa-\cup b-conn}$ are $N$ =
    100 points, $M$ = 100 realisations, area $S$ = 2 pixels, $Rmax$ = 30 pixels.}
    \label{fig:seg:WS_sto_mark_ball_region_U_connex}
\end{figure}

\begin{table}
\begin{tabular}{@{\vspace{-0.3cm}}p{\columnwidth}}
\hline \vspace{-0.3cm} \caption{\small Regionalized random
balls-germs: each connected class may be hit several times and the
union of connected balls is made in each connected class of the
pre-segmentation $mrk_i^{\kappa-\cup b-conn}(x)$}\label{alg_balls_germs_region_U_balls_c}\\
\hline
    \begin{algorithmic}[1]
    \STATE Given $N$ the number of drawn germs $m$, $\{C_k\}$ the
    set of all the connected components of the transform
    classification $\widehat{\kappa}$, $S$ the minimal area of a
    connected component $C_k$, and a boolean array of size equals to
    the number of connected component $C_k$ (the array values are equal to \emph{marked}
 or \emph{not marked})
    \STATE Set the image of germs $mrk_i^{\kappa-\cup b-conn}(x)$
    equals to zero
    \STATE Set the background class and the void class $C_0$ to \emph{marked}
    \STATE Set the class $C_k$ of which the area is lower than $S$ to \emph{marked}
    \FORALL{ drawn germs $m$ from 1 to $N$ }
        \IF{$C_k$, such as $m \in C_k$, is \emph{not marked}}
                \STATE $r = \mathcal{U}[1,Rmax]$
                \STATE $mrk_i^{\kappa-\cup b-conn}(x) = (\mathrm{B}(m,r) \cap C_k) \cup mrk_i^{\kappa-\cup
                b-conn}(x)$
        \ENDIF
    \ENDFOR
    \STATE Label each connected regions in the image of markers
  \end{algorithmic}\\
\\ \hline
\end{tabular}
\end{table}

After computing the marginal pdf of contours $mpdf$ with random
balls germs $mrk_i^{\kappa-\cup b-conn}(x)$, from the representation
of pixels in the parameters space, the pdf are segmented by a
hierarchical WS with a volume criterion (fig.
\ref{fig:seg:WS_sto_mark_ball_region_U_connex}). The classification
used to build the pdf is the LDA classification in the PCA space of
the training set of parameters $\ctrain$.

In \citet{Noyel_ECMI_2008}, the pdf built with these germs seemed to
give better results than others. Therefore, the results shown use
these germs.

In order to better understand the process to build the regionalized
random balls-germs $mrk_i^{\kappa-\cup b-conn}$, in figure
\ref{fig:seg:realisations_germs} some realisations of germs and
their associated contours are presented. For the watershed
segmentation, a morphological gradient is used on each channel of
the PCA space of the training set of parameters. The random
balls-germs are conditioned by the LDA classification in this space.

\begin{figure}
\centering
\begin{tabular}{c}
    \begin{tabular}{@{}ccc@{}}
        \footnotesize $i=1$ &
        \footnotesize $i=5$ &
        \footnotesize $i=9$\\
        \includegraphics[width=0.3\columnwidth]{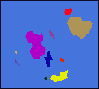}&
        \includegraphics[width=0.3\columnwidth]{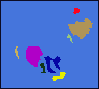}&
        \includegraphics[width=0.3\columnwidth]{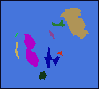}\\

        \includegraphics[width=0.3\columnwidth]{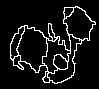}&
        \includegraphics[width=0.3\columnwidth]{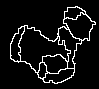}&
        \includegraphics[width=0.3\columnwidth]{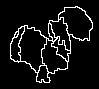}\\
     \end{tabular}\\
        \\
        \includegraphics[width=0.3\columnwidth]{serim447_marqueurs_classif_LDA_filtres_color_fond_noir}\\
        \footnotesize $\widehat{\kappa}^{LDA,4}_{\ctrain}$\\
\end{tabular}
    \caption{Top: some realisations of contours necessary to build the $mpdf$ and,
     second line: the regionalized random ball-germs
$mrk_i^{\kappa-\cup b-conn}(x)$ by the transformed classification
LDA $\widehat{\kappa}^{LDA,4}_{\ctrain}$ (with $N$ = 100 points, $M$
= 100 realisations and $Rmax$ = 30 pixels).}
    \label{fig:seg:realisations_germs}
\end{figure}

\FloatBarrier
\subsection{Validation of the method: application to computer aided detection
of tumours}

After presenting the way to compute the stochastic WS with
regionalized random balls-germs, we are going to apply it to
computer aided detection of tumours on several DCE-MRI series.

In order to detect potentially tumourous areas, the DCE-MRI series
are first segmented by stochastic WS. Then the regions of the
segmentation are classified in potentially tumourous (or not
tumourous) areas. The whole analysis flowchart is presented in
figure \ref{fig:seg:flowchart_seg}. It combines the different parts
introduced earlier in this paper:
\begin{itemize}
  \item pre-processing stage: a noise reduction by FCA and model fitting
  \item training stage of the classifier: the LDA classifier is
  trained on some reference pixels selected on a reference image of
  the parameters.
  \item classification stage: a LDA after normalising the histograms
  of the maps of the parameters which model the spectra. The
  histograms are normalised in order to match the histogram of the
  reference image.
  \item segmentation stage: a stochastic WS with regionalized random-balls
  germs $mrk_i^{\kappa-\cup b-conn}(x)$ conditioned by the
  classification.
\end{itemize}

\begin{figure}[!htb]
\centering
        \includegraphics[width=0.7\columnwidth]{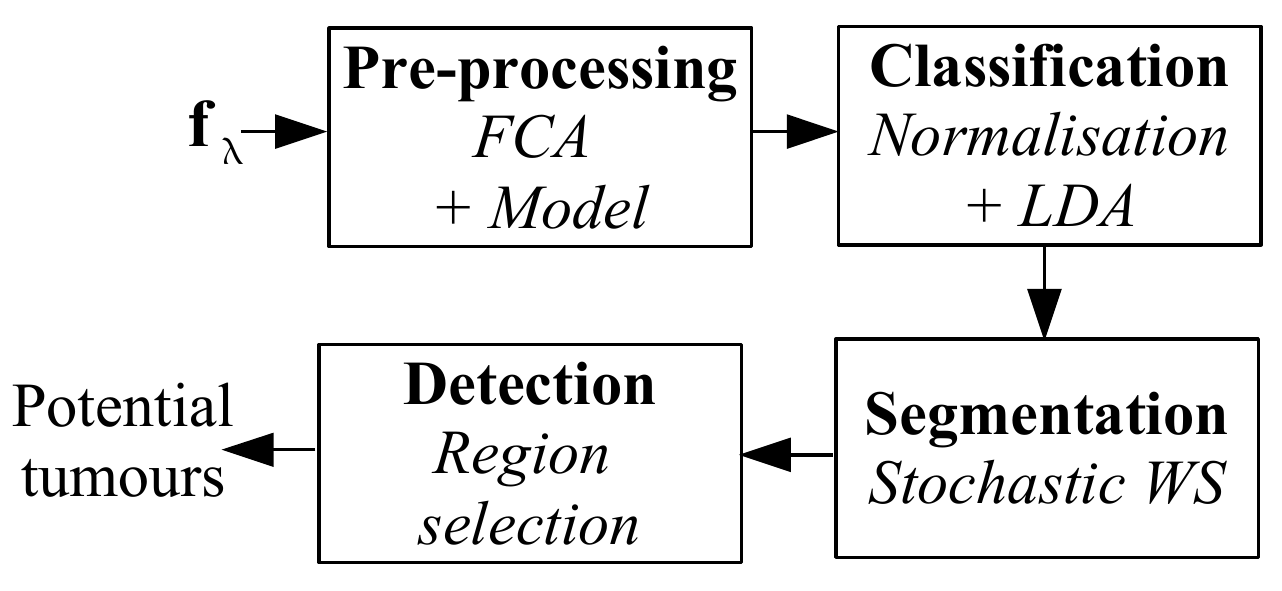}
    \caption{Flowchart of tumour detection.}
    \label{fig:seg:flowchart_seg}
\end{figure}

Starting from the segmentation by stochastic watershed, the
detection of potential tumours depends on two criteria which have
been empirically determined according to a prior knowledge on
DCE-MRI series:
\begin{enumerate}
  \item a positive mean slope parameter $a$ because the contrast
  agent tends to accumulate in these areas during the acquisition
  \item a mean intercept $b$ higher than given a threshold (800) after
  histogram normalisation. With this parameter, the areas of the
  background with a small positive slope are removed from the
  detection.
\end{enumerate}

So that medical doctors can evaluate the pertinence of the detected
zones, confidence maps on the parameters were built. For each zone,
some coefficients of variation were computed. These coefficients
$\beta$ are defined as the ratio between the standard deviation,
$\sigma$, and the mean, $mean$, of the parameters for the considered
region:

\begin{equation}\label{eq:seg:coef_variations}
    \beta_a = \frac{\sigma_a}{E[a]} \text{ \qquad and \qquad } \beta_b =
\frac{\sigma_b}{E[b]}
\end{equation}

Then, the confidence maps were thresholded: for $\beta_a$ at 5 and
for $\beta_b$ at 1. For coefficients  close to zero, the considered
region is more likely classified as cancerous. A look up table is
applied on the confidence maps: in blue is the highest risk ($\beta
= 0$) and in red is the lowest risk ($\beta_a \geq 5$ or $\beta_b
\geq 1$).

The detection $det(x)$ and the confidence maps for the series
``serim447'' are in figure \ref{fig:seg:det_tumour_serim447}. We
notice that for the largest potentially cancerous zone, the risk is
high (in blue). This corresponds to the tumour specified by the
medical doctors. On the other hand, the smallest zone in red, for
which the risk is low, is not a tumour. Therefore, our detection
method works.

\begin{figure}
\centering
\begin{tabular}{@{}c@{}}
    \begin{tabular}{ccc}
        \includegraphics[width=0.3\columnwidth]{souris_classif_LDA_train_complet_image_parametres}&
        \includegraphics[width=0.3\columnwidth]{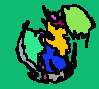}&
        \includegraphics[width=0.3\columnwidth]{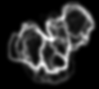}\\
        \footnotesize $\kappa^{LDA,4}_{\mathbf{p}}$ &
        \footnotesize $\widehat{\kappa}^{LDA,4}_{\mathbf{p}} = mrk^{\kappa}$ &
        \footnotesize $mpdf(\mathbf{p} , mrk^{\kappa})$\\

        \includegraphics[width=0.3\columnwidth]{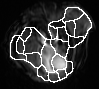}&
        \includegraphics[width=0.3\columnwidth]{serim447_thread_100_verite_small}&
        \includegraphics[width=0.3\columnwidth]{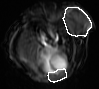}\\
        \footnotesize $sg^{R-vol}(mpdf)$&
        \footnotesize $ref$ &
        \footnotesize $det(x)$\\
    \end{tabular}\\
    \begin{tabular}{@{}c@{ }c@{ }c@{ }c@{}}
        \includegraphics[width=0.3\columnwidth]{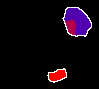}&
        \includegraphics[width=0.06\columnwidth]{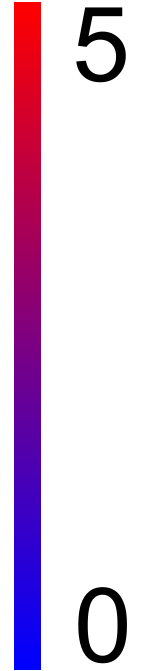}&
        \includegraphics[width=0.3\columnwidth]{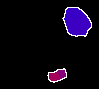}&
        \includegraphics[width=0.06\columnwidth]{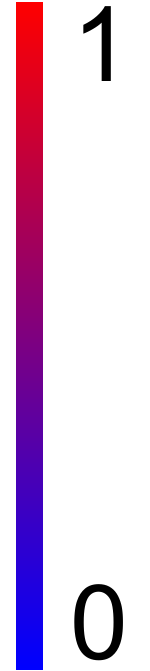}\\
        \footnotesize $\be_a$ &
        \footnotesize &
        \footnotesize $\be_b$ &\\
    \end{tabular}
\end{tabular}
    \caption{Detection of potentially cancerous areas. LDA classification
    in the parameter space
    $\widehat{\kappa}^{LDA,4}_{\mathbf{p}}$, morphological transform
    of the classification used to condition the germs of the pdf $mrk^{\kappa}$,
    $mpdf$ with random balls germs regionalized by the classification
    $mpdf(\mathbf{p} , mrk^{\kappa})$ ($N = 100$ points, $M = 100$
    realisations, area $S = 10$ pixels, $Rmax = 30$ pixels), segmentation by volumic
    WS in $R$ = 20 regions $sg^{R-vol}(mpdf)$, reference $ref$, detection $det(x)$
    of potentially cancerous zones
     ($mean(a) > 0$ and $mean(b) > 800$),
    confidence maps for the slope $\be_a$ and for the intercept $\be_b$.}
    \label{fig:seg:det_tumour_serim447}
\end{figure}

The same approach has been applied to 25 series of images. In the
figure \ref{fig:seg:det_tumours_others} some results are presented
for 6 series. The potentially cancerous detected zones correspond to
the references given by the doctors. In the series \textquotedblleft
serim450\textquotedblright\ and \textquotedblleft
serim457\textquotedblright\ some potentially cancerous zones with a
higher risk are even detected while they were not selected in the
reference. In the others series, similar results have been obtained.
All the tumours marked by clinicians have been detected.

\begin{figure*}
\begin{center}
    \begin{tabular}{@{}c@{ }c@{ }c@{ }c@{ }c@{ }c@{ }c@{}}
        \footnotesize $ref(x)$ &
        \footnotesize $mpdf(x)$ &
        \footnotesize $det(x)$ &
        \footnotesize $\beta_a(x)$&
        \footnotesize &
        \footnotesize $\beta_b(x)$ &
        \footnotesize \\
        \includegraphics[width=0.3\columnwidth]{serim406_thread_100_verite}&
        \includegraphics[width=0.3\columnwidth]{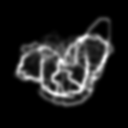}&
        \includegraphics[width=0.3\columnwidth]{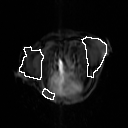}&
        \includegraphics[width=0.3\columnwidth]{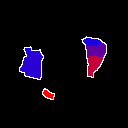}&
        \includegraphics[width=0.067\columnwidth]{echelle_carte_confiance_pente}&
        \includegraphics[width=0.3\columnwidth]{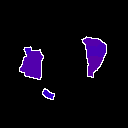}&
        \includegraphics[width=0.067\columnwidth]{echelle_carte_confiance_ordonnee}\\
        \multicolumn{7}{c}{\footnotesize \emph{«~serim406~»}}\\
        \includegraphics[width=0.3\columnwidth]{serim415_thread_100_verite}&
        \includegraphics[width=0.3\columnwidth]{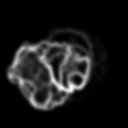}&
        \includegraphics[width=0.3\columnwidth]{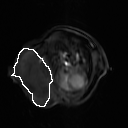}&
        \includegraphics[width=0.3\columnwidth]{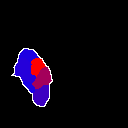}&
        \includegraphics[width=0.067\columnwidth]{echelle_carte_confiance_pente}&
        \includegraphics[width=0.3\columnwidth]{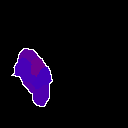}&
        \includegraphics[width=0.067\columnwidth]{echelle_carte_confiance_ordonnee}\\
        \multicolumn{7}{c}{\footnotesize \emph{«~serim415~»}}\\
        \includegraphics[width=0.3\columnwidth]{serim450_thread_100_verite}&
        \includegraphics[width=0.3\columnwidth]{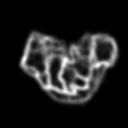}&
        \includegraphics[width=0.3\columnwidth]{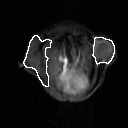}&
        \includegraphics[width=0.3\columnwidth]{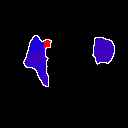}&
        \includegraphics[width=0.067\columnwidth]{echelle_carte_confiance_pente}&
        \includegraphics[width=0.3\columnwidth]{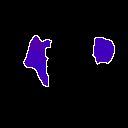}&
        \includegraphics[width=0.067\columnwidth]{echelle_carte_confiance_ordonnee}\\
        \multicolumn{7}{c}{\footnotesize \emph{«~serim450~»}}\\
        \includegraphics[width=0.3\columnwidth]{serim457_thread_100_verite}&
        \includegraphics[width=0.3\columnwidth]{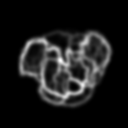}&
        \includegraphics[width=0.3\columnwidth]{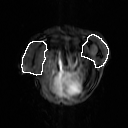}&
        \includegraphics[width=0.3\columnwidth]{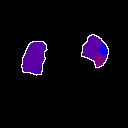}&
        \includegraphics[width=0.067\columnwidth]{echelle_carte_confiance_pente}&
        \includegraphics[width=0.3\columnwidth]{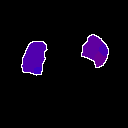}&
        \includegraphics[width=0.067\columnwidth]{echelle_carte_confiance_ordonnee}\\
        \multicolumn{7}{c}{\footnotesize \emph{«~serim457~»}}\\
        \includegraphics[width=0.3\columnwidth]{serim461_thread_100_verite}&
        \includegraphics[width=0.3\columnwidth]{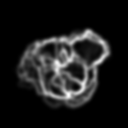}&
        \includegraphics[width=0.3\columnwidth]{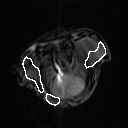}&
        \includegraphics[width=0.3\columnwidth]{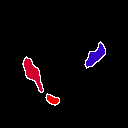}&
        \includegraphics[width=0.067\columnwidth]{echelle_carte_confiance_pente}&
        \includegraphics[width=0.3\columnwidth]{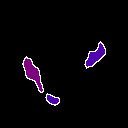}&
        \includegraphics[width=0.067\columnwidth]{echelle_carte_confiance_ordonnee}\\
        \multicolumn{7}{c}{\footnotesize \emph{«~serim461~»}}\\
        \includegraphics[width=0.3\columnwidth]{serim1441_thread_100_verite}&
        \includegraphics[width=0.3\columnwidth]{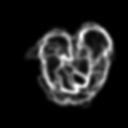}&
        \includegraphics[width=0.3\columnwidth]{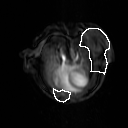}&
        \includegraphics[width=0.3\columnwidth]{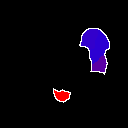}&
        \includegraphics[width=0.067\columnwidth]{echelle_carte_confiance_pente}&
        \includegraphics[width=0.3\columnwidth]{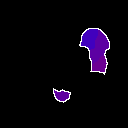}&
        \includegraphics[width=0.067\columnwidth]{echelle_carte_confiance_ordonnee}\\
        \multicolumn{7}{c}{\footnotesize \emph{«~serim1441~»}}\\
    \end{tabular}
    \caption{Detection results: Reference $ref$, marginal pdf $mpdf(\mathbf{p} , mrk^{\kappa^{LDA}})(x)$,
    detection $det(x)$ of potentially cancerous zones
    in the parameters space after histogram normalisation ($mean(a) > 0$ and $mean(b) >
    800$), confidence maps on the slope $\be_a$ and on the intercept $\be_b$.}
    \label{fig:seg:det_tumours_others}
\end{center}
\end{figure*}

\section{Conclusion}

In this paper, an automatic method of detection of potentially
cancerous zones on DCE-MRI series is presented. The results have
been tested on a limited number of images. They are very promising
and in agreement with the references given by medical doctors.

Our method is composed of four stages. In the pre-processing stage,
a dimensionality reduction and a noise reduction are first performed
with the Factor Correspondence Analysis and the pixel-based spectrum
modeling. These operations preserve the spatial contours of the
image. Then in the classification stage Linear Discriminant Analysis
is performed on a subset of training pixels. In the third stage, the
image is segmented by stochastic watershed with random-balls markers
regionalized by the previous classification. The originality of this
approach is the combination of the spatial and temporal information
to produce a ``multivariate gradient'' representing the probability
density function of contours. These probability maps are segmented
by stochastic watershed which is very useful when segmenting the low
contrasted regions corresponding to tumours, since it regularises
the contours. The last stage is a detection of potentially tumourous
zones by statistical criteria. A confidence maps is associated to
the selected zones. These maps show the risk of the regions to be
cancerous.

A method of computer aided detection of potentially cancerous zones
on DCE MRI sequences of small animal has been detailed. It seems to
be very promising for low contrasted data sets. Further systematic
tests should be performed, in order to validate the method on a
larger data sets. In the future, some physical models could be
fitted on the temporal series in place of the actual model.

\FloatBarrier
\appendix

\section{Appendix}
\label{sec:appendix}

\subsection{Experimental conditions}
\label{sec:appendix:exp}

As explained in \cite{Brochot_2006}, here are the details about
experimental conditions: animals used and MRI examination.

\paragraph{Animals}

Experiments were performed on nude nu/nu male mice (Laboratoire Iffa
Credo, L’Arbresle, France), in full compliance with the National
Institutes of Health recommendations for animal care. Approximately
1.5 $\times$ $10^{6}$ PC-3 human tumour cells were implanted
subcutaneously into the flank of each mouse, as described in
\cite{Pradel_2003}. For imaging, the animals were anesthetised by a
peritoneal injection of ketamine (Rompun, Bayer, Leverkusen,
Germany) and xylazine (Imalg\`ene, M\'erial, Lyon, France).

\paragraph{MRI examination}

MRI examination was performed using a 1.5-T system (Sigma, General
Electrics, Milwaukee, WI, USA) and a custom small-animal dedicated
coil. A sagittal 2D T1-weighted spin echo sequence (TE 11 ms, TR 400
ms, FOV 8$\times$8 cm, 256$\times$128 matrix, 1 NEX) was used to
check adequate positioning of the animal and to select the axial
plane level containing the left ventricle cavity and one or two
flank tumours. The dynamic acquisition was performed using a
single-slice T1-weighted 2D fast spoiled gradient recalled (FSPGR)
sequence: TR 15 ms, TE 2.2 ms, flip angle of 608, bandwidth 31.25
kHz, 256$\times$76 matrix for the asymmetric FOV of 7$\times$3 cm, 5
mm slice. The single slice was positioned at the level selected by
the previous sagittal sequence, and dynamic acquisition was
performed with 10 baseline images and after a caudal vein bolus
injection of 0.045 mmol Gd/kg of a macromolecular contrast agent
(Vistarem, Guerbet, Aulnay-Sous-Bois, France).

\subsection{Noise reduction}

In this section, more details are given about our method of noise
reduction based on two sequences of FCA and reconstruction (see
section \ref{sec:filter:DA:noise}). This approach needs the
subtraction of a constant $\varepsilon$ as shown in equations
\ref{eq:filter:DA:FCA2}, \ref{eq:filter:DA:translation} and
\ref{eq:filter:DA:valeur_minimum}.

The idea of subtracting a constant from the data is based on the
fact that after one sequence of FCA-reconstruction some values of
the reconstructed image $\fr^{(1)}$ turn out negative. This is due
to the fact that only a limited number of the factorial axes are
kept for the reconstruction. These negative values, however, have no
physical meaning. They are corrected in the first reconstructed
image. Then a second sequence of FCA-reconstruction is applied
because the data set has been modified. In order to be consistent,
the constant is added after the second sequence of
FCA-reconstruction also.

The data are classified by k-means on the factor pixels of the first
FCA $\mathbf{c}^{\mathbf{f}^{(1)}}_{\al}$ and on the factor pixels
of the second FCA $\mathbf{c}^{\mathbf{f}^{(2)}}_{\al}$. One can
notice, in figure \ref{fig:filter:DA:mouse_kmeans_1FCA_2FCA}, that
the classification is better with two FCA than with one FCA.

\begin{figure}[!htb]
\centering
\begin{tabular}{cc}
   \includegraphics[width=0.3\columnwidth]{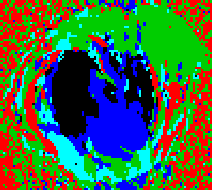}&
   \includegraphics[width=0.3\columnwidth]{serim447_kmeans_5_classes_2AFC}\\
   \footnotesize (a) & \footnotesize (b)
\end{tabular}
   \caption{Classifications \emph{kmeans} into 5 classes in the space of the factor pixels
   (a) of FCA 1, $\mathbf{c}^{\mathbf{f}^{(1)}}_{\al}$, and (b) of
   FCA 2, $\mathbf{c}^{\mathbf{f}^{(2)}}_{\al}$.}
   \label{fig:filter:DA:mouse_kmeans_1FCA_2FCA}
\end{figure}

Starting from the sixteen retained factorial axes of the first FCA
(see section \ref{sec:filter:DA:selection_axes}), the image is
partially reconstructed and a second sequence of FCA-reconstruction
is applied with a spectral translation (equation
\ref{eq:filter:DA:FCA2}).

In order to verify the importance of two sequences of
FCA-reconstruction, the SNR are estimated: on the channels of the
original image $\f$, on the channels of the first reconstructed
image $\fr^{(1)}$ and on the channels of the second reconstructed
image $\fr^{(2)}$. The SNR are also estimated on the factor pixels
of the first FCA $\mathbf{c}^{\mathbf{f}^{(1)}}_{\al}$ and of the
second FCA $\mathbf{c}^{\mathbf{f}^{(2)}}_{\al}$ (fig.
\ref{fig:filter:DA:souris_RSB_im_axes_2AFC}). After performing the
first sequence of FCA-reconstruction, an improvement of the SNR is
noticed in the image space. However, the second sequence of FCA does
not improve the SNR in the image space.

In the factor space, the SNR is improved after the second FCA in
comparison with the SNR, in the factor space, after the first FCA.

Why is it necessary to apply two FCA to filter the noise in the
factor space, while only one FCA is necessary in the image space?
During the reconstruction stage the factor pixels are weighted by
their inertia. Therefore the weight of the noise is reduced because
it appears on the factor pixels with a small inertia. Moreover, the
image reconstruction is based on the product of the marginal
frequencies, $\nu_{i.}\nu_{.j}$, which corresponds to the
reconstruction of the barycentre. This barycentre gives the general
appearance of the image. However, in order to remove the noise on
the factor pixels two FCA are necessary.

\begin{figure}
\begin{center}
\begin{tabular}{c}
  \includegraphics[width=0.7\columnwidth]{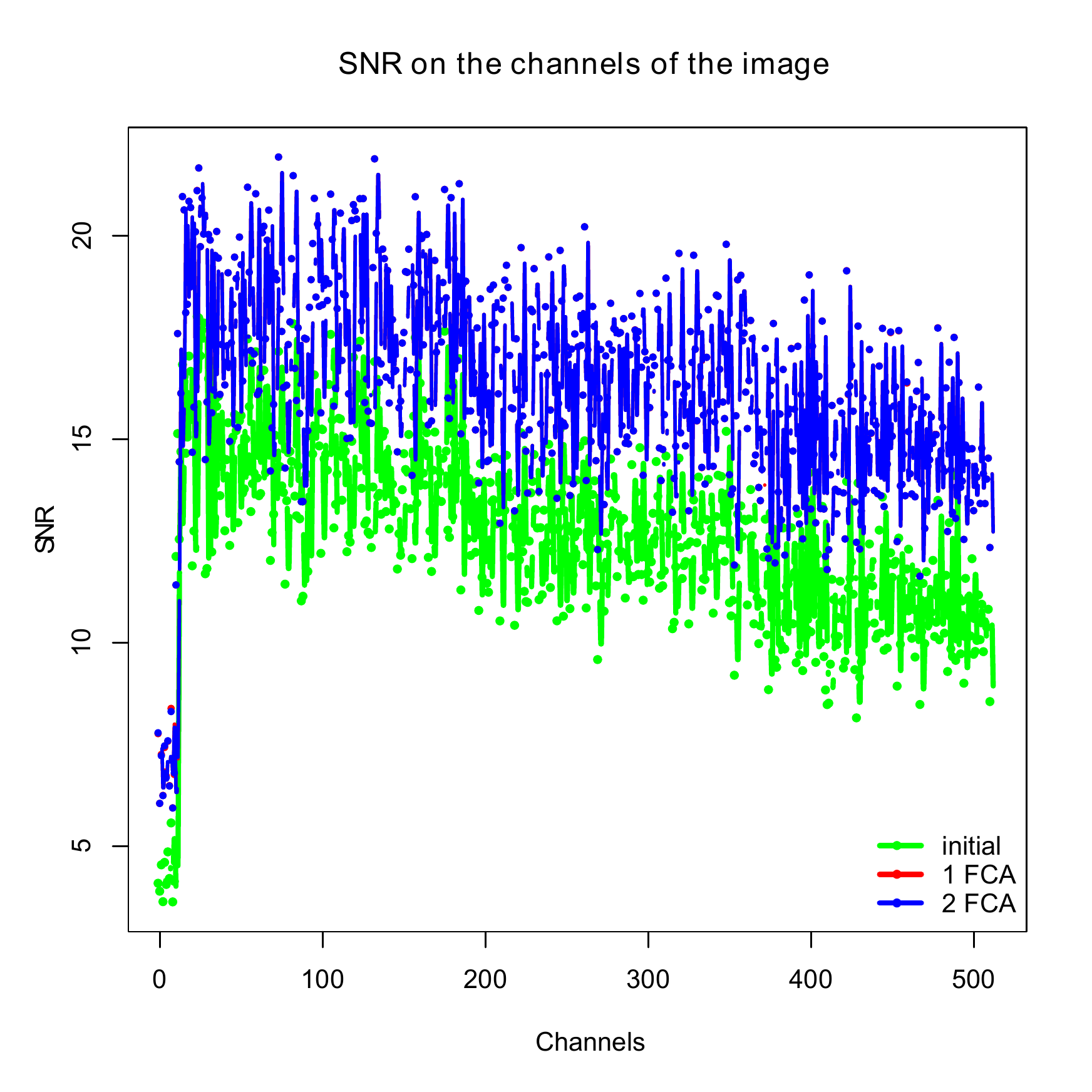}\\
  \footnotesize (a)\\
  \includegraphics[width=0.7\columnwidth]{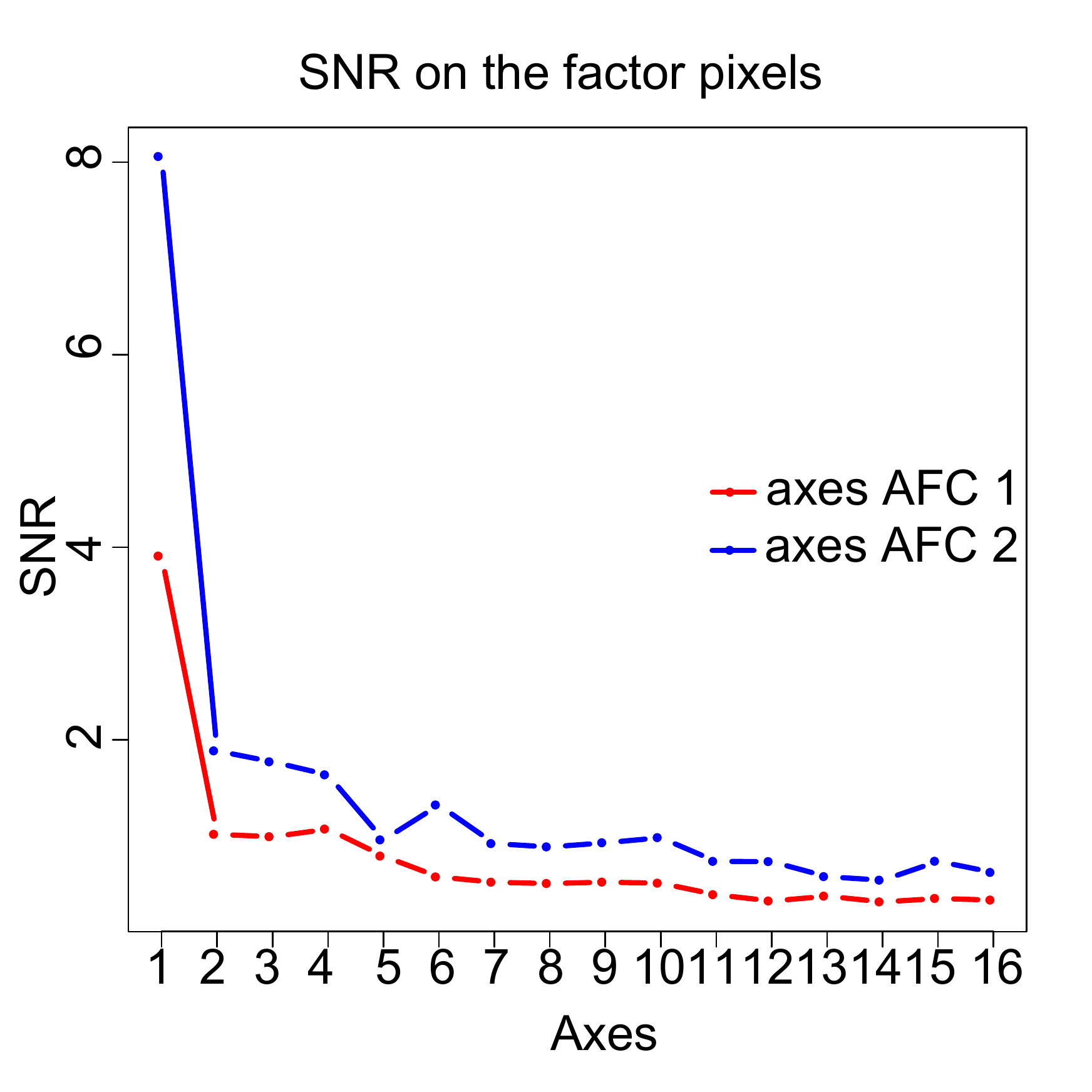}\\
  \footnotesize (b)\\
\end{tabular}
  \caption{(a) SNR on the channels of the image. The red curve and the blue curve are superimposed.
  (b) SNR on the factor pixels for different sequences of FCA-reconstruction.}
  \label{fig:filter:DA:souris_RSB_im_axes_2AFC}
\end{center}
\end{figure}

A hyperspectral signal to noise ratio between the original image
(with noise) and the reconstructed image (filtered) may be defined
as the ratio between the sum of the variance of the signal for each
channel and the sum of the variance of the noise for each channel:
\begin{equation}\label{eq:filter:DA:_SNR_hyper}
  SNR_{hyper}(\f) = \frac{\sum_{j=1}^L var(\widehat{f}_{\la_j})}
  {\sum_{j=1}^L var(f_{\la_j}-\widehat{f}_{\la_j})}
\end{equation}

\bibliography{refs}

\end{paper}
\end{document}